\documentclass[english]{revtex4-2}
\usepackage[T1]{fontenc}
\usepackage[latin9]{inputenc}
\usepackage{geometry}
\usepackage{amsmath}
\usepackage{amsthm}
\usepackage{graphicx}
\usepackage{float}
\usepackage{color}
\usepackage{natbib}

\PassOptionsToPackage{normalem}{ulem}
\usepackage{ulem}
\usepackage[font=small,labelfont=bf,
   justification=justified,
   format=plain]{caption} 
\usepackage[font=small,labelfont=bf,
   justification=justified,
   format=plain]{subcaption}

\makeatletter

\newcommand{\be}{\begin{equation}}
\newcommand{\ee}{\end{equation}}

\def\tb{\textcolor{black}}

\numberwithin{equation}{section}

\makeatother

\usepackage{babel}
\begin{document}
\title{Entrainment in dry and moist thermals}
\author{G. R. Vybhav}
\email{vybhav@jncasr.ac.in}
\affiliation{Engineering Mechanics Unit, Jawaharlal Nehru Centre for Advanced Scientific Research, Jakkur, Bengaluru 560064}
\author{S. Ravichandran}
\email{ravichandran@su.se}
\affiliation{Nordita, KTH Royal Institute of Technology and Stockholm University, Stockholm SE 10691}

\begin{abstract}
We study entrainment in dry thermals in neutrally and unstably
stratified ambients, and moist thermals in dry-neutrally stratified ambients
using direct numerical simulations (DNS). We find, in 
agreement with results of Lecoanet and Jeevanjee \tb{[J. Atmos. Sci. 76(12), 3785-3801, (2019)]}
that turbulence plays a minor role in entrainment in dry thermals in a neutral ambient for Reynolds numbers {$Re \lesssim 10^4$}.
We then show that the net entrainment rate increases when the
buoyancy of the thermals increases, either by condensation heating or because of an
unstably stratified ambient. This is in contrast with the findings of Morrison et al. \tb{[J. Atmos. Sci. 78(3), 797-816, (2021)].}
We also show that the role of turbulence is greater in
these cases than in dry thermals and, significantly, that the combined action of
condensation heating and turbulence creates intense small scale vorticity,
destroying the {coherent} vortex ring that is seen in dry and moist laminar thermals. These
findings suggest that fully resolved simulations at Reynolds numbers significantly
larger than the mixing transition Reynolds number $Re=10^4$ are necessary
to understand the role of turbulence in the entrainment in growing cumulus clouds,
which consist of a  series of thermals rising and decaying in succession.
\end{abstract}
\maketitle

\section{Introduction} \label{sec:intro}

General circulation models (GCMs) that are used to predict the weather and
climate solve the  Navier-Stokes equations for the evolution of the 
properties of the Earth's atmosphere. These calculations are computationally
intensive, especially at high resolutions. With present computational resources, 
the resolutions that can be achieved in GCMs are such that 
neighbouring horizontal grid-points are separated by tens if not hundreds 
of kilometres \cite{Rio2019}. While {approaches} like 
superparameterisation \cite{Randall2003,Grabowski2004,Majda2007},
as a step towards global cloud resolving models that would obviate cumulus 
parameterisation, have been proposed, these are too computationally 
expensive at present. Individual clouds and most cloud systems, therefore, 
exist within a single grid box in current GCM simulations, and the processes
occurring in clouds have to be parameterised. \\

Among the processes in clouds that need to be parameterised, the mixing of the fluid
in the rising cloud with the ambient atmospheric air is the most important, since 
this controls the amounts of water substance in the cloud. The mixing process by
which ambient fluid becomes part of the cloud flow is termed `entrainment', and the 
process by which fluid leaves the cloud flow is termed `detrainment'.
The parameterisation of entrainment in growing cumulus clouds has 
proved to be among the hardest problems to tackle in the general circulation 
models used today, since model outputs (e.g. the climate impact or regional 
rainfall patterns) depend sensitively on the entrainment parameterisation \cite{deRooy2013}.\\

The most common approach adopted in parameterising entrainment in cumulus 
clouds is to model (ensembles of) these clouds as one of a few types of basic
flow \cite{deRooy2013}, typically plumes or thermals. (A review of other more 
recent approaches to modelling cumulus entrainment may be found in 
\cite{Rio2019}.)
A steady state plume model for an ensemble of cumulus clouds is the most
common model used in GCMs today \cite{Arakawa2004,deRooy2013,Rio2019}, despite
the evidence from airborne measurements as well as large eddy
simulations (LES) that cumulus clouds more often resemble a series of thermals \cite{Zhao2005a,Zhao2005b,Damiani2006,Romps2015a,Romps2015b,HernandezDeckers2016,HernandezDeckers2018,Moser2017,Morrison2020,Peters2020}. {(See also Ref. \cite{Romps2021} where observations and LES are found to disagree about whether shallow cumulus clouds show predominantly `bubble-like' or `plume-like' behaviour.)} Cumulus convection
can therefore be thought of as a succession of thermals, each rising further
than its predecessor, entraining and mixing with the ambient air, and 
decaying. Understanding the fluid dynamics of thermals, therefore, may help 
improve the modelling of cumulus entrainment.\\

Laboratory experiments of (dry) thermals have been undertaken
beginning in the 1950s \cite{Morton1956,Scorer1957,Turner1962,Simpson1969,Johari1992}, and have 
informed the understanding of entrainment. Studies using direct numerical 
simulations (DNS) of the entrainment in dry thermals are relative rare 
\citep[e.g.][hereafter LJ19]{Lecoanet2019} and \cite{McKim2020}. LJ19 show from DNS
of laminar and turbulent dry thermals at two Reynolds numbers differing by a factor
of $10$ that the entrainment coefficients only differ by about $20\%$. They argue, following \cite{Scorer1957} that entrainment in dry thermals is driven not by
turbulence but by buoyancy. \\

The assumptions that enable the analysis in
\cite{Lecoanet2019,McKim2020} break down (see section 
\ref{sec:theory}) when the thermals 
are influenced not only by their initial buoyancy but also by off-source 
heating by condensation, as occurs in atmospheric clouds. The entrainment in 
such thermals naturally occurring in a moist convecting atmosphere has been studied 
in cloud-resolving simulations and LES 
\citep[e.g.][]{Romps2010a,Romps2015a,HernandezDeckers2016,HernandezDeckers2018,Morrison2020, Peters2020}. Morrison et al. \citep[][]{Morrison2021} study individual
moist thermals using axisymmetric and 3D large eddy simulations (LES). In the 
initial few diameters from release, the authors find, moist thermals entrain 
less than dry thermals. After the initial stages, the entrainment rates in 
moist and dry thermals are indistinguishable. These results, as the authors
note, are in contrast to earlier findings by the same authors and others
that moist thermals do not grow in size \cite{HernandezDeckers2016,Romps2015a}.
We also note that the axisymmetric and LES results in \cite{Morrison2021}
seem to disagree about whether moist entrainment is the same as or
different from dry entrainment in the initial stages (see, e.g. their
figure 13 compared to their figures 3 and 5), with (noisy) LES results
finding no difference between dry and moist thermals.\\

Here, we aim to definitively answer the question of whether moist thermals
indeed entrain ambient fluid at rates different from dry thermals.
To this end, we perform direct numerical simulations (DNS) resolving the smallest
scales of the flow in thermals rising, studying cases with moist thermodynamics
as well as with unstable ambient stratification. \\

The rest of the paper is organised as follows. In section \ref{sec:setup}, we 
describe the geometry of the problem we study, and write down the governing 
equations in nondimensional form. We then list the controlling parameters and the
initial and boundary conditions.  We also briefly describe the numerical solver used
to perform the DNS. In section \ref{sec:theory}, we outline the theoretical 
arguments of LJ19 and list the cases we study to examine when these arguments break
down. In section \ref{sec:results}, we present results from the cases listed and
discuss these results in light of earlier results in the literature. We conclude in 
section \ref{sec:conclusion}.

\section{Governing Equations and Numerical Method}\label{sec:setup}

Our domain is a three-dimensional cuboidal volume with dimensions $(L_x,L_y,L_z)$ in the three space directions. The horizontal directions are $x$ and $y$, and gravity (of constant magnitude $g$) points in the $-z$ direction. We make the Boussinesq approximation \cite{Spiegel1960}, so that the fluid velocity $\mathbf{u}$ is incompressible. The ambient temperature $T_\infty$ is a function of the height $z$, and water vapour at a (constant) relative humidity $s_\infty$ exists in the ambient; since the saturation vapour pressure is a steeply varying function of the temperature, the ambient vapour mixing ratio $r_{v,\infty}$ decreases rapidly with height.

The governing equations are the Navier-Stokes equations subject to the incompressibility condition, and advection-diffusion equations for the temperature and vapour and liquid mixing ratios. We nondimensionalise these equations using the initial temperature anomaly $\Delta T$ as the scale for temperature differences, the diameter of the thermal $b_0$ as the length scale, the buoyancy velocity $U_b = \left( b_0 g \Delta T / T_0 \right)^{1/2}$ as the velocity scale, and the saturation mixing ratio {$\tilde{r}_{s,0} = \tilde{r}_s(T_0)$} at the temperature $T_0$  as the scale for water quantities. The nondimensional equations become
\begin{eqnarray}
\frac{D\mathbf{u}}{Dt} &  =  &  -\nabla p + \frac{1}{Re} \nabla^2 \textbf{u} + B \hat{e}_z, \label{eq:momentum}\\
\nabla \cdot \textbf{u} & = & 0, \label{eq:continuity}\\
\frac{D \theta}{Dt} &  = & \frac{1}{Re} \nabla^2 \theta + L_2 C_d + w \left(\Gamma_0 - \Gamma_u \right), \label{eq:temperature} \\
\frac{D r_v}{Dt} &  = &\frac{1}{Re} \nabla^2 r_v - C_d, \label{eq:vapour}\\
\frac{D r_l}{Dt} &  = & \frac{1}{Re} \nabla^2 r_l + C_d, \label{eq:liquid}
\end{eqnarray}
where $\textbf{u} = (u,v,w)$ is the fluid velocity, $p$ is the dynamic pressure, 
\be
{\theta = \frac{T - T_\infty}{\Delta T} \label{eq:theta_def}}
\ee
is the temperature and the {normalised} vapour and liquid mixing ratios are {$r_v = \tilde{r}_v/\tilde{r}_{s,0}$ and $r_l=\tilde{r}_l/\tilde{r}_{s,0}$} respectively. {Quantities without the $\tilde{}$ are $O(1)$, since they are normalised using $\tilde{r}_{s,0}$.}
Similar equations have been used in
\citep[e.g.][]{hernandez2013minimal,Ravichandran2020b_mammatus}.
The nondimensional Reynolds number is $Re$ (defined in equation \ref{eq:Reynolds}), and the Prandtl and Schmidt numbers are implicitly assumed to be
equal to unity. The nondimensional dry adiabatic lapse rate is $\Gamma_u$ and
the lapse rate in the ambient is $\Gamma_0$ (see table \ref{tab:list_params}). The thermodynamic constant $L_2$ is defined in equation \ref{eq:L1_L2}.
Thus, the buoyancy 
\begin{equation}
    B = \theta + r_0 \left( \chi (r_v - r_{v,\infty} ) - r_l \right), \label{eq:buoyancy}
\end{equation}
where $r_0=\tilde{r}_{s,0}/\left(\Delta T / T_0 \right)$ is a density ratio and $\chi = M_a/M_w - 1$ is derived from the ratio of molecular masses of air ($M_a$) and water vapour ($M_w$); the nondimensional condensation rate
\begin{equation}
    C_d = \frac{1}{\tau_s} \left( \frac{r_v}{r_s} - 1 \right), \label{eq:condensation}
\end{equation}
where the local {normalised} saturation vapour mixing ratio $r_s \equiv \tilde{r}_s / \tilde{r}_{s,0} = \textrm{exp}\left(L_1 ( \theta - \Gamma_0 z ) \right)$, and $L_1$ is a constant defined below; and the buoyancy and condensation rate are both functions of 
the local values of $\left(\theta,r_v,r_l\right)$. The timescale for condensation or
evaporation, $\tau_s$, is a function of droplet size and liquid water mixing ratio,
and, when $\tau_s$ is large, variations of $\tau_s$ can be imporant \citep[see, e.g.][]{Ravichandran2020b_mammatus}. Here, we assume that the droplets making up
the liquid content $r_l$ are very small, so that $\tau_s \ll 1$ and the system is 
always  close  to equilibrium. 
The governing parameters are the Reynolds number 
\be
    Re = \frac{U_b b}{\nu}, \label{eq:Reynolds}
\ee
and the ambient lapse rate $\Gamma_0$ and the ambient relative humidity $s_\infty$. The thermodynamic constants
\be
    L_1 = \frac{L_v \Delta T}{R_v T_0^2} ; \textrm{and}\ \ \ \ \ \ \ \ \ L_2 = \frac{L_v {\tilde{r}_{s,0}}}{C_p \Delta T} \label{eq:L1_L2} 
\ee
and the saturation mixing ratio {$\tilde{r}_{s,0}$}
are fixed when the {temperature $T_0$ at the initial height $z_0$} and the temperature scale $\Delta T$ are 
chosen. We choose $T_0=300$K and $\Delta T = 10$K, 
giving  $L_1 \approx 0.58$ and $L_2 \approx 4.8$ and the saturation mixing ratio ${\tilde{r}_{s,0}} = 0.02$. We also choose a length scale $b=100$m, such that the
nondimensional dry adiabatic lapse rate $\Gamma_u = 0.098$. These values are typical of cumulus clouds in the tropics.
Note that this means that the viscosity in our problem (equation \ref{eq:Reynolds}) is artificially large. We will, however, only report our results in nondimensional terms. We study the evolution of dry thermals in {dry-unstably} stratified and {moist thermals in dry-neutrally (i.e. moist-unstably) stratified} ambients. A list of the cases studied is given in table \ref{tab:list_params}.

\begin{table}
\begin{centering}
\begin{tabular}{|c|c|c|c|c|c|c|}
\hline 
Case & $L_{x}\times L_{y}\times L_{z}$ & $N_{x},N_{y},N_{z}$ & $dt$ & $Re$ & $s_{\infty}$ & $\Gamma_{u}-\Gamma_{0}$\tabularnewline
\hline 
\hline 
DNL & {$14^{2}\times28$} & $384^{2}\times768$ & $0.002$ & $630$ & $0$ & $0$\tabularnewline
\hline 
DNT & {$14^{2}\times28$} & $768^{2}\times1536$ & $0.002$ & $6300$ & $0$ & $0$\tabularnewline
\hline 
DNT12 & {$14^{2}\times28$} & $1024^{2}\times2048$ & $0.001$ & $12600$ & $0$ & $0$\tabularnewline
\hline 
MNL & {$14^{2}\times28$} & $768^{2}\times1536$ & $0.002$ & $630$ & $0.8$ & $0$\tabularnewline
\hline 
MNT & {$14^{2}\times28$} & $1024^{2}\times2048$ & $0.0015$ & $6300$ & $0.8$ & $0$\tabularnewline
\hline 
DUL & $12^{2}\times24$ & $768^{2}\times1536$ & $0.002$ & $630$ & $0$ & $-0.02$\tabularnewline
\hline 
DUT & $12^{2}\times24$ & $1024^{2}\times2048$ & $0.0015$ & $6300$ & $0$ & $-0.02$\tabularnewline
\hline 
\end{tabular}
\par\end{centering}
\caption{\label{tab:list_params} List of cases presented, along with the domain size, resolution, and time step used in the simulations. These values are all nondimensional (see text). Note that the term $\Gamma_u - \Gamma_0$ is the nondimensional Brunt-V\"ais\"ala frequency. The Prandtl ($Pr$) and Schmidt ($Sc$) numbers are taken to be unity (close to atmospheric values). The grid resolutions chosen resolve the smallest scales for the respective cases. We run all the simulations until the thermal hits the boundary at $z=L_z$. The letters in the abbreviations stand for D: dry, M: moist, N:neutral, U:unstable, L:laminar and T:turbulent, with the $12$ denoting $Re=12600$.}
\end{table}

Equations \ref{eq:momentum} - \ref{eq:liquid} are solved using the finite volume 
solver \emph{Megha-5}, which uses second-order central differences in space and a 
second-order Adams-Bashforth timestepping scheme. Simple open flow boundary 
conditions \cite{orlanski1976simple} are imposed at the boundaries at $x=\pm L_x/2$,
$y=\pm L_y / 2$ and $z=L_z$, while the lower boundary at $z=0$ is no-slip. The 
solver has been extensively validated and used in studies of free-shear flows, flows
with moist thermodynamics, and combinations thereof \cite{Ravichandran2020a_cumulus,
Ravichandran2020b_mammatus, Diwan2020, Singhal2021, Ravichandran2021_asperitas}, and
we refer readers to these earlier published works for details of the implementation.

\subsection{Initial Conditions} \label{sec:initial_conds}
The thermal is initialised as a spherical patch with $\theta=1$  and {$r_v=s_\infty r_s(\theta=0,z)$,} so that the temperature in the thermals is greater than the local ambient
temperature, and the vapour mixing ratio equal to the ambient value.
The initial velocity is zero. We set $z_0=1.5$ in all cases reported
here; the choice of $z_0$ does not affect our results.
In both the laminar simulations at $Re=630$ and the turbulent simulations 
$(Re=6300)$, noise is added to the initially spherical patch. We find, as in LJ19, that simulations with different instantiations of noise can produce significantly different results. Our results are therefore ensemble-averages over $5$ runs at the same parameters for the turbulent simulations; for the laminar simulations where the effects of noise are smaller, we do not perform ensemble averages. Larger ensembles produced similar results, and we have verified that the type of noise used does not affect the results: Gaussian and uniformly distributed white noise lead to the same average results.

\subsection{Tracking of thermals} \label{sec:tracking}
\begin{figure}
    \centering
    \includegraphics[width=0.3\columnwidth]{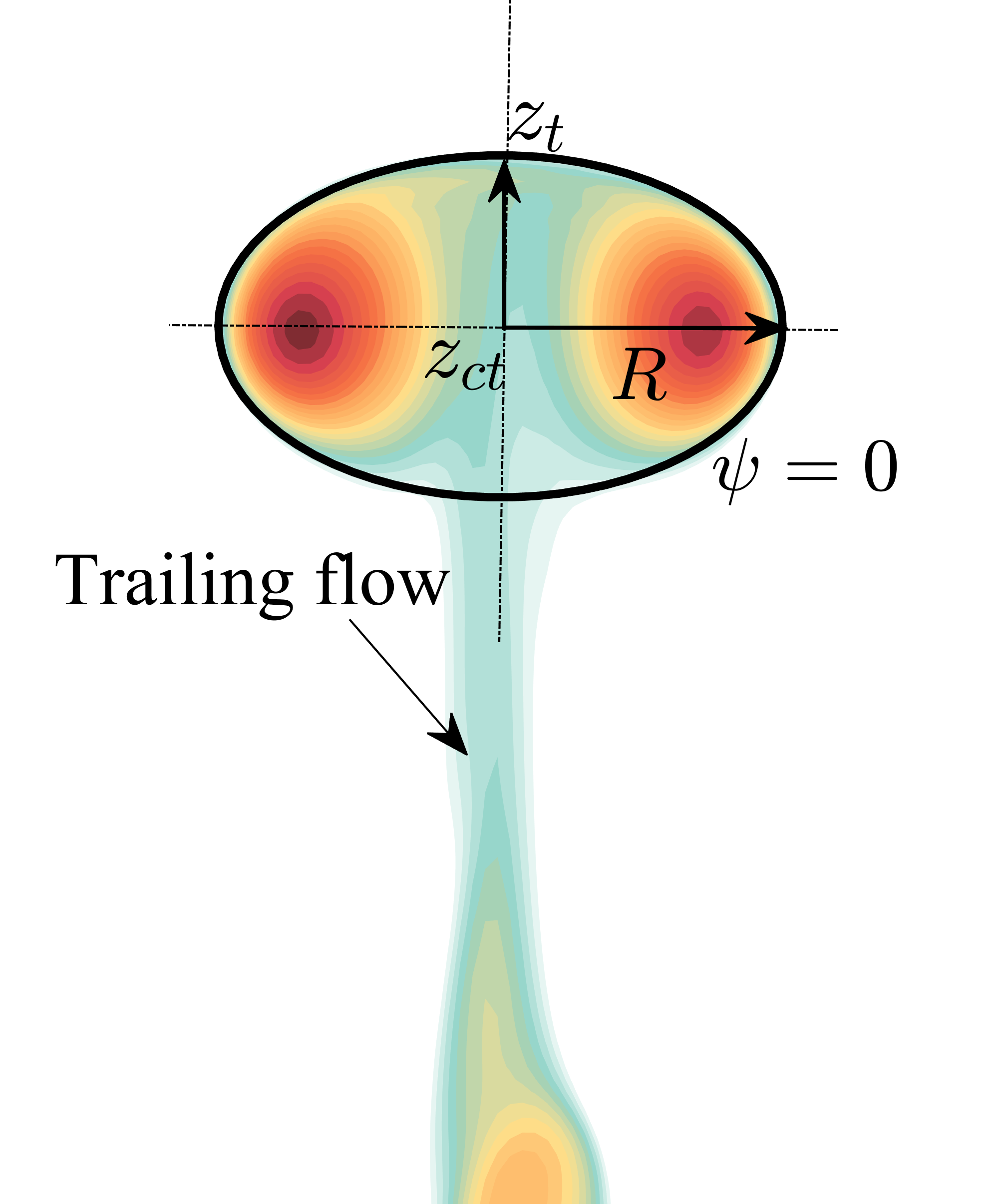}
    \caption{A typical (dry) thermal in our simulations, with the filled
    contours representing a typical distribution of the temperature $\theta$ {coloured logarithmically}. The radius of the thermal $R$ is defined as the distance measured from the axis of symmetry ($r=0$) to the widest location of the curve $\psi=0$. The location of the centroid of the thermal $z_{ct}$ is defined in Appendix A. The  vertical locations of maximum radius and centroid of the thermal do not, in general, coincide. The thermal top height ($z_t$) is defined as the location of the steepest gradient of the azimuthally averaged temperature. The volume of the thermal is taken to be the volume bounded by the dividing streamline $\psi=0$ in a frame of reference moving with the velocity $w_{th}$ of the thermal (see section \ref{sec:tracking} in the text). The maximum flow velocity in the region identified as the thermal is labelled $w_{max}$. The trailing flow suggests detrainment, which is small for dry thermals (see LJ19), but larger for moist thermals. }
    \label{fig:schematic}
\end{figure}

The \emph{net} entrainment rate is calculated from the rate of change of the volume 
$V$ of the thermal,
\be
\epsilon_{net} = \frac{1}{V} \frac{dV}{dz} \label{eq:entrainment_rate}.
\ee
For a dry spherical thermal, $\epsilon_{net}$ can be written down analytically 
{ \citep[see][]{McKim2020}}. In order to find this rate numerically for a
thermal of arbitrary shape, the volume of the thermal has to be consistently 
defined. Several methods have been used in the literature. In our 
simulations, we track the thermals as in, e.g., 
\cite{Romps2015a,Lecoanet2019,Morrison2021}, as described presently.

We first calculate the velocity at which the thermal rises vertically by calculating
the time derivative of the location of the centroid $z_{ct}$ of the thermal. Then, 
in a frame of reference moving with the thermal, we compute the azimuthally averaged
flow velocity and thus the streamfunction $\psi$ of the flow. The volume bounded by 
the dividing streamline $\psi=0$ is 
assumed to be the volume of the thermal, as shown in figure 
\ref{fig:schematic}. The \emph{net} entrainment rate is then 
calculated using equation \ref{eq:entrainment_rate}. These calculations are 
detailed in Appendix A.  \\

Since the entrainment rate thus defined is expected to vary as $1/R$, we also define the entrainment efficiency
\be
e = \epsilon_{net} R \label{eq:entrainment_efficiency}
\ee

\section{Theoretical Preliminaries} \label{sec:theory}

Since there are several governing nondimensional parameters and combinations of 
ambient conditions, we outline our approach to studying entrainment in 
thermals, and our reasons for studying the specific cases reported
in section \ref{sec:results}.\\

For completeness, we briefly recount the arguments of 
LJ19 following {Turner \cite{Turner1957}}. Buoyant vortex rings are known to expand in
a flow and, since thermals `spin up' into vortex rings, the impulse
\begin{equation}
{I=\frac{1}{2}\rho\int \mathbf{r}\times\mathbf{\omega}\ dV}\label{eq:impulse_general}
\end{equation}
of a dry thermal may be written as the impulse of a vortex ring with circulation
\begin{equation}
{\Gamma= \int \omega_\phi dr dz,}\label{eq:circulation}
\end{equation}
{where the integral is over the cross-sectional area of the vortex ring,} and radius $R$, giving
\begin{equation}
I=\pi\rho\Gamma R^{2},\label{eq:impulse_thermal}
\end{equation}
where $\rho$ is the fluid density. {Note that the symbol $\Gamma$ (no subscript) is used for the circulation, and $\Gamma_{0,d}$ are used for the nondimensional (constant, imposed) ambient lapse rates.}
Since the impulse of a flow can only increase by the action of an external force, here the buoyancy of the thermal, this impulse increases with time at a rate
\be
\frac{dI}{dt} = \pi\rho R^2 \frac{d\Gamma}{dt} + \pi R^2 \Gamma \frac{d\rho}{dt} +  \pi\rho\Gamma \frac{dR^2}{dt} = F,\label{eq:dIdt}
\ee
where $F$ is the volume-integrated buoyancy force on the 
thermal. In dry thermals, $F$ is {nearly constant since detrainment is about two orders of magnitude smaller than entrainment}. Furthermore, since $\Gamma$ is also constant for 
vortex rings, and $\rho$ is constant (under the Boussinesq approximation in a 
neutrally stratified ambient), the $d{R^2}/dt$ term alone has to balance the
buoyancy force term on the right hand side and we have
\be
\pi \Gamma \rho \frac{dR^2}{dt} = F . \label{eq:dry_thermal_entrainment}
\ee
Since the preceeding arguments are independent of the level of turbulence,
entrainment in dry thermals cannot be turbulent in nature. As mentioned in
section \ref{sec:intro}, LJ19 show that increasing the Reynolds number from 
$Re=630$ to $Re=6300$ changes the entrainment rate by only about $20\%$.
Further evidence for the role of buoyancy in entrainment in dry thermals is 
found in \cite{McKim2020} where switching off buoyancy also 
reduces the entrainment rate to about $1/3$rd of its value.\\

Clearly, this analysis breaks down \citep[see, e.g., the arguments
in][]{Morrison2021} if either of the first two terms in the expansion of $dI/dt$ in 
equation \ref{eq:dIdt} is nonzero. This can occur \\
\\
(i) by the off-source heating of the thermal by condensation which (a) generates 
vorticity inside the thermal so the first term is nonzero; and (b) changes the  
density of the thermal, so the second term can be nonzero. \\
\\
(ii) by detrainment which in dry thermals is very small compared to entrainment,
but could be  significant in moist thermals, thus changing the density $\rho$ of the thermal. \\
\\
(iii) when the (effective) radius of the vortex core is no longer small compared to 
the radius $R$ of the vortex ring, so that the {definition of the impulse 
equation \ref{eq:impulse_thermal}  in terms of the vortex strength $\Gamma$ (eq. \ref{eq:circulation}) is no longer accurate} \citep[see, e.g., the discussion in ][]{McKim2020}.  LJ19 comment that in
their simulations at the higher $Re=6300$, this requirements are not met as well as 
in their laminar simulations. This is also evident in our turbulent simulations at 
$Re=6300$, as discussed in section \ref{sec:dry_unstrat}, where the thermals appear
to become more elongated. 

{The terms in Eq. \ref{eq:dIdt} are explicitly computed in section \ref{sec:entrainment}. We note, however, that under the Boussinesq approximation which we make here, the contributions due to changing density cannot strictly be computed. Density changes with altitude can be accommodated using the anelastic approximation \cite{anders2019}, while fully compressible simulations (such as in \cite{Morrison2021}) are needed to account for local density changes}.\\

To test possibilities (i) and (ii), we study the
evolution of moist thermals in a (dry-) neutrally stratified ambient (section 
\ref{sec:moist_unstrat}) and the evolution of a dry thermal in an dry-unstably 
stratified ambient (section \ref{sec:dry_unstable}). Regarding the role of 
turbulence (iii), we note that in general, shear-dominated
(buoyancy or momentum-driven) flows are known to undergo a transition to a higher
mixing rate for a large scale Reynolds number $Re\geq \mathcal{O}(10^{4})$
\citep[][]{DIMOTAKIS2000}, and the Reynolds number in LJ19 is smaller
than this mixing transition Reynolds number. If the assumption  that the vortex 
core radius is small no longer holds for sufficiently large Reynolds numbers, 
turbulence could play a greater role in the entrainment. We test this notion by
studying a dry thermal at a Reynolds number of $Re = 12,600 > 10^{4}$ (section \ref{sec:high_Re}).
    
{A possible source of error in this exercise is the inconsistent definition of the radius $R$. LJ19 (consistently) define $R$ to be the radius of the \emph{vortex ring} and use this in their calculations; whereas Morrison et al. \citep{Morrison2021}, while defining $R$ to be radius of the vortex ring, use the radius of the thermal in their calculations. We examine the results of this switch in \ref{sec:entrainment} (see Fig. \ref{fig:impulse_budget})}
\\

\section{Results and Discussion} \label{sec:results}

\subsection{Dry thermals {in a dry-neutral ambient}} \label{sec:dry_unstrat}

We begin with results from simulations of dry thermals in a neutral ambient. Our 
results are in good agreement with those of  \citep[][]{Lecoanet2019}, and therefore
serve both as validation of our methods of numerical analysis as well as
independent verification of the results of LJ19.

\begin{figure}
     \centering
     \begin{subfigure}[b]{0.32\textwidth}
         \centering
         \includegraphics[width=\textwidth]{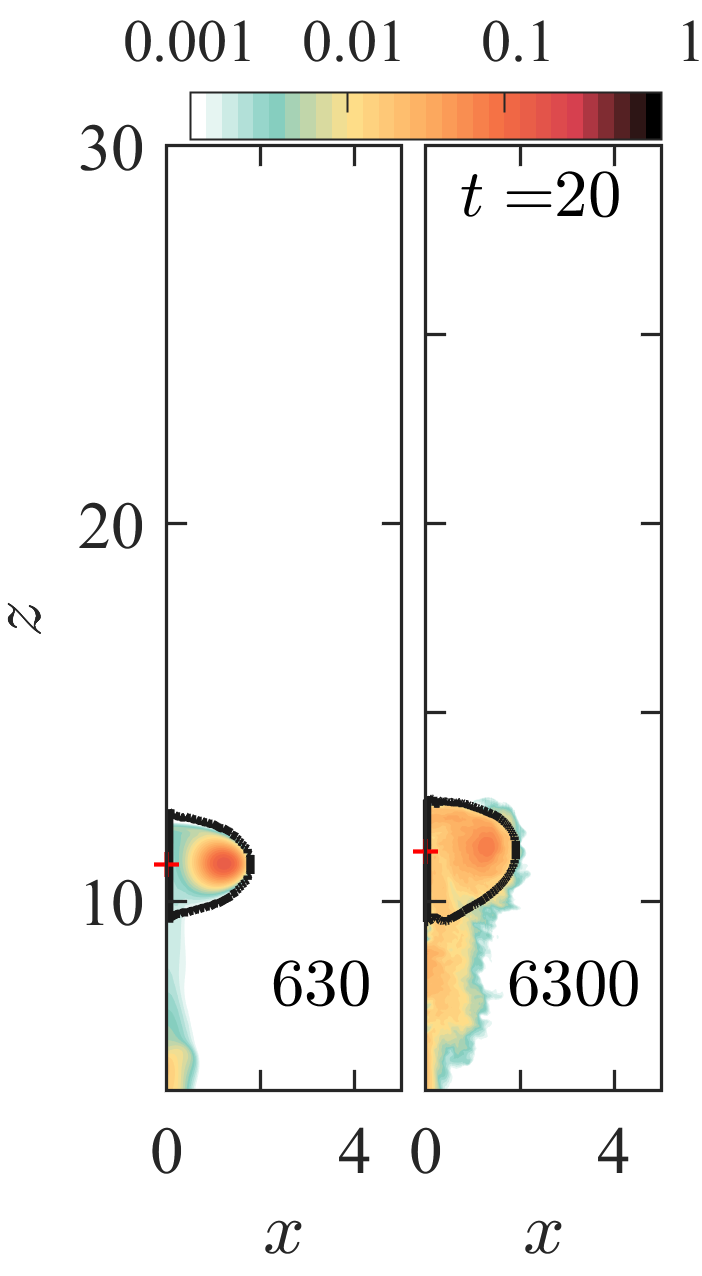}
     \end{subfigure}
     \hspace{0.1cm}
        \begin{subfigure}[b]{0.32\textwidth}
         \centering
         \includegraphics[width=\textwidth]{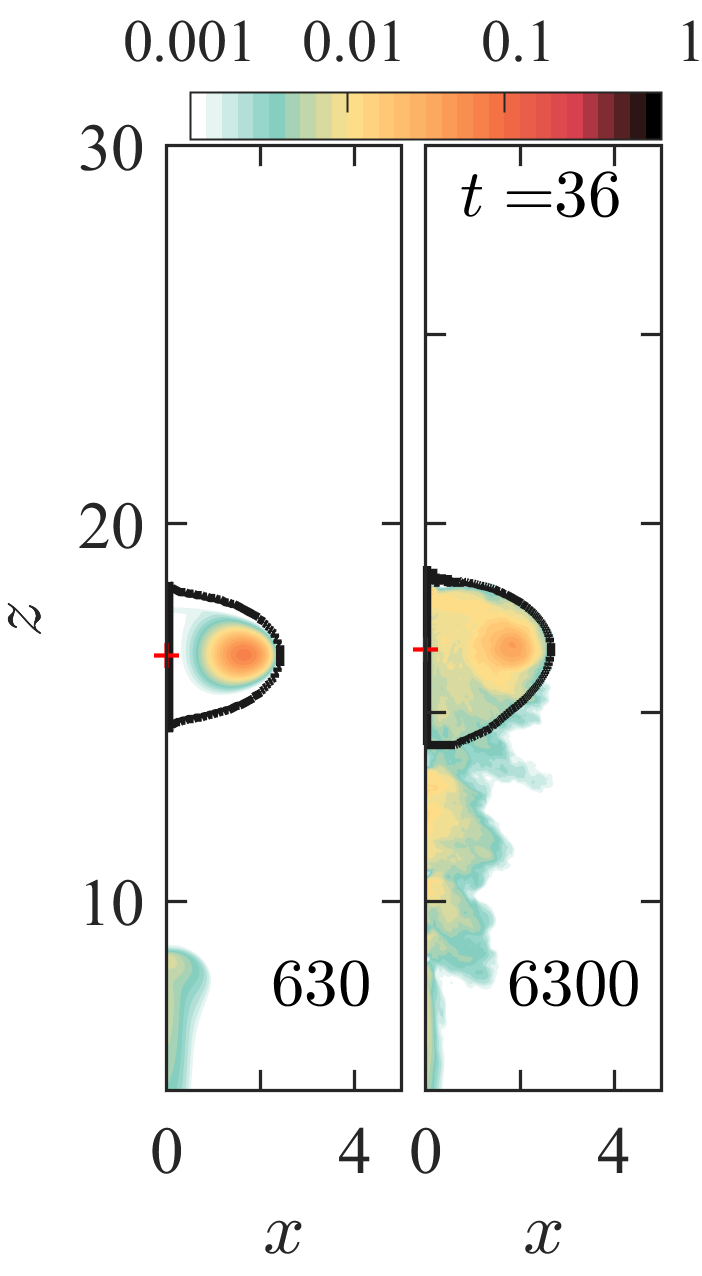}
     \end{subfigure}
    \hspace{0.1cm}
     \begin{subfigure}[b]{0.32\textwidth}
         \centering
         \includegraphics[width=\textwidth]{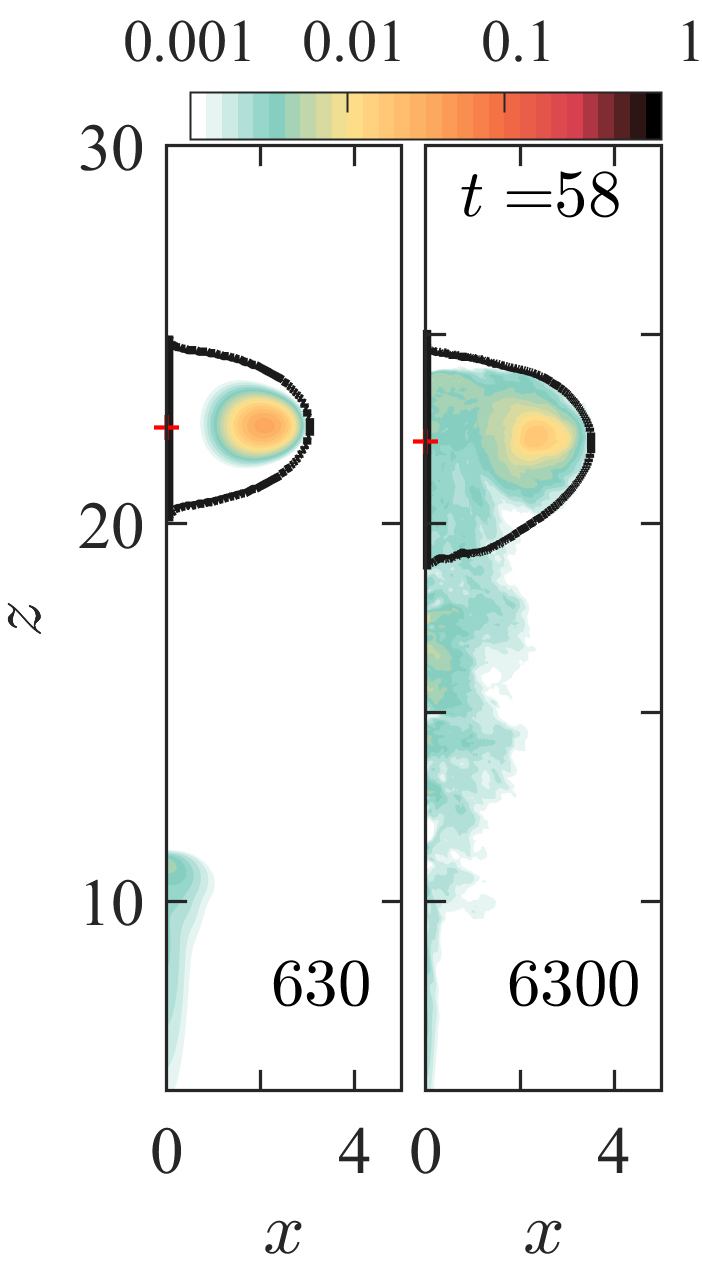}
     \end{subfigure}
     \caption{The azimuthally averaged temperature $\theta$ in laminar ($Re=630$, left panels) and turbulent ($Re=6300$, right panels) dry thermals in a neutrally stratified ambient at the times indicated. The thermals were initialised with the same level of noise. The colour maps are logarithmically scaled and the same for all three times shown. The laminar thermals have spun up into vortex rings as expected. The turbulent thermals are noticeably      different in shape, with more diffuse cores and more prominent tails. }
     \label{fig:dry_unstrat}
\end{figure}

Figure \ref{fig:dry_unstrat} shows the evolution of laminar ($Re=630$) and
turbulent ($Re=6300$) dry thermals in our simulations. These may be compared with
Figure 1 in LJ19. Vortex rings with cores where the temperature 
contrast is concentrated are seen for both laminar and turbulent cases
\citep[see, e.g.,][]{Zhao2013}. We note the differences in shape between the
laminar and turbulent cases, as well as the fact that the tails in the
turbulent case are more prominent (and hence that the detrainment is 
marginally larger), which are also in agreement with LJ19.\\

Our results are thus broadly consistent with LJ19. The net entrainment rate
obeys the relation  $\epsilon \sim R^{-1}$ for $R<2$. The values for laminar and
turbulent thermals differ by less than $10\%$, whereas LJ19 report a $20\%$
difference. Despite this, it is clear that the entrainment rate 
is not a strong function of $Re$ at least for $Re < 10^4$ 
\citep[][]{DIMOTAKIS2000}.\\

\begin{figure}
     \centering
     \begin{subfigure}[b]{0.45\textwidth}
         \centering
         \includegraphics[width=\textwidth]{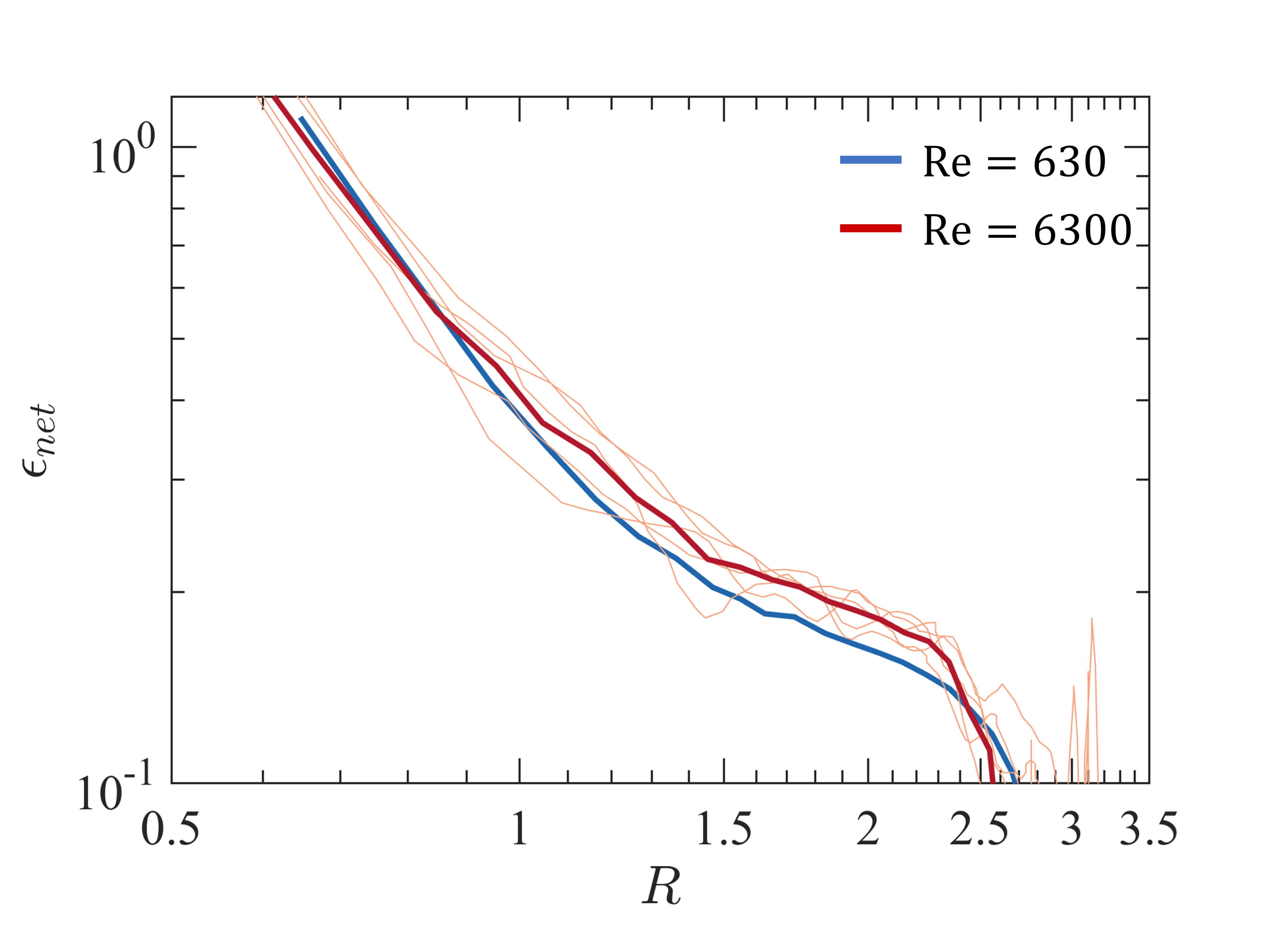}
         \caption{}
     \end{subfigure}
     \hspace{0.1cm}
        \begin{subfigure}[b]{0.45\textwidth}
         \centering
         \includegraphics[width=\textwidth]{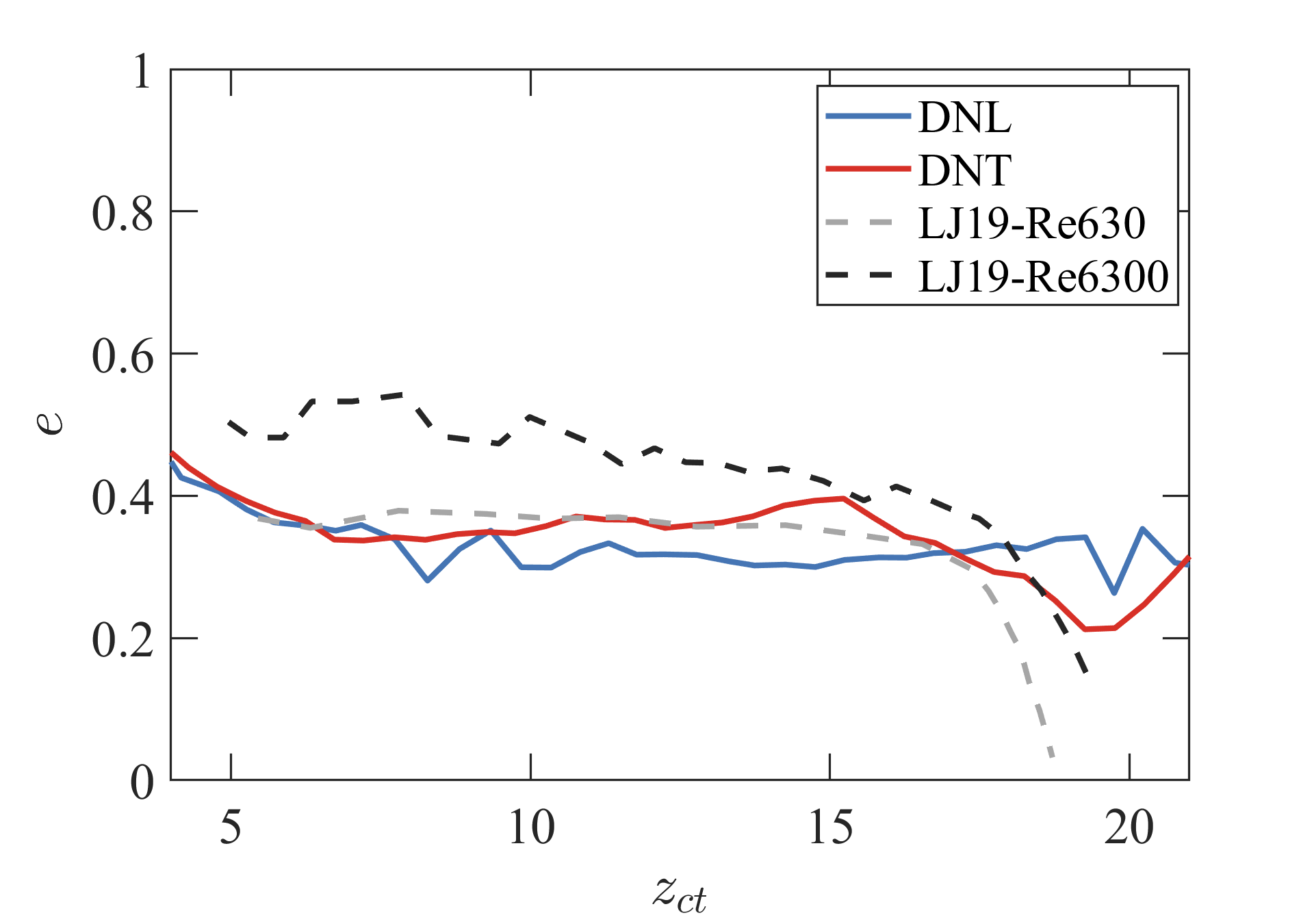}
         \caption{}
     \end{subfigure}
     \caption{The (a) entrainment rate and (b) entrainment efficiency, defined in
     equations \ref{eq:entrainment_rate} and \ref{eq:entrainment_efficiency}, for laminar (blue, $Re=630$) and turbulent (red, $Re=6300$) dry thermals in a
     neutral ambient ($\Gamma_0 = \Gamma_u$)d. The thin red lines are the five 
     individual runs for the turbulent thermals with different instantiations
     of noise, while the average is plotted as a thick line. In (b), the curves from LJ19 are also plotted for comparison. Turbulent entrainment rates are marginally higher than laminar entrainment rates, and the difference is smaller than in LJ19. There is a dip in the value of $e$ after $z_{ct}$>16, more pronounced in the turbulent cases, which is also seen in LJ19.
     \label{fig:dry_unstrat_eps_e}}
\end{figure}

The net entrainment $\epsilon_{net}$ and entrainment efficiency $e=\epsilon_{net} R$
are plotted in figure \ref{fig:dry_unstrat_eps_e}(a,b) respectively. These are
comparable to Figures 5 and 7 in LJ19. The entrainment efficiency remains
essentially constant for $z_{ct}<17$ for both laminar and turbulent thermals. There
is a slow decrease in entrainment efficiency above $z_{ct}>17$, also noted by LJ19,
perhaps because of boundary effects.\\

The evolution of the thermal location, the maximum flow velocity in the thermal, the
thermal radius and the thermal volume are plotted in figures
\ref{fig:dry_unstrat_V_w_R} (a-d) respectively in section \ref{sec:high_Re},
and the curves for $Re=630$ and $Re=6300$ therein may be compared with Figure 4 in
LJ19.

\subsection{Moist {thermals in a dry-neutral ambient}} \label{sec:moist_unstrat}

Having validated our numerical method and analysis on dry thermals, we apply
the same methods to moist thermals in a {dry-}neutrally stratified ambient.\\

\begin{figure}
     \centering
     \begin{subfigure}[b]{0.32\textwidth}
         \centering
         \includegraphics[width=\textwidth]{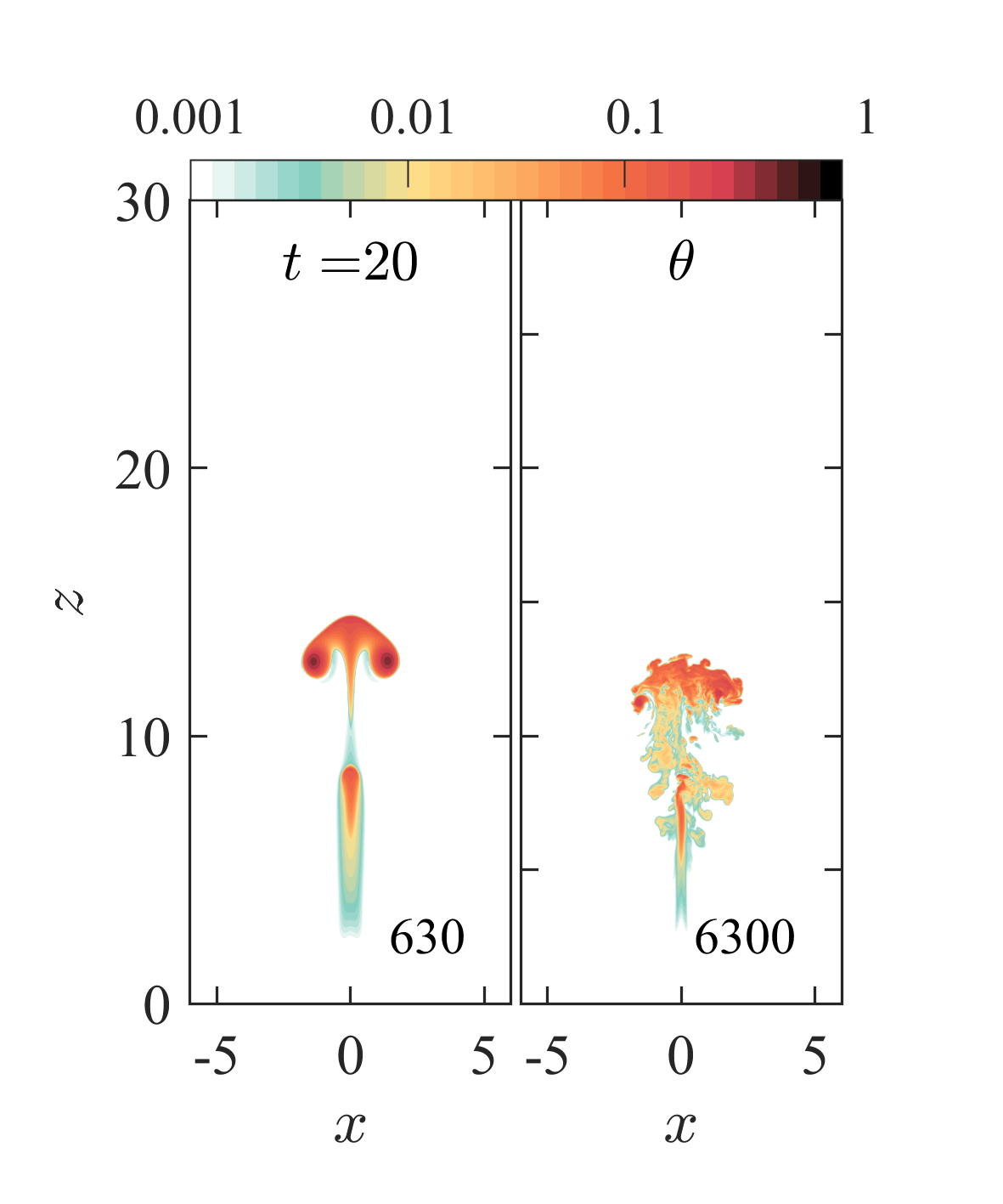}
         \caption{}
     \end{subfigure}
     \hspace{0.1cm}
        \begin{subfigure}[b]{0.32\textwidth}
         \centering
         \includegraphics[width=\textwidth]{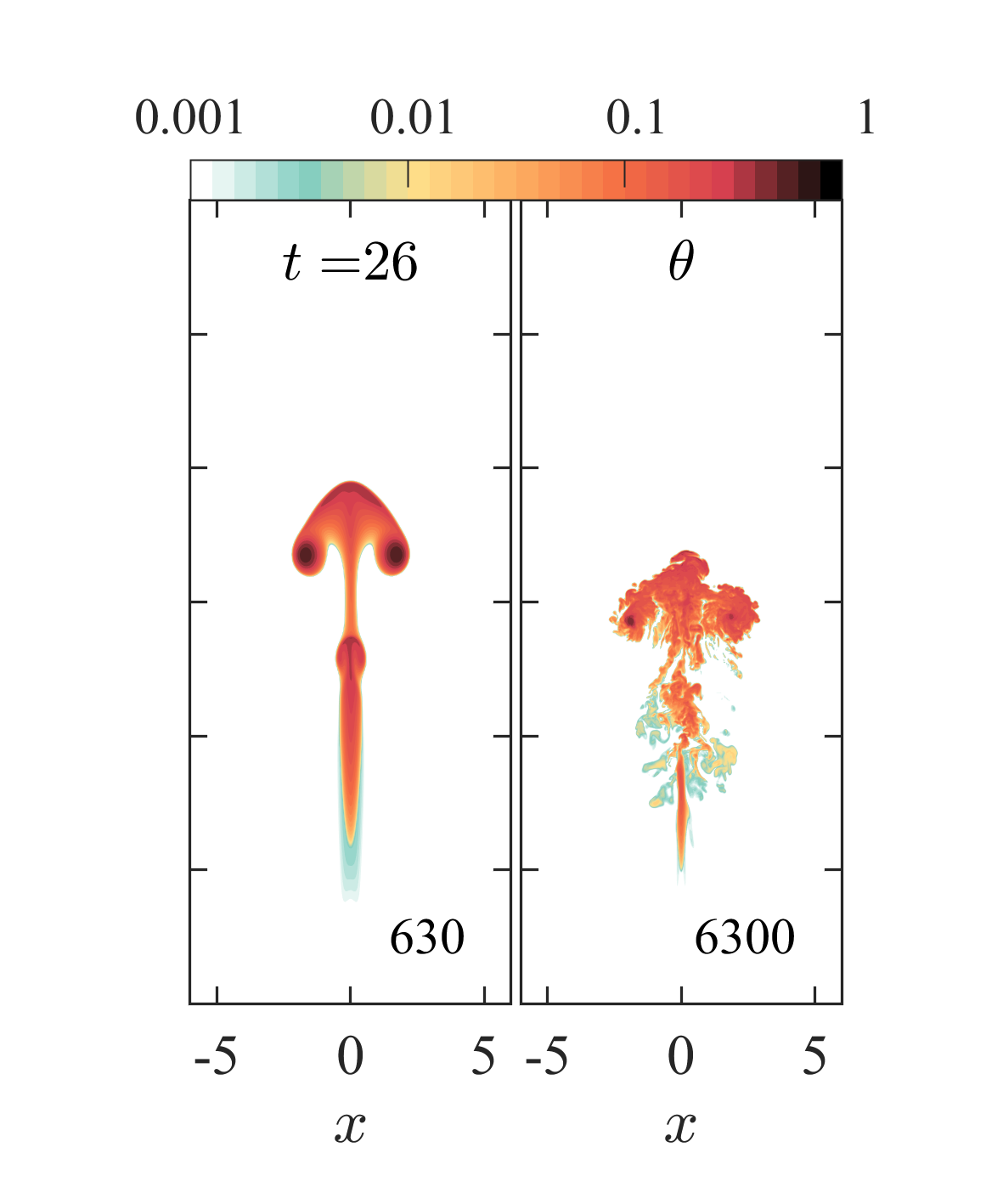}
         \caption{}
     \end{subfigure}
    \hspace{0.1cm}
     \begin{subfigure}[b]{0.32\textwidth}
         \centering
         \includegraphics[width=\textwidth]{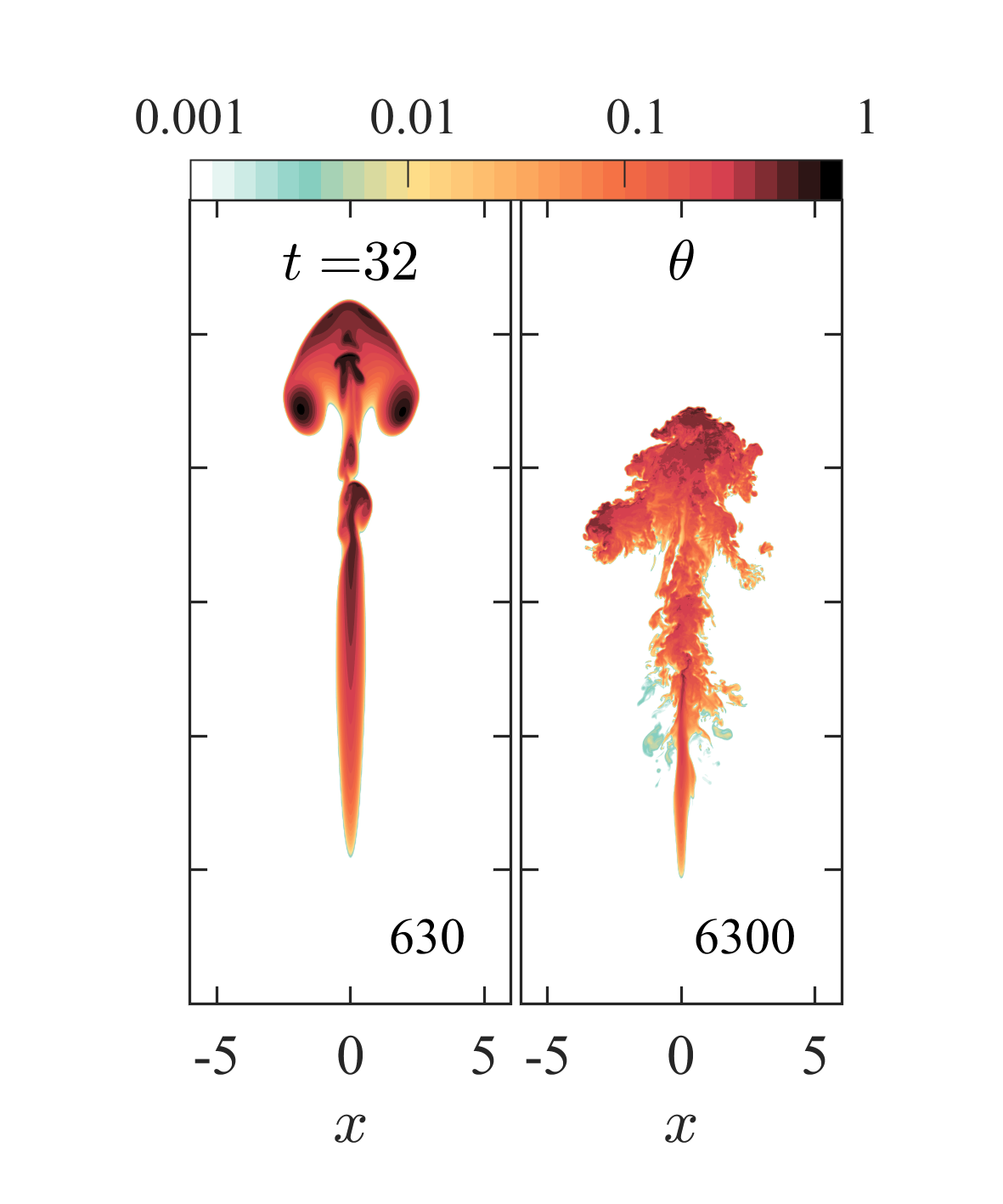}
         \caption{}
     \end{subfigure}
     \caption{Instantaneous 2D slices at $y=0$ of the temperature $\theta$ in laminar ($Re=630$, left panels) and turbulent ($Re=6300$, right panels) moist thermals at the times indicated. As in figure \ref{fig:dry_unstrat}, a logarithmic colour scale is used. Laminar moist thermals assume a distinct `arrowhead' shape where the vortex ring remains. See figure \ref{fig:moist_unstrat_u} for corresponding plots of the vertical velocity.}
     \label{fig:moist_unstrat_T}
\end{figure}

\begin{figure}
     \centering
     \begin{subfigure}[b]{0.29\textwidth}
         \centering
         \includegraphics[width=\textwidth]{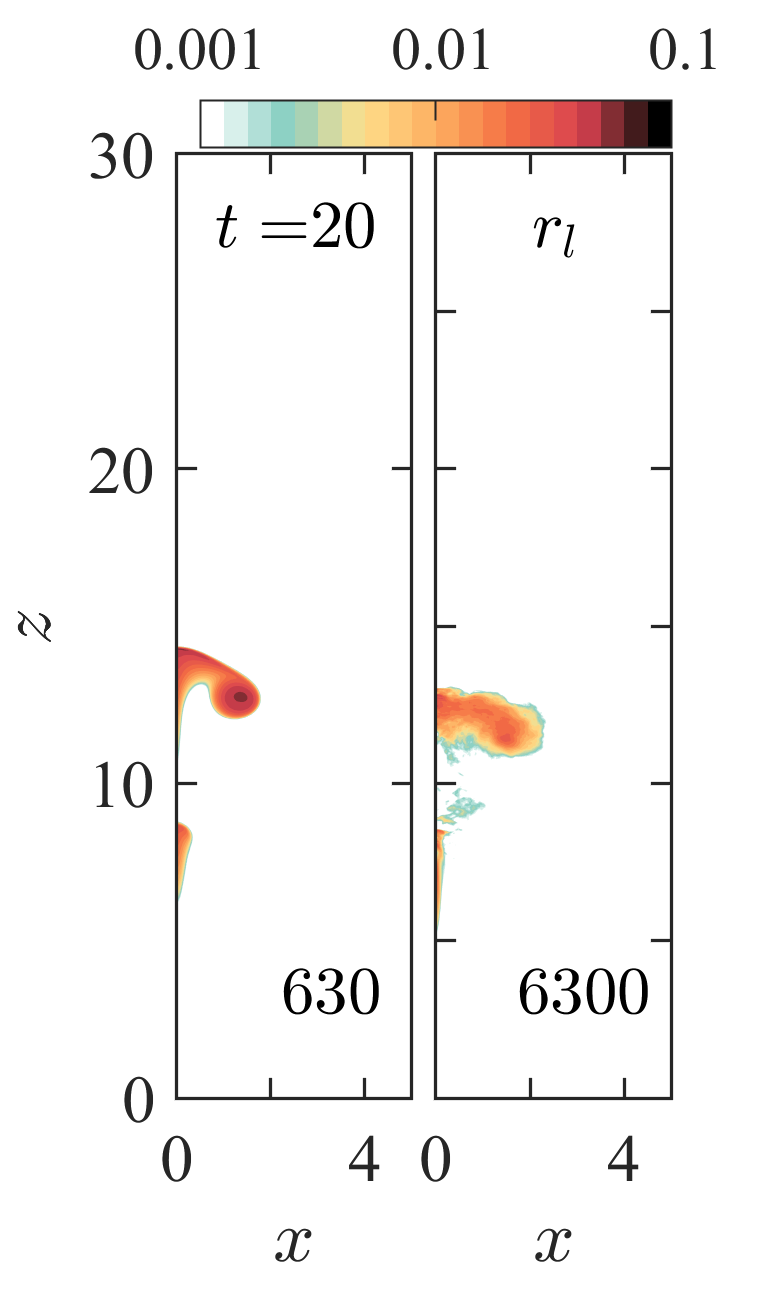}
         \caption{}
     \end{subfigure}
     \hspace{0.1cm}
        \begin{subfigure}[b]{0.29\textwidth}
         \centering
         \includegraphics[width=\textwidth]{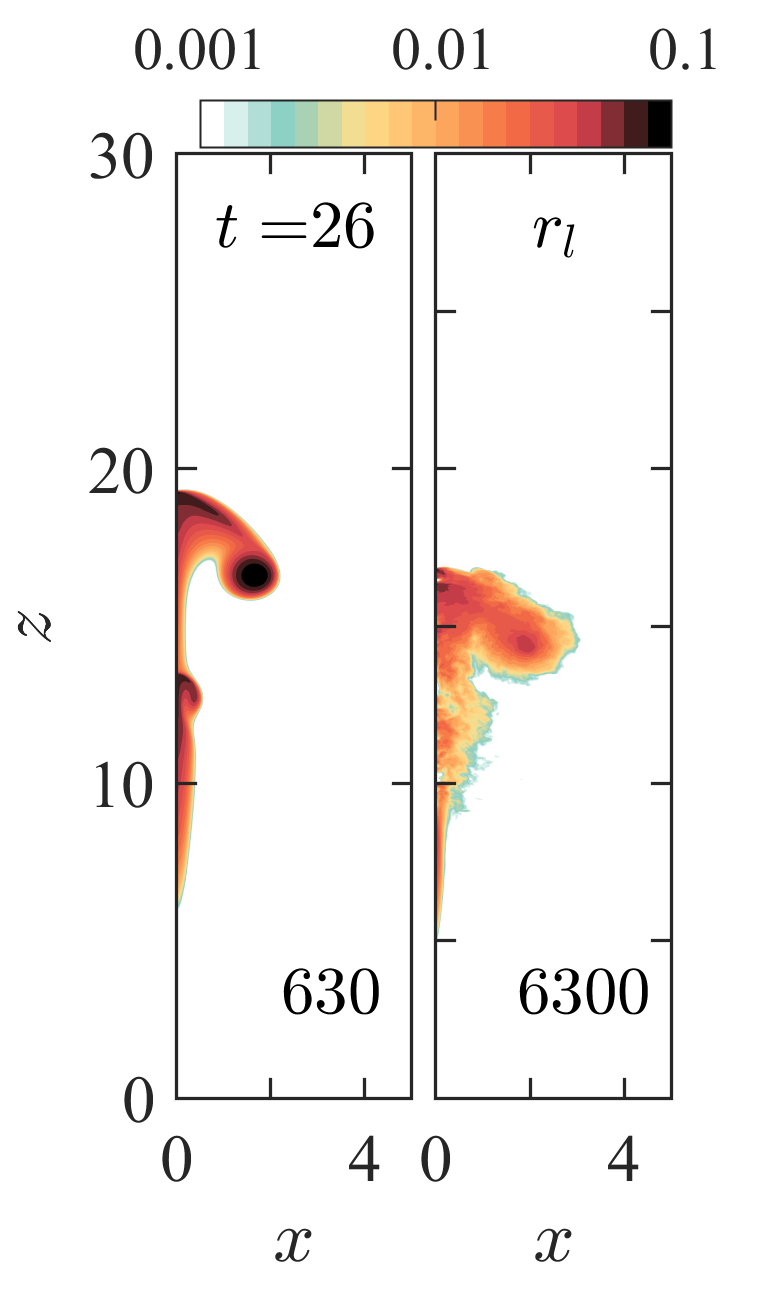}
         \caption{}
     \end{subfigure}
    \hspace{0.1cm}
     \begin{subfigure}[b]{0.29\textwidth}
         \centering
         \includegraphics[width=\textwidth]{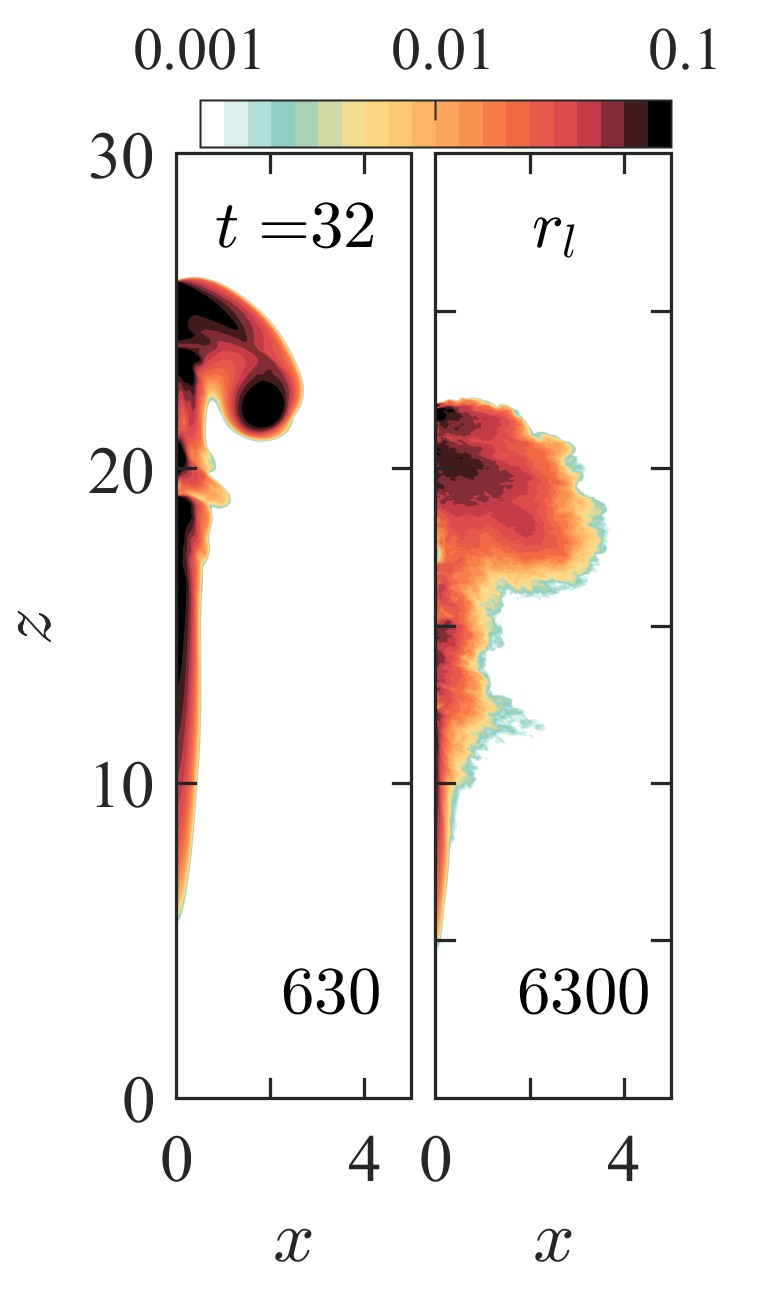}
         \caption{}
     \end{subfigure}
     \caption{Azimuthally averaged liquid mixing ratio $r_l$ for laminar ($Re=630$, left panels) and turbulent ($Re=6300$, right panels) moist thermals in a dry-neutrally stratified ambient at the same times as in figure \ref{fig:moist_unstrat_T}. (a) initially, moist
     thermals behave similarly as dry thermals, forming vortex rings whose cores have large $r_l$ in both laminar and turbulent cases. (b) these vortex rings morph into the `arrowhead' shape with large $r_l$ both in the vortex cores as well as the tip of the arrow. (c) at $t=32$, the distribution of $r_l$ is very different in the laminar and turbulent cases; in the laminar thermal, large $r_l$ values occur in the vortex ring as well as the thermal top, whereas the vortex ring is destroyed in the turbulent thermal and the maximum in $r_l$ occurs only at the thermal top. Note that a logarithmic colour scale is used. }
     \label{fig:moist_unstrat_liq}
\end{figure}

\begin{figure}
     \centering
     \begin{subfigure}[b]{0.28\textwidth}
         \centering
         \includegraphics[width=\textwidth]{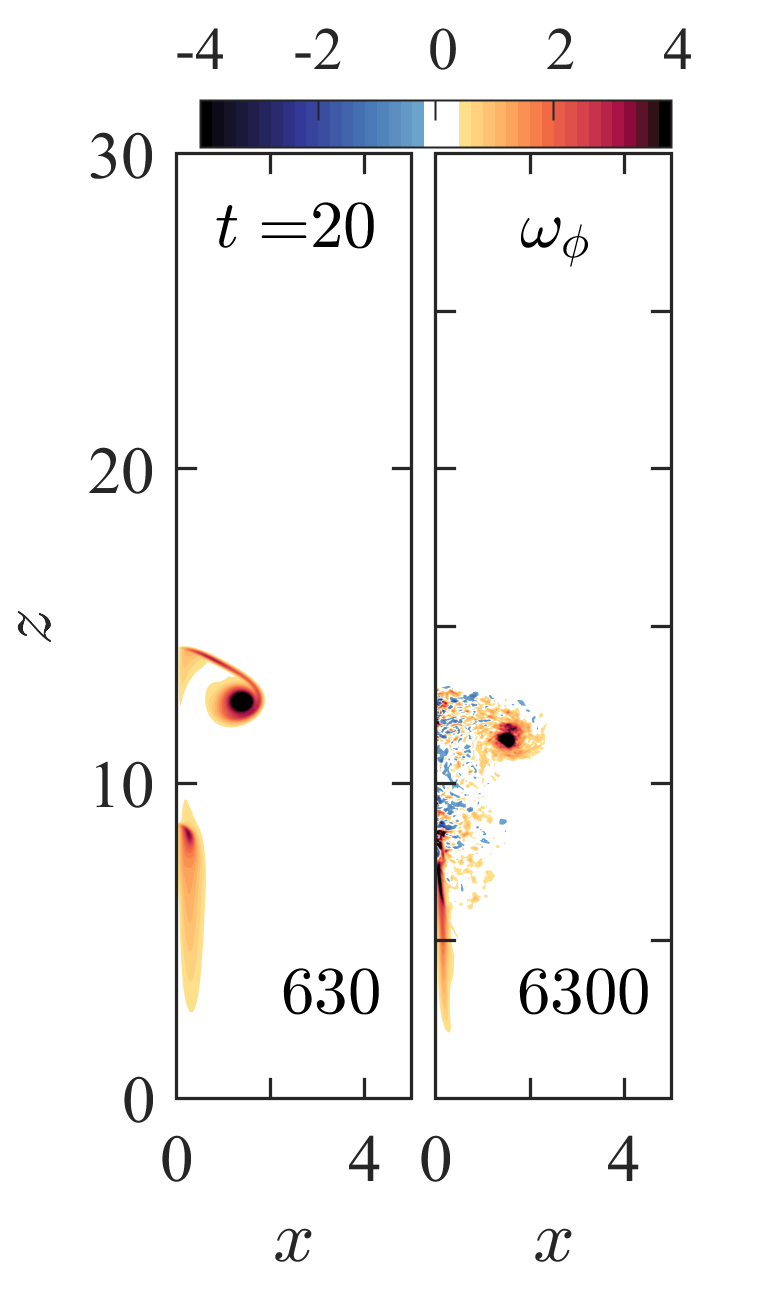}
         \caption{}
     \end{subfigure}
     \hspace{0.1cm}
        \begin{subfigure}[b]{0.28\textwidth}
         \centering
         \includegraphics[width=\textwidth]{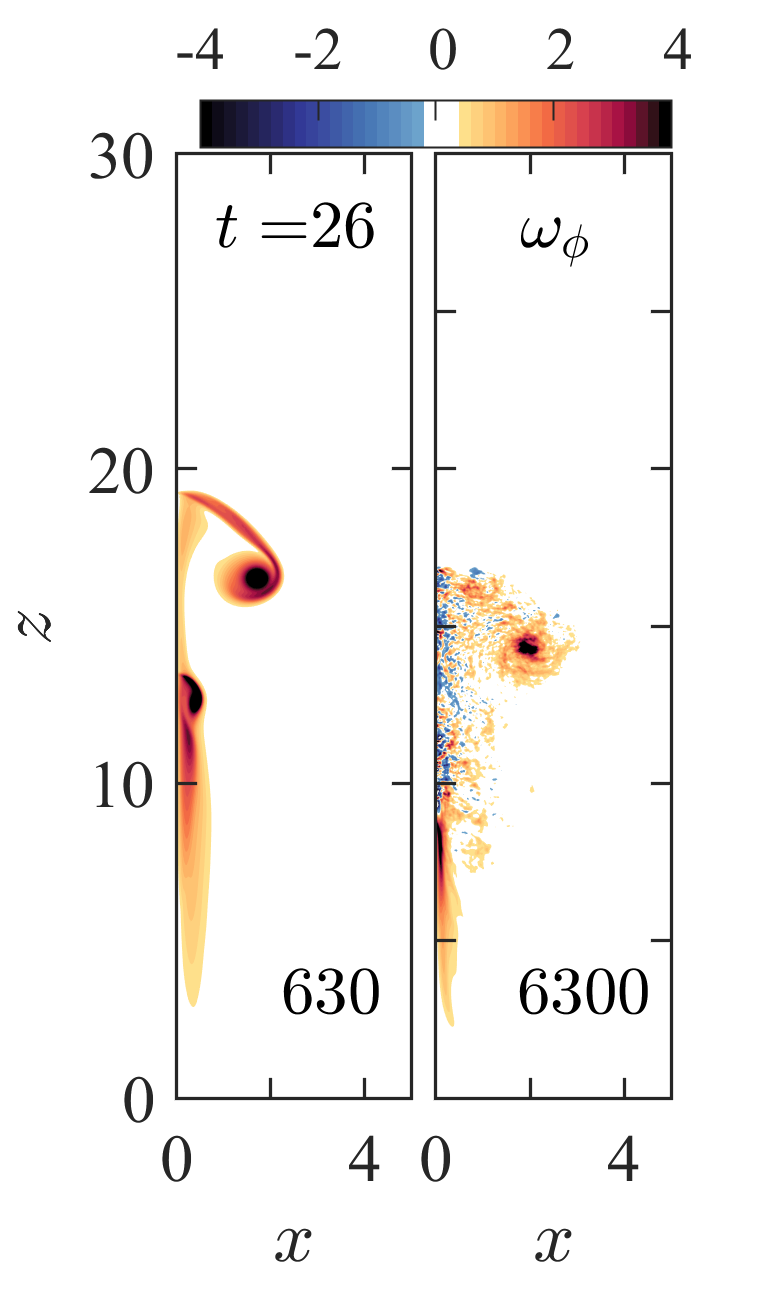}
         \caption{}
     \end{subfigure}
    \hspace{0.1cm}
     \begin{subfigure}[b]{0.28\textwidth}
         \centering
         \includegraphics[width=\textwidth]{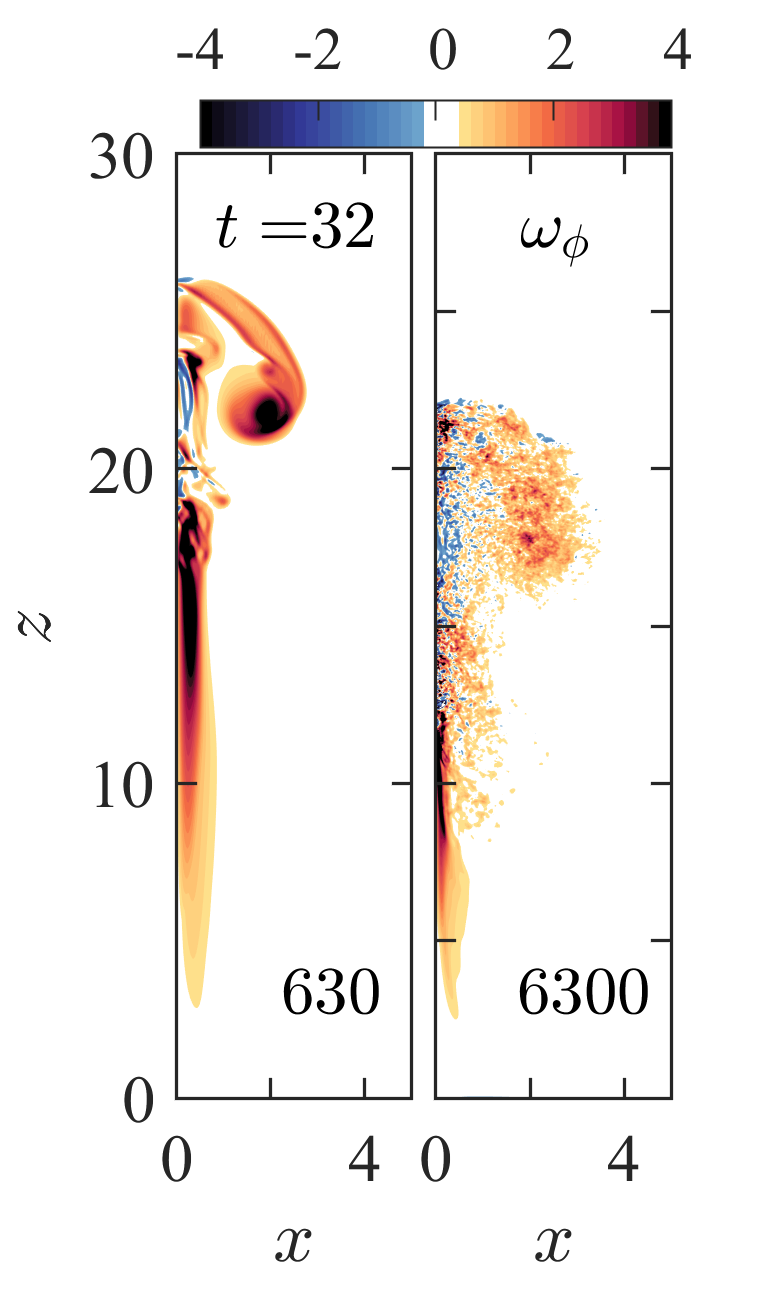}
         \caption{}
     \end{subfigure}
     \caption{Azimuthally averaged azimuthal vorticity $\omega_\phi$ for laminar ($Re=630$, left panels) and turbulent ($Re=6300$, right panels) moist thermals in a dry-neutral ambient at the same times as in figure \ref{fig:moist_unstrat_T}. (a) The vorticity is initially concentrated in
     the vortex ring that forms. (b) The vorticity in the ring increases in
     the laminar thermal, while the ring becomes more diffuse in the turbulent thermal. (c) significant negative vorticity can be seen along the axis of symmetry in both laminar and turbulent thermals, and the vortex ring in the turbulent thermal is even more diffuse. }
     \label{fig:moist_unstrat_omgt}
\end{figure}

\begin{figure}
    \centering
    \includegraphics[width=0.9\columnwidth]{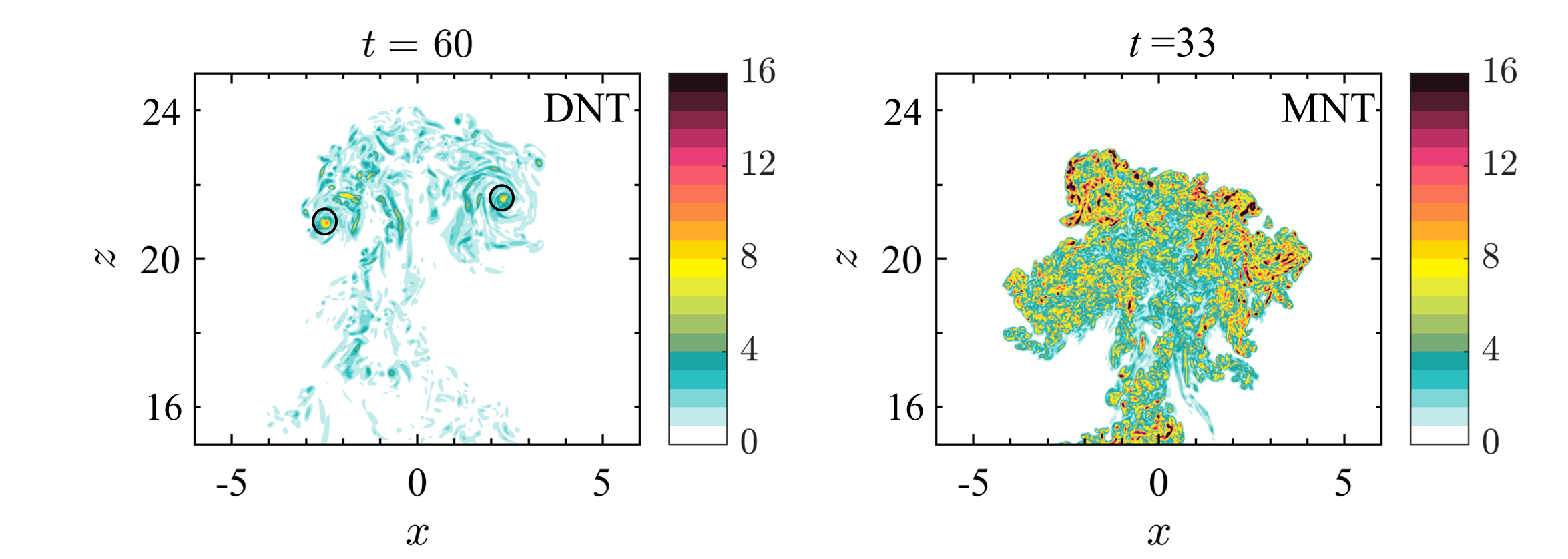}
    \caption{{\label{fig:omg_mag_dnt_mnt} Vorticity magnitude ($|\omega| = \sqrt{\omega_x^2+\omega_y^2+\omega_z^2}$) in (a) DNT and (b) MNT thermals. The black circles in (a) denote the location of core of the vortex ring in the DNT thermal. In the MNT thermal in (b), the vortex ring has disintegrated and the vorticity is space-filling, and the maximum $|\omega|$ is $\approx5$ times higher than in (a).}}
\end{figure}

{Laminar and turbulent moist thermals are initialised as described in
section \ref{sec:initial_conds}, and instantaneous slices of the temperature 
$\theta$ and azimuthally averaged liquid water mixing ratio $r_l$ are shown in
figures \ref{fig:moist_unstrat_T} and \ref{fig:moist_unstrat_liq} respectively.
Unlike dry thermals which take the form of vortex rings (section
\ref{sec:dry_unstrat}), the laminar moist thermals develop
a distinct `arrowhead' shape which includes the vortex ring; in
turbulent moist thermals, the coherent vortex ring and arrowhead shape are destroyed, and the flow features are small-scale (figures \ref{fig:moist_unstrat_omgt} and \ref{fig:omg_mag_dnt_mnt}).}

{Condensation heating and the resulting increase of buoyancy in the MNL 
thermal leads to significantly larger values of the azimuthal vorticity (figure \ref{fig:moist_unstrat_omgt}) compared to the DNL thermal (not shown). The other components of the vorticity are of much smaller magnitude in these laminar cases where the rings are coherent. In both dry and moist turbulent thermals, on the other hand, the flow is more isotropic and the different components of the vorticity are of similar magnitude, and we find that the combined action of condensation heating in concert with turbulence lead to maximal vorticity magnitudes $|\omega|$ $4-5$ times larger in the MNT thermals than in the DNT thermals (figure \ref{fig:omg_mag_dnt_mnt}).}

\begin{figure}
     \centering
     \begin{subfigure}[b]{0.32\textwidth}
         \centering
         \includegraphics[width=\textwidth]{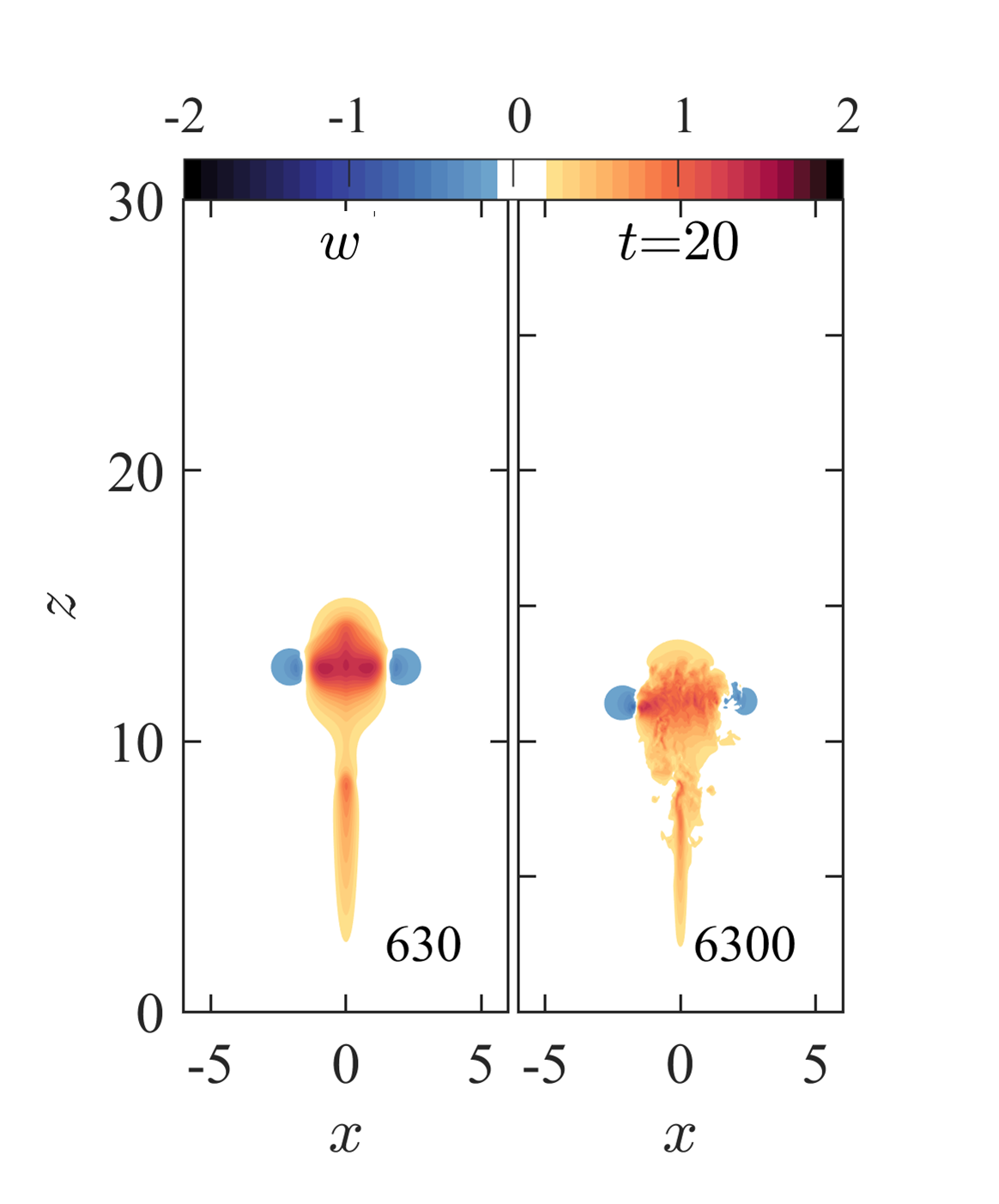}
         \caption{}
     \end{subfigure}
     \hspace{0.1cm}
        \begin{subfigure}[b]{0.32\textwidth}
         \centering
         \includegraphics[width=\textwidth]{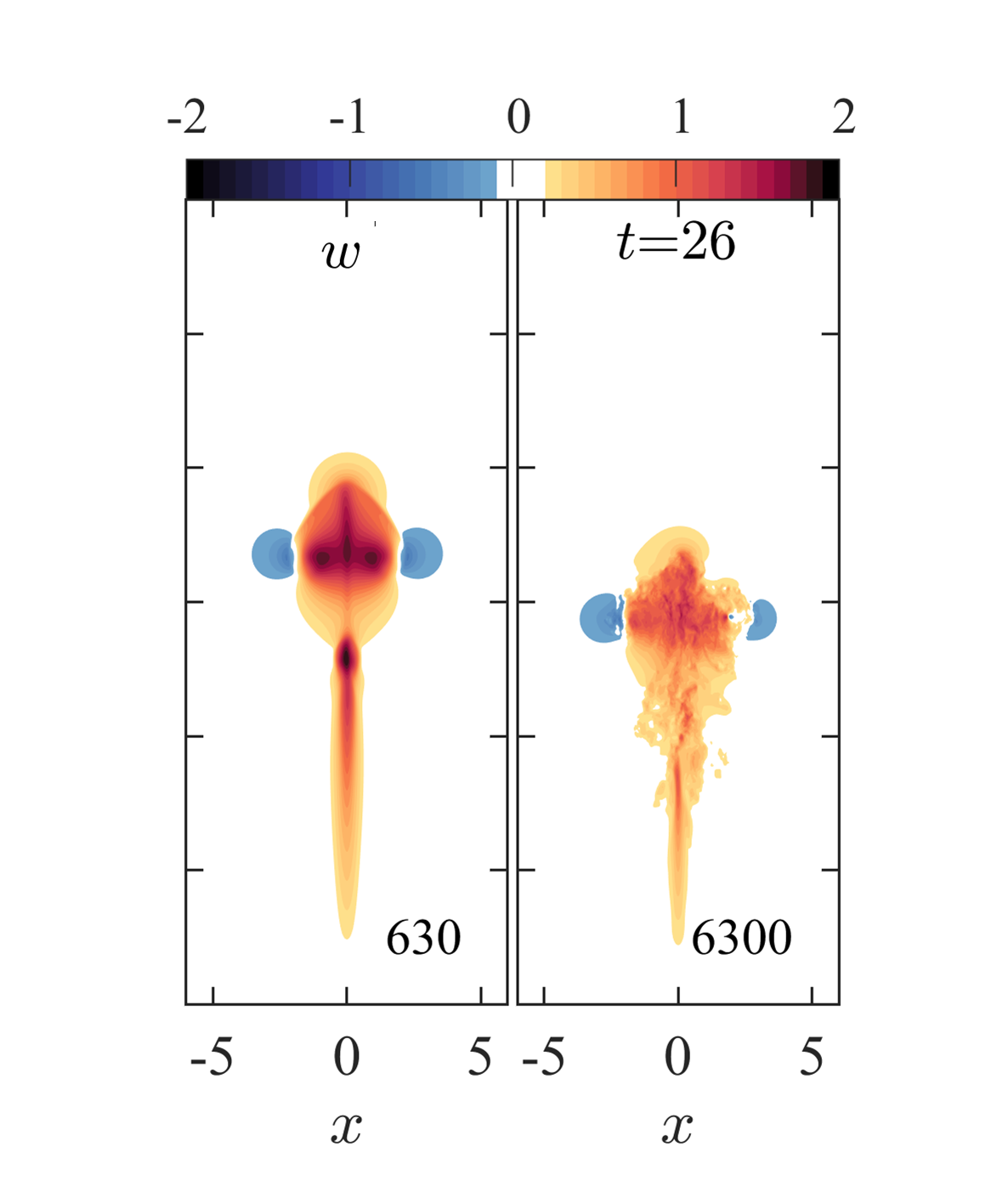}
         \caption{}
     \end{subfigure}
    \hspace{0.1cm}
     \begin{subfigure}[b]{0.32\textwidth}
         \centering
         \includegraphics[width=\textwidth]{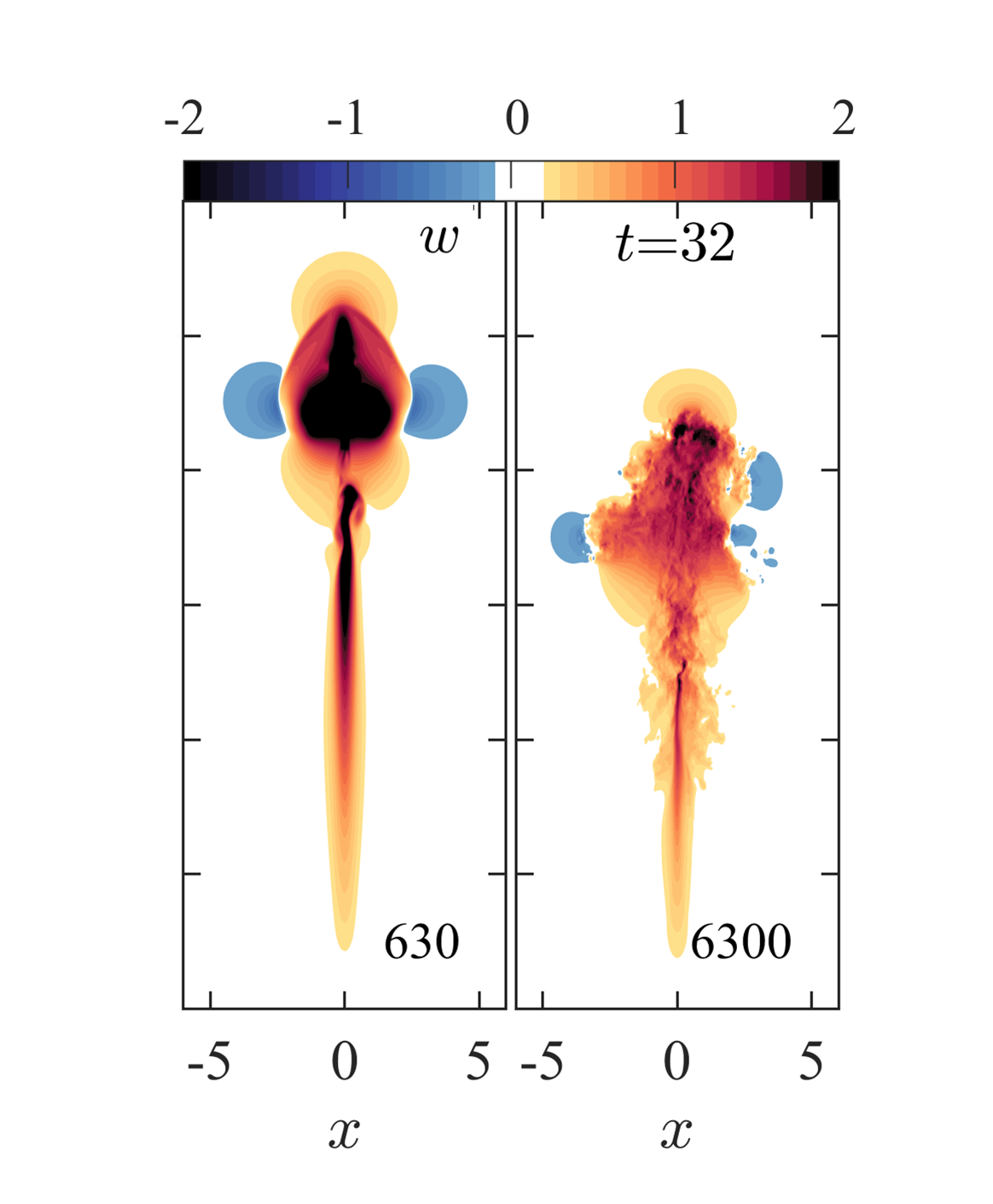}
         \caption{}
     \end{subfigure}
     \caption{Instantaneous 2D slices at $y=0$ of the vertical velocity $w$ in
    laminar ( $Re=630$, left panels) and turbulent ($Re=6300$, right panels) moist thermals at the same
     times as in figure \ref{fig:moist_unstrat_T}. The vertical velocities for moist
     thermals are significantly higher than for dry thermals. (a) The flow is initially axisymmetric, with maximum velocities at the vortex rings. (b) As
     the thermal is heated by condensation, the laminar case retains its symmetry
     but the turbulent case can be seen to show departures from symmetry. Large velocities occur both in the rings and in thermal centre.(c) Velocities have increased significantly in both laminar and turbulent thermals. The laminar thermal continues to be axisymmetric and the vortex ring continues to become stronger, whereas the vortex ring is destroyed in the turbulent case. C.f. figures \ref{fig:moist_unstrat_T} and \ref{fig:moist_unstrat_omgt}.
     \label{fig:moist_unstrat_u}}
\end{figure}

{Furthermore, the coherent vortex core present in the dry thermals (even for larger Re; see Section \ref{sec:high_Re}) is destroyed in the MNT thermal; thus, turbulence is a necessary but not sufficient condition for the destruction of the vortex ring and the emergence of intense small-scale vorticity in the flow. In highly turbulent flows, the magnitude of the vorticity fluctuations relative to the mean vorticity is known to increase with the flow Reynolds number as $\omega^\prime \sim Re^{1/2}$ \citep[e.g.][]{Narasimha2012}, and our observations are consistent with this. Such small scale vorticity may be responsible for the crinkly edges seen in cumulus clouds \cite{Narasimha2012}. This generation of intense small scale vorticity is also responsible for the 
smaller vertical mean velocities seen in plots of the vertical velocity $w$ in figure \ref{fig:moist_unstrat_u}.}


{From figures \ref{fig:moist_unstrat_T} -- \ref{fig:moist_unstrat_u},} we see, as also reported in Ref. \cite{Morrison2021}, that a condensing thermal in
a {dry-}neutrally stratified ambient is highly unstable and leads to large 
vertical velocities; {and thus the effective Reynolds numbers in the
turbulent moist thermal are a factor $4-5$ larger than the nominal
$Re=6300$.}\\

\begin{figure}[H]
     \centering
     \begin{subfigure}[b]{0.47\textwidth}
         \centering
          \includegraphics[width=\textwidth]{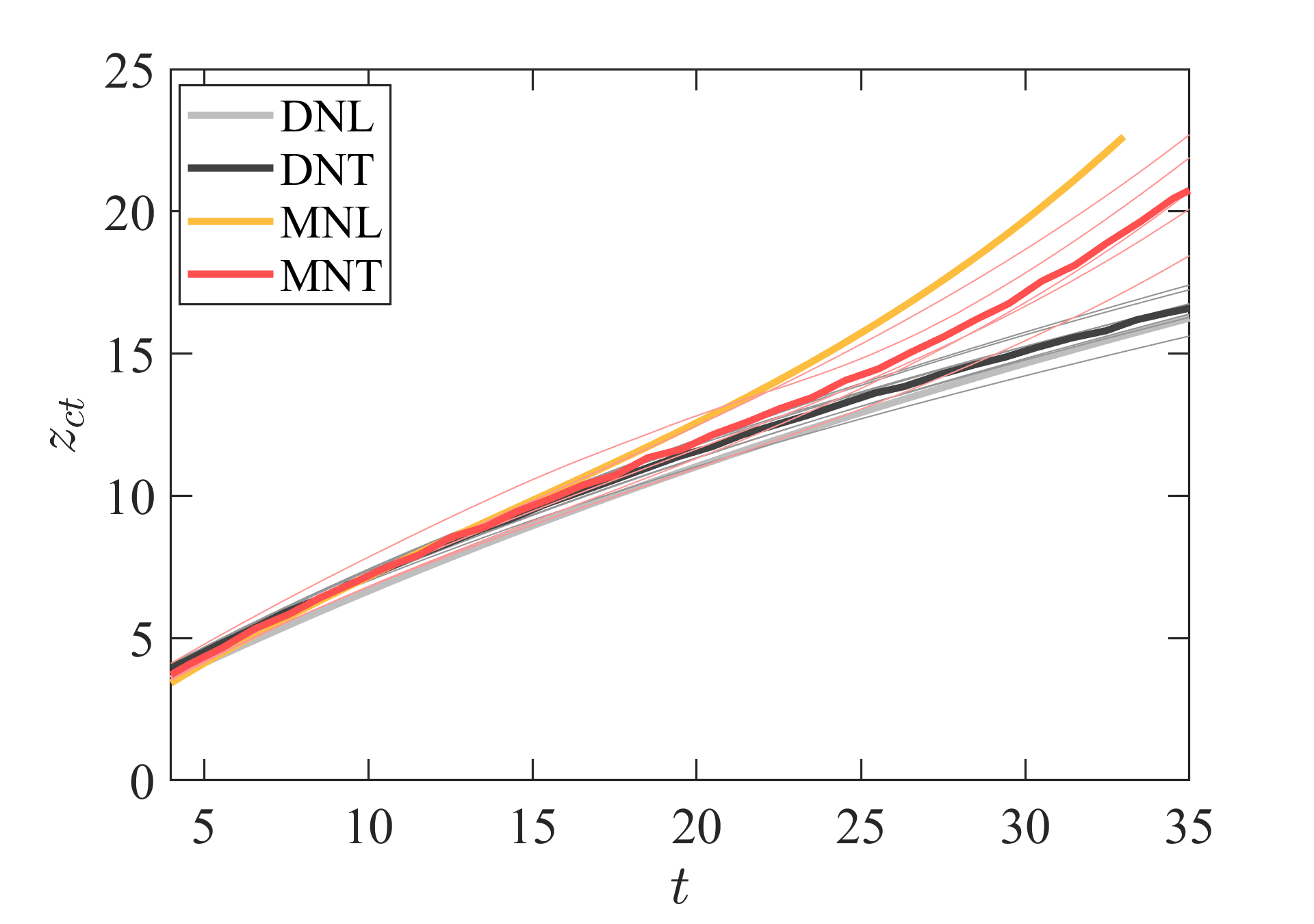}
          \caption{}
     \end{subfigure}
     \hspace{0.1cm}
        \begin{subfigure}[b]{0.47\textwidth}
         \centering
        \includegraphics[width=\textwidth]{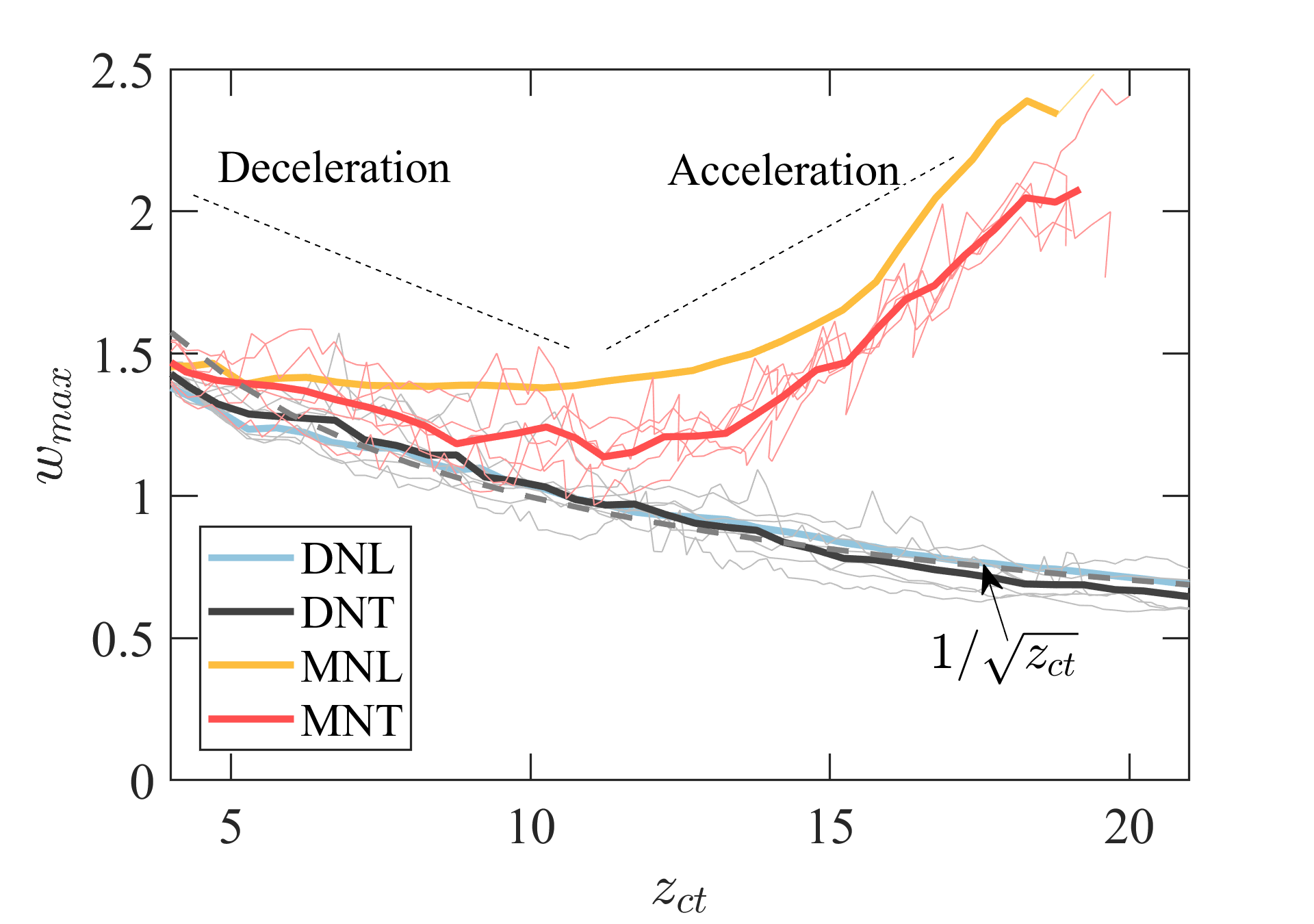}
        \caption{}
     \end{subfigure}
     \hspace{0.1cm}
        \begin{subfigure}[b]{0.47\textwidth}
         \centering
         \includegraphics[width=\textwidth]{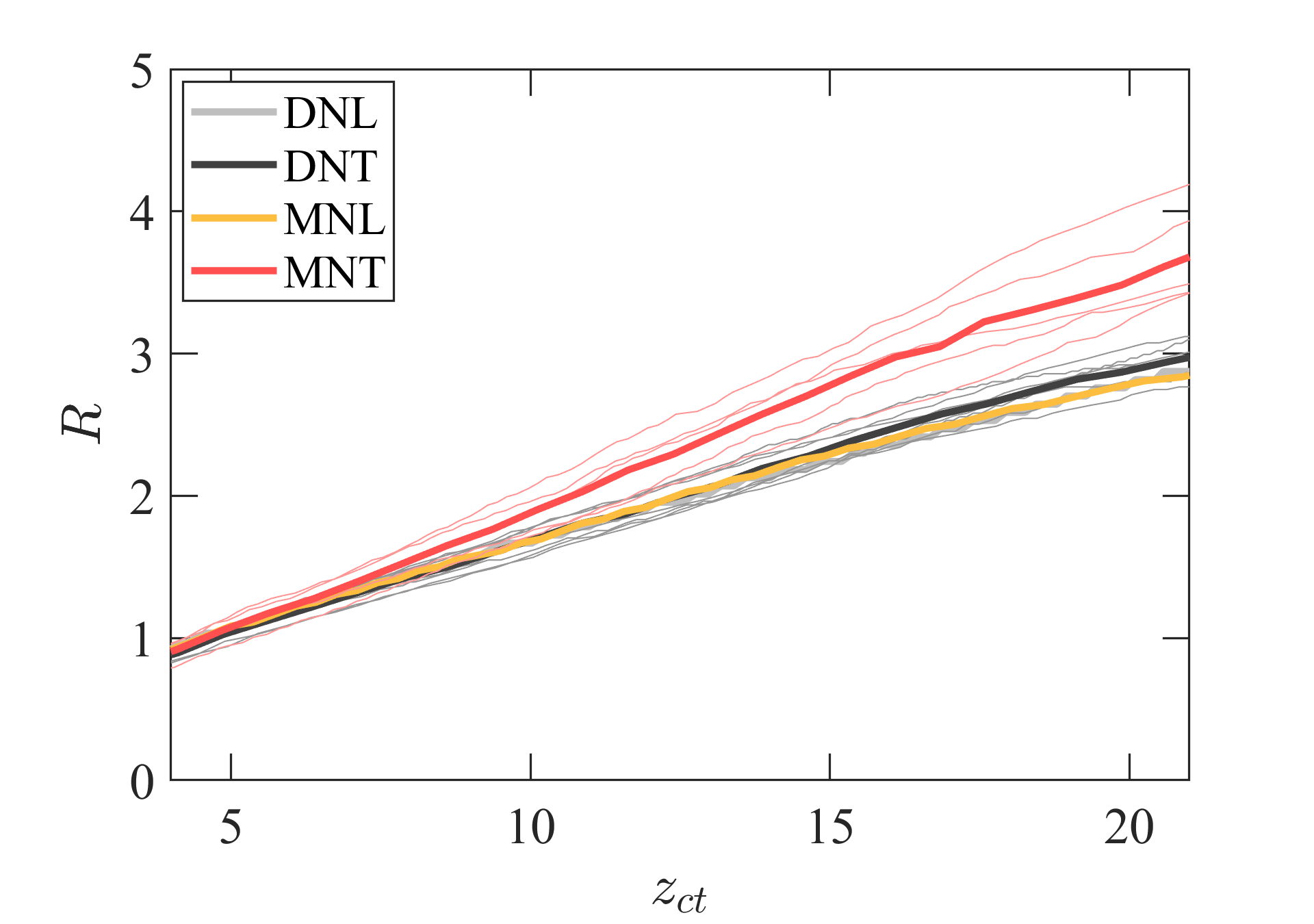}
         \caption{}
     \end{subfigure}
         \hspace{0.1cm}
        \begin{subfigure}[b]{0.47\textwidth}
         \centering
         \includegraphics[width=\textwidth]{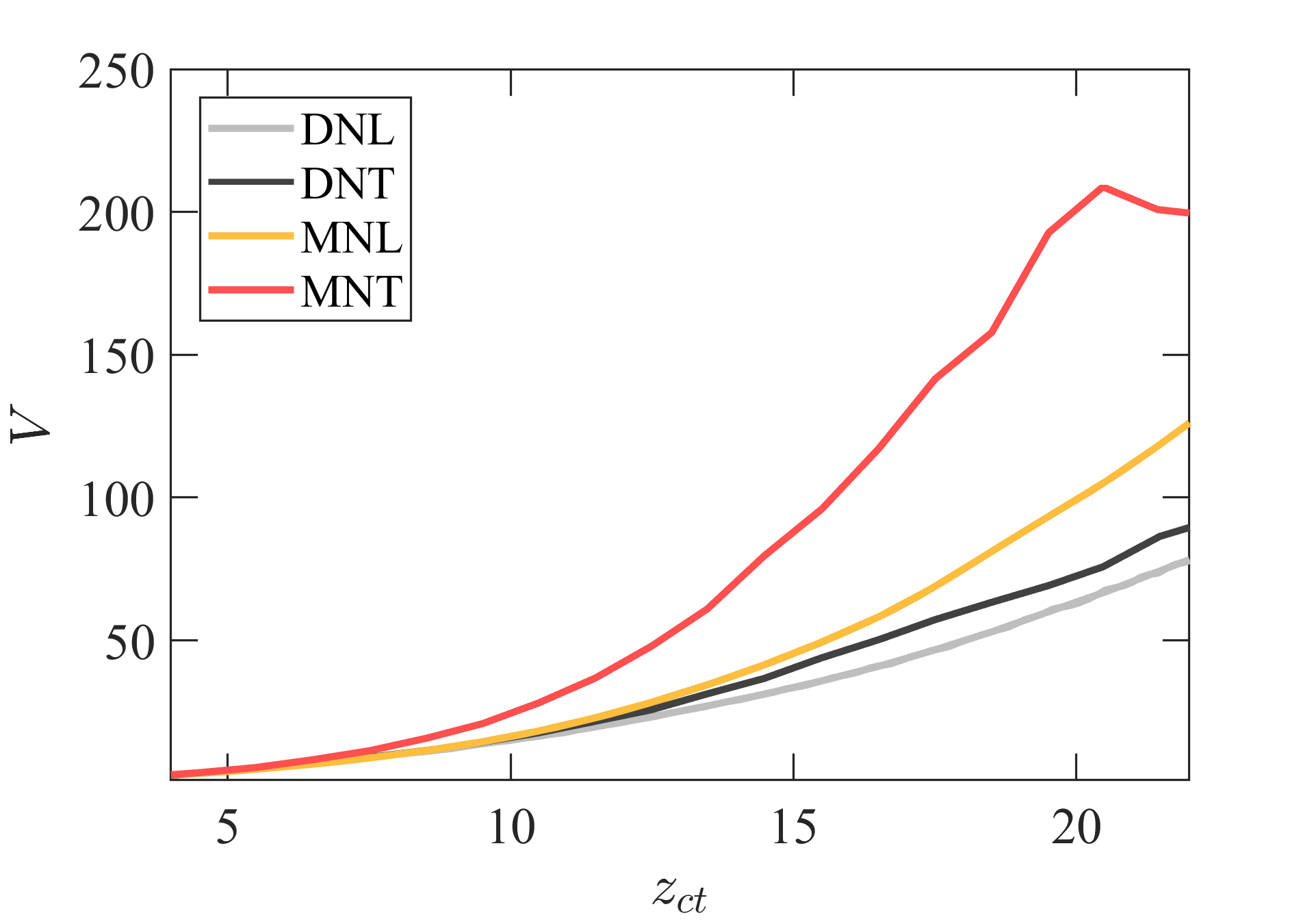}
         \caption{}
     \end{subfigure}
     \caption{The (a) location $z_{ct}$, (b) maximum velocity $w_{max}$ (c) radius $R$, and (d) volume $V$ of laminar ($Re=630$) and turbulent ($Re=6300$) moist thermals in a (dry-) neutrally stratified ambient.The x-axis starts from four for all the sub figures. The curves for
     dry thermals in the same ambient are shown for comparison. In each case,
     individual runs are plotted as thin lines while ensemble averages are plotted
     as thick lines. Moist thermals begin to accelerate for $z_{ct}\gtrsim 10$ due to condensation heating and, as a result, grow to larger radii, have greater volumes and rise faster than dry thermals. It is worth noting that, even though the volumes of the moist thermals grow much faster than their dry counterparts,
     the radius of the moist laminar thermal grows similarly to the dry cases. This is due to a change in the shape of the thermal (see, e.g., figure \ref{fig:psi_moist_unstrat}; see also figure \ref{fig:thermal_volume}).}
     \label{fig:moist_unstrat_V_w_R}
\end{figure}

The location, maximum flow velocity, radius and volume of moist thermals
are compared with dry thermals in figure \ref{fig:moist_unstrat_V_w_R}. {We see that the velocities in moist thermals are larger than the velocties in dry thermals by a factor of $2-3$ whereas the buoyancy is larger by an order of magnitude (compare figures \ref{fig:dry_unstrat} and \ref{fig:moist_unstrat_T}). This is consistent with the idea that the velocity scale is set by the square root of the buoyancy scale (see Sec. \ref{sec:setup}).}

{We also find that moist thermals have \emph{larger} volumes than the 
corresponding dry cases, that the thermal radii in the moist turbulent case
are larger than the radii of dry thermals of the same {nominal} $Re$, and that in both dry and moist thermals, turbulence leads to larger thermal volumes and radii and smaller velocities. However, the influence of turbulence is significantly larger in moist thermals.}

{The larger entrainment rate in moist thermals is in contrast to the findings in Ref. \cite{Morrison2021} from axisymmetric and 3D LES that moist thermals have \emph{smaller} volumes and radii than dry thermals. 
The reasons for this difference are unclear. Morrison et al. argue in Ref. \cite{Morrison2021} that the radial distribution of the heat added to the thermals (by condensation) results in the smaller entrainment in moist thermals in the initial stages (e.g. see their figure 6). We note, however, that the 
`arrowhead' shape that laminar moist thermals take (figure \ref{fig:moist_unstrat_T}) permits the same  mechanism of
outward expansion operant in dry thermals that is shown schematically in figure 9 of Ref. \citep[][]{Morrison2021}.}

{Similarly,} the large increase of vorticity due to the action of (external)
buoyancy addition  has been noted in earlier numerical studies 
\citep[e.g.][]{Basu1999}, and is thought to be responsible for the increased 
mixing of the fluid inside the flow seen in Ref. \cite{Bhat1996}. These studies
also report a  \emph{decrease} in entrainment attributed to the destruction of the coherent toroidal vortical structures due to the external heat addition, which we do not see. We instead find that laminar moist thermals rise
faster than turbulent moist thermals, and grow to smaller radii and volumes
(and therefore have smaller entrainment rates).

{Entrainment in shear flows may occur through the action of
large scale coherent structures (through `engulfment'), or intense small 
scale vortices (through `nibbling'), or a combination \citep{Westerweel2005,Philip2012}. {The role of buoyancy in entrainment (LJ19 and Section \ref{sec:dry_unstrat}) suggests the former, while the fact that increased turbulence leads to greater entrainment in moist thermals suggests that the latter may not be negligible. The relative importance of these mechanisms (whether there is a transition beyond some critical $Re$, say) is a subject of ongoing study.}}

{Turbulence is known to lead to increased scalar mixing \emph{across} a shear layer  \cite[e.g.][]{Baker1984}. In clouds, this mixing of the saturated cloud parcel with the unsaturated ambient air can lead to evaporative cooling and a subsiding shell of colder fluid at the edges of the cloud \cite{heus2008}. This subsiding shell is richer in vapour that the ambient, and may subsequently be re-entrained into the cloud  \citep{Romps2010a}, thus altering the rate of dilution of the flow \cite{Hannah2017}. As a result, trailing thermals often encounter different properties from their predecessors which dissipated more quickly, which is an important factor in the behaviour of thermal chains \cite{Morrison2020a,Peters2020}.}\\

{

We noted that due to the addition of buoyancy through condensation heating and the resulting increase in both their vertical velocity and radius, the turbulent moist thermals in figures \ref{fig:moist_unstrat_T}--\ref{fig:moist_unstrat_u} have effective Reynolds numbers $4-5$ times larger than the nominal $Re=6300$. 
In section \ref{sec:dry_unstable}, we study the addition of buoyancy through unstable stratification instead of condensation heating. 

Furthermore, in order to delineate the effects of buoyancy addition from the effects of turbulence, we study thermals with a nominal $Re=12600$ in section \ref{sec:high_Re}.}

\subsection{Dry thermals {in a dry-unstable ambient}} \label{sec:dry_unstable}

\begin{figure}
     \centering
     \begin{subfigure}[b]{0.32\textwidth}
         \centering
         \includegraphics[width=\textwidth]{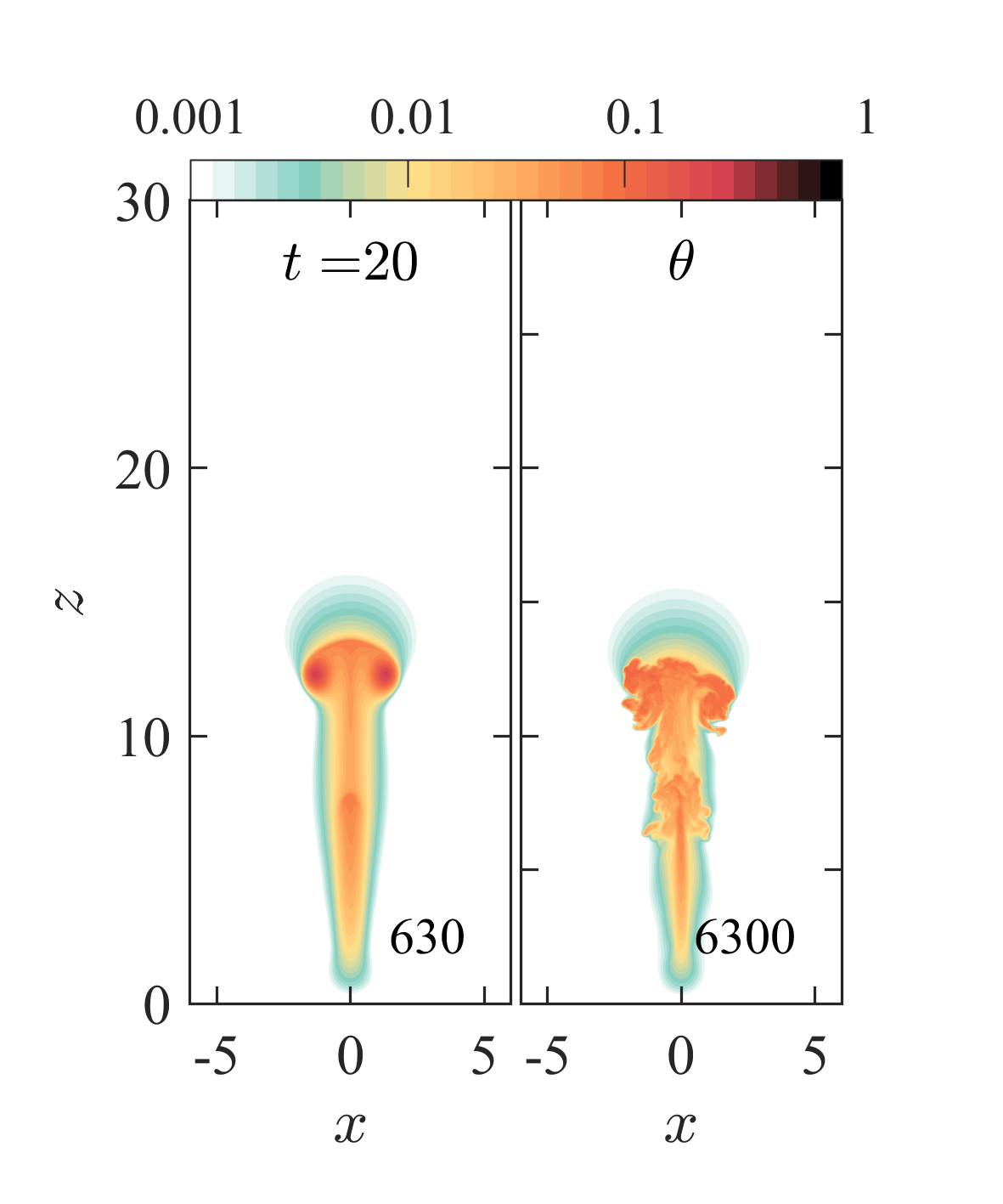}
         \caption{}
     \end{subfigure}
     \hspace{0.1cm}
        \begin{subfigure}[b]{0.32\textwidth}
         \centering
         \includegraphics[width=\textwidth]{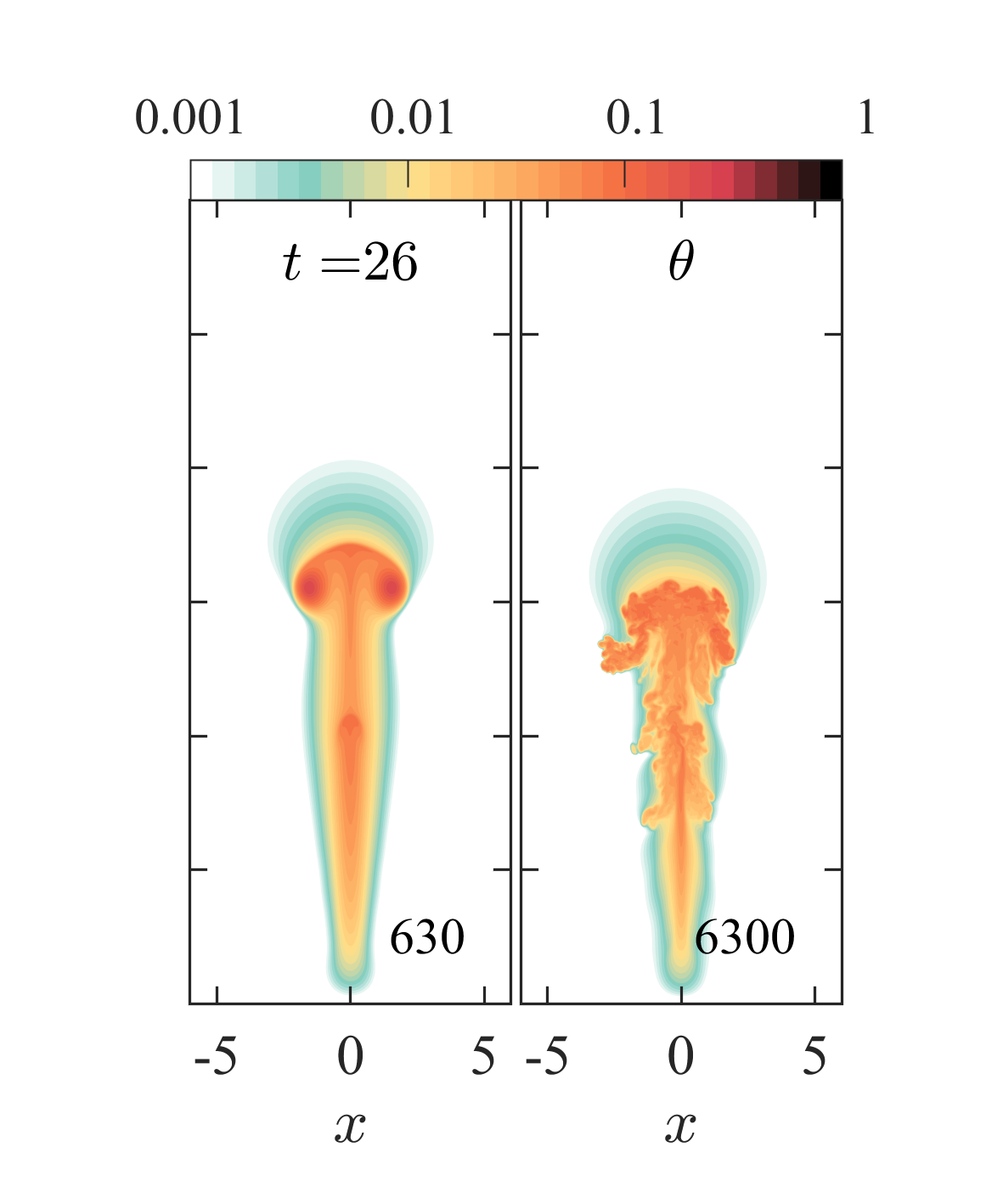}
         \caption{}
     \end{subfigure}
    \hspace{0.1cm}
     \begin{subfigure}[b]{0.32\textwidth}
         \centering
         \includegraphics[width=\textwidth]{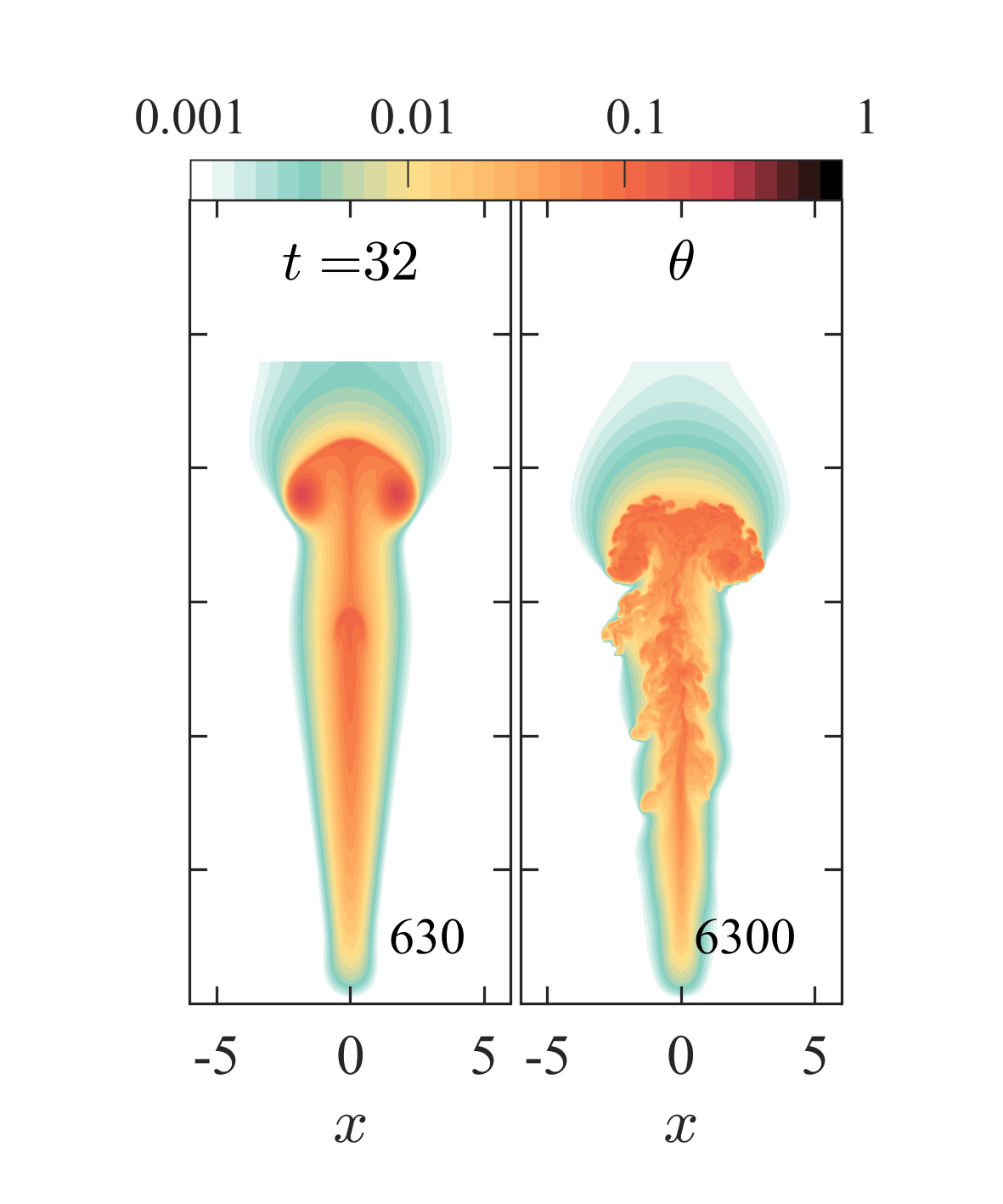}
         \caption{}
     \end{subfigure}
     \caption{As in figures \ref{fig:dry_unstrat} and \ref{fig:moist_unstrat_T}, but
     for dry thermals in an unstably stratified ambient.}
     \label{fig:dry_unstable_T}
\end{figure}

\begin{figure}
     \centering
     \begin{subfigure}[b]{0.32\textwidth}
         \centering
         \includegraphics[width=\textwidth]{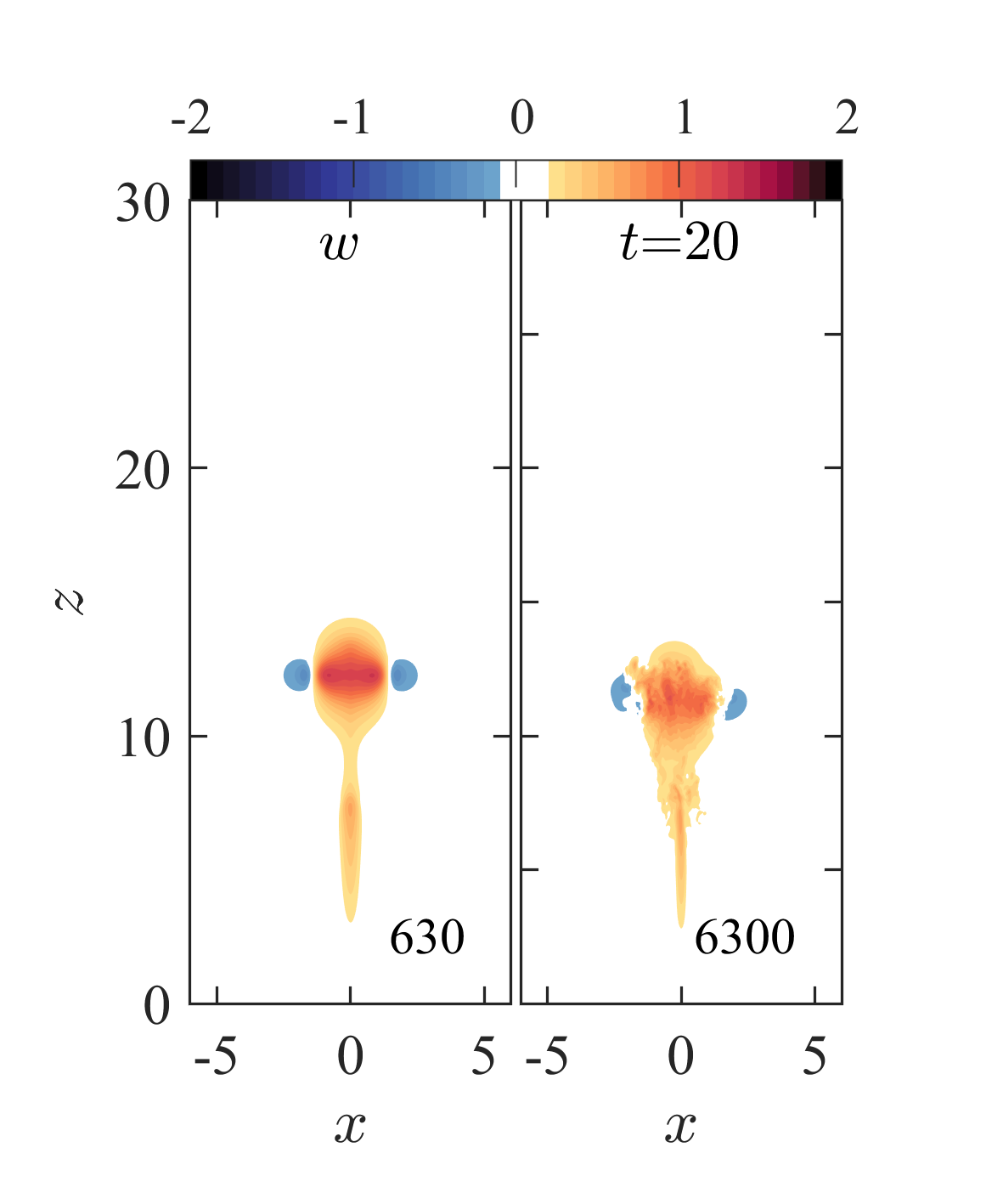}
         \caption{}
     \end{subfigure}
     \hspace{0.1cm}
        \begin{subfigure}[b]{0.32\textwidth}
         \centering
         \includegraphics[width=\textwidth]{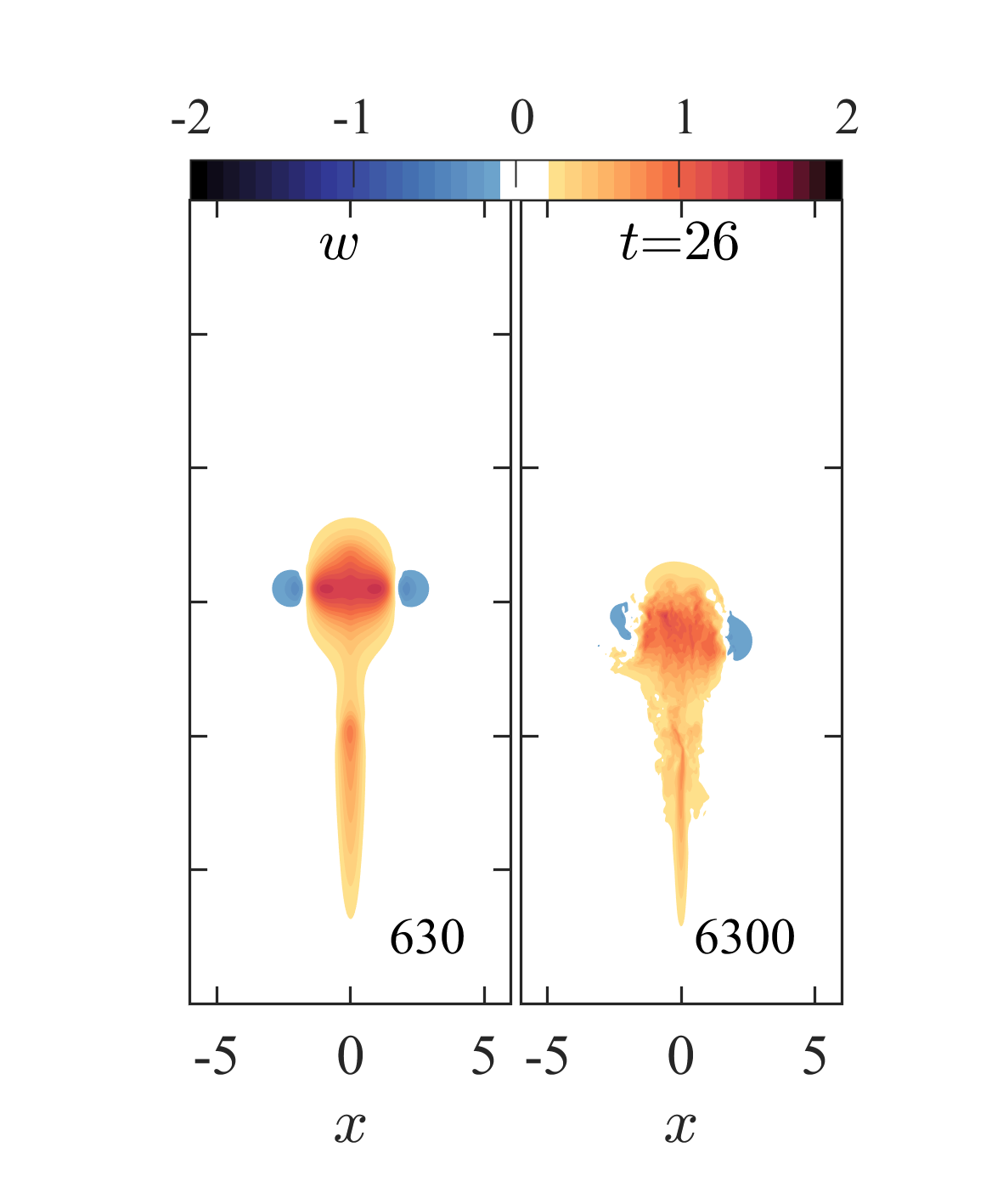}
         \caption{}
     \end{subfigure}
    \hspace{0.1cm}
     \begin{subfigure}[b]{0.32\textwidth}
         \centering
         \includegraphics[width=\textwidth]{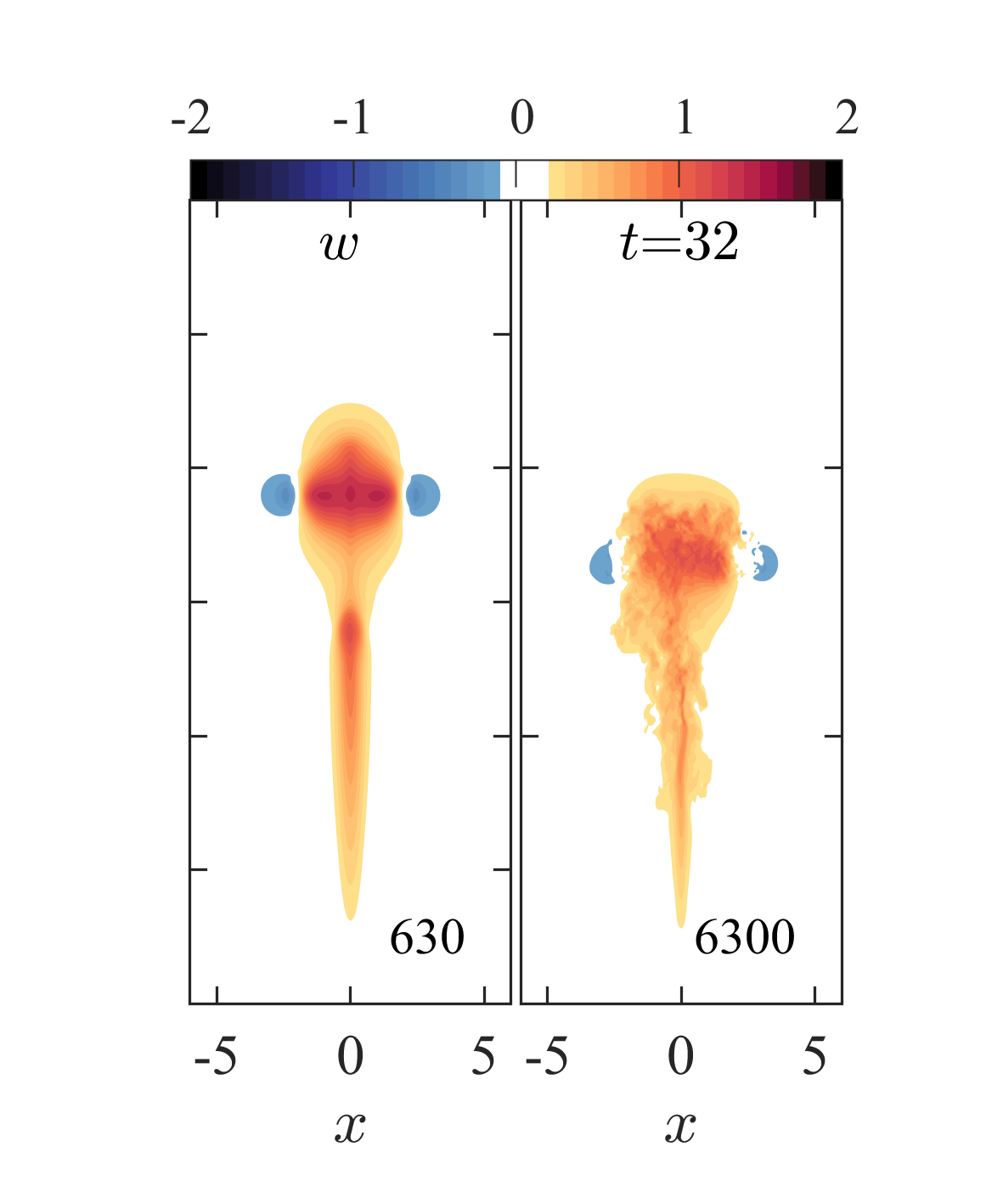}
         \caption{}
     \end{subfigure}
     \caption{As in figure \ref{fig:moist_unstrat_u}, but for dry thermals in an unstably stratified ambient. }
     \label{fig:dry_unstable_u}
\end{figure}

In moist thermals, the buoyancy of the thermal increases due to condensation 
heating. The buoyancy of the thermal can also increase if the ambient is unstably stratified, i.e. when the lapse rate in the ambient is greater than the dry adiabatic lapse rate. The evolution of $\theta$ and $w$ for a dry thermal in an ambient with $\Gamma_0 - \Gamma_u = 0.02$ is shown in figures \ref{fig:dry_unstable_T} and \ref{fig:dry_unstable_u} respectively.
We see that the flow resembles the starting plumes described by Turner \cite{Turner1962}, with heads that are thermals and plumes rising behind them.  We note that the shape of the thermals in figures \ref{fig:dry_unstable_T} resembles the `arrowhead' shape in figures \ref{fig:moist_unstrat_T}.  Since the ambient is inherently unstable, even small upward velocities that the ambient air in the wake of the thermal are amplified.

\begin{figure}
     \centering
     \begin{subfigure}[b]{0.47\textwidth}
         \centering
         \includegraphics[width=\textwidth]{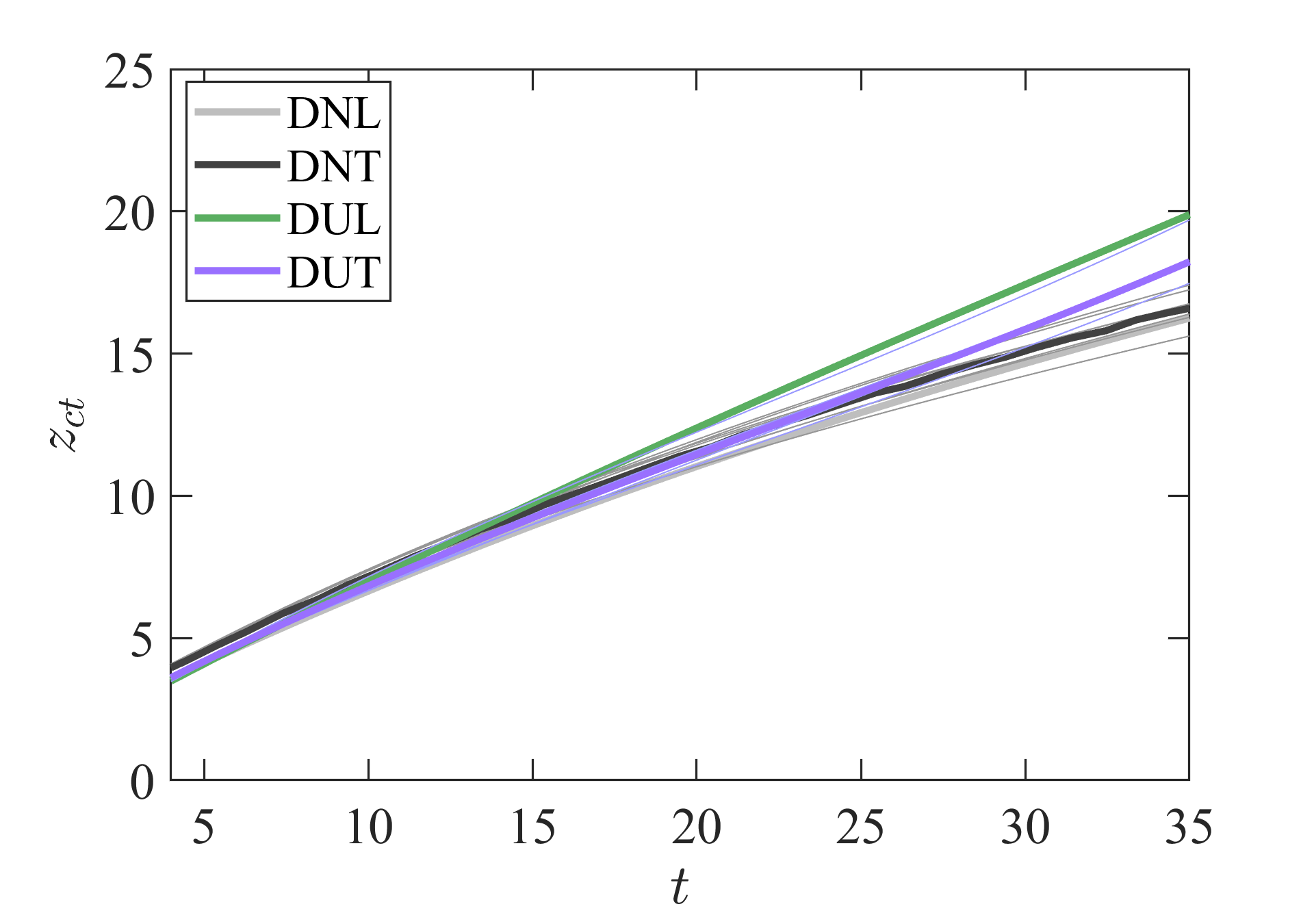}
         \caption{}
     \end{subfigure}
     \hspace{0.1cm}
        \begin{subfigure}[b]{0.47\textwidth}
         \centering
          \includegraphics[width=\textwidth]{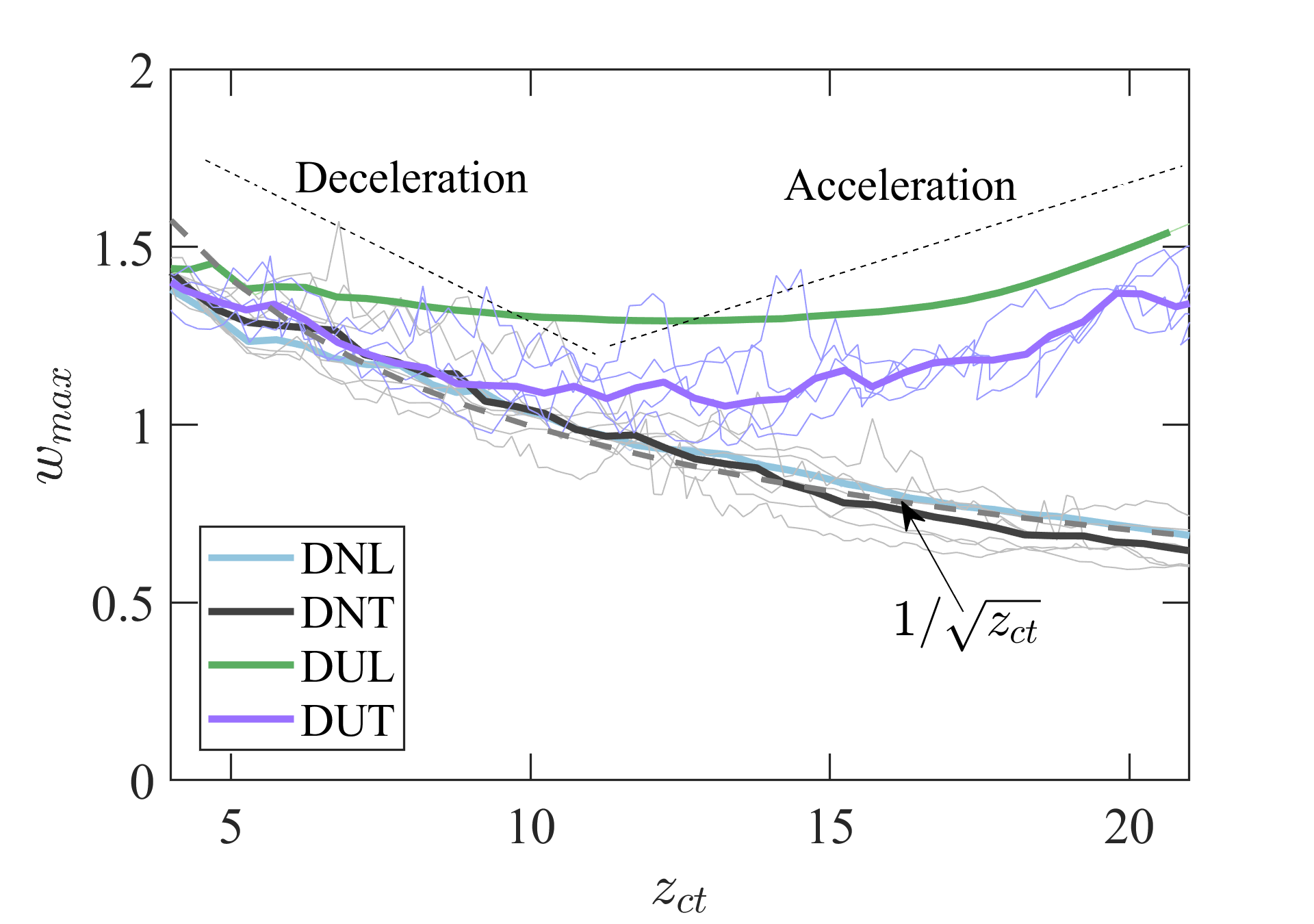}
          \caption{}
     \end{subfigure}
     \hspace{0.1cm}
        \begin{subfigure}[b]{0.47\textwidth}
         \centering
        \includegraphics[width=\textwidth]{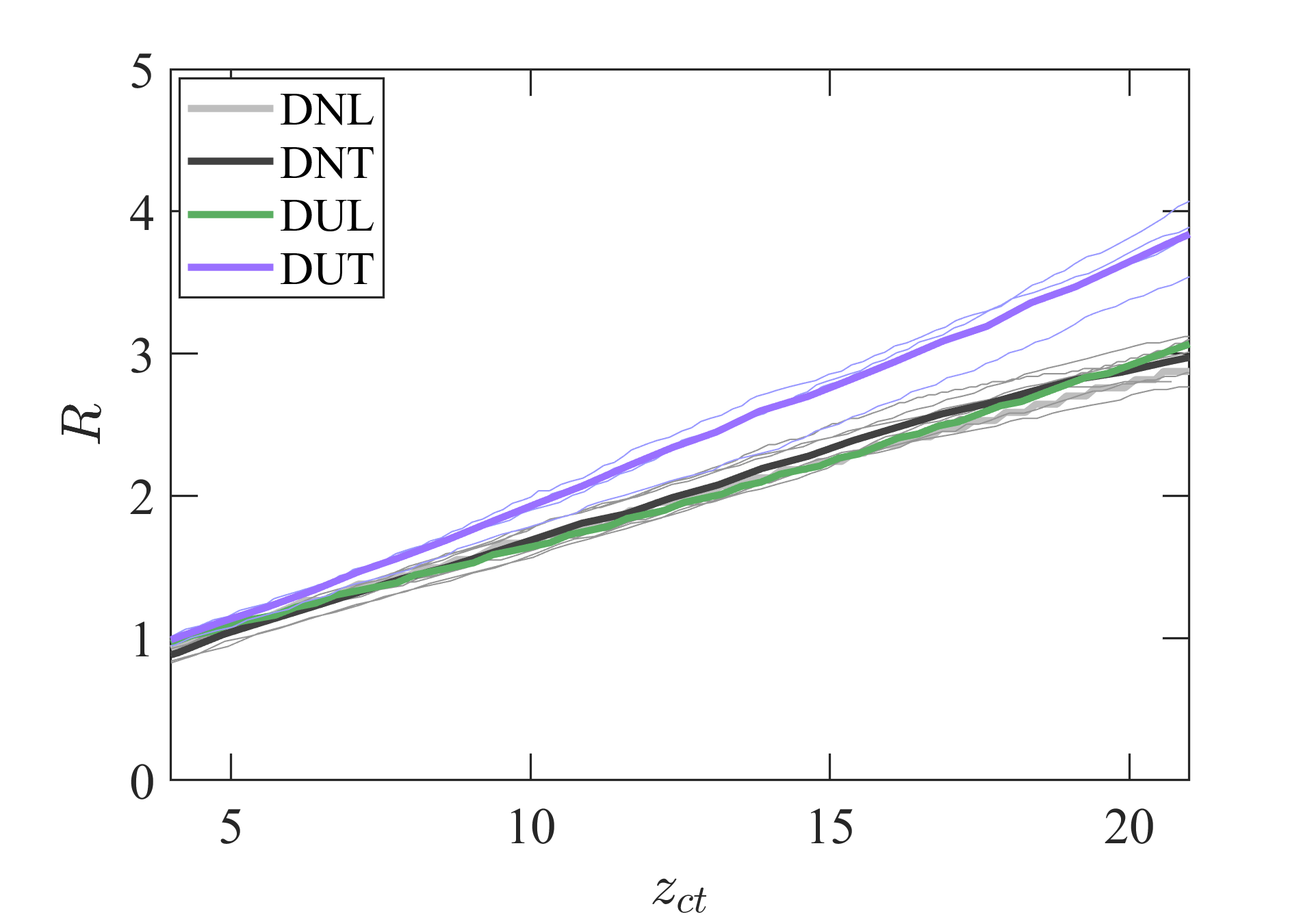}
        \caption{}
     \end{subfigure}
          \hspace{0.1cm}
        \begin{subfigure}[b]{0.47\textwidth}
         \centering
        \includegraphics[width=\textwidth]{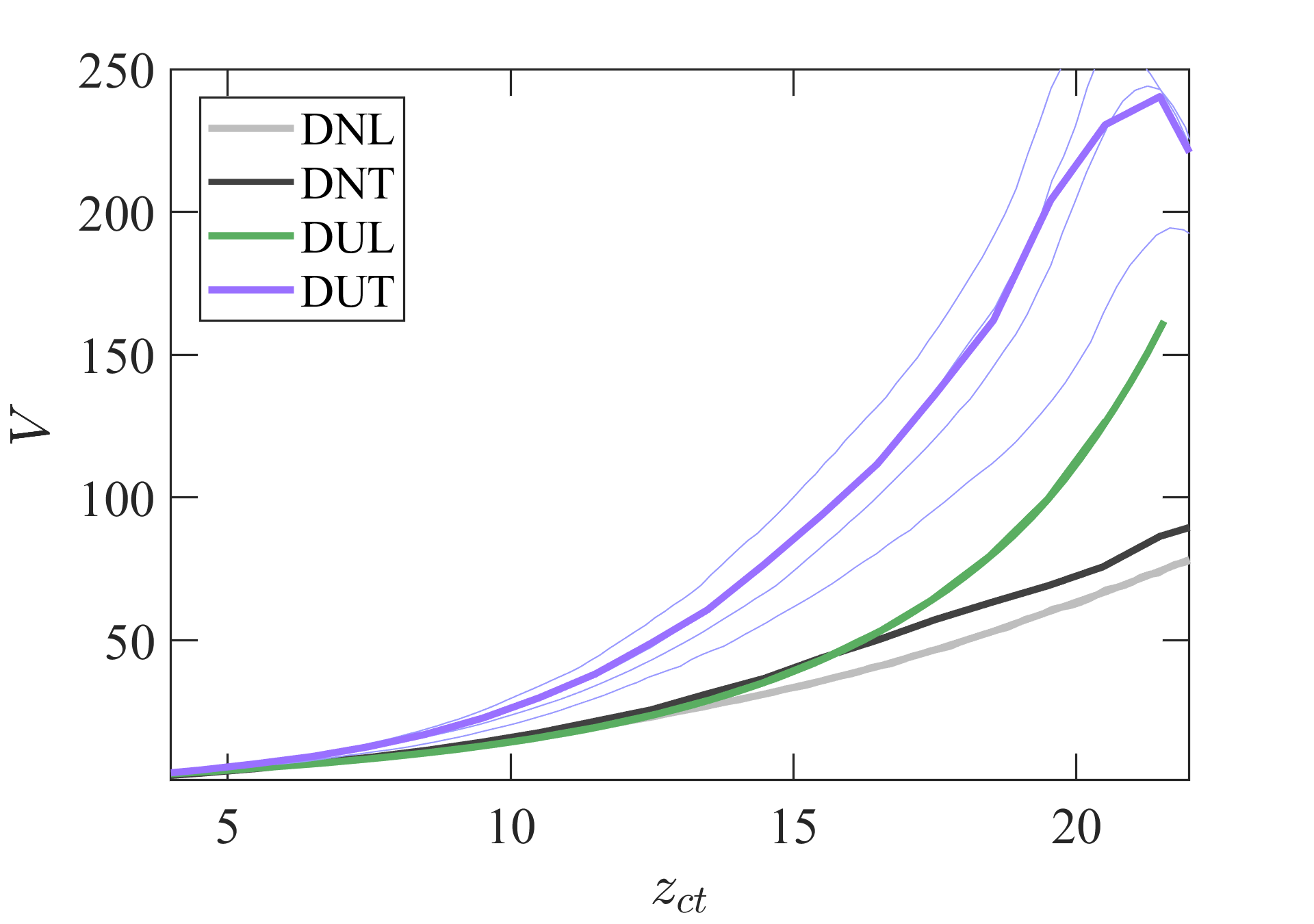}
        \caption{}
     \end{subfigure}
     \caption{The (a) location $z_{ct}$, (b) vertical velocity $w_{th}$ (c) radius $R$, and (d) volume $V$ of laminar ($Re=630$) and turbulent ($Re=6300$) dry thermals in an unstably stratified ambient with $\Gamma_0-\Gamma_u=0.02$ compared with dry thermals in a neutrally stratified ambient. Similarly to moist thermals in a (dry-) neutrally stratified ambient, the accelerating thermals here rise faster, grow to greater radii, and have larger volumes than dry thermals in a neutrally stratified ambient. Also note that the laminar thermals rise faster but have smaller radii and volumes than the turbulent thermals. The changing shape of the thermal accounts for the increase in thermal volume in the laminar case even though the thermal radius evolves
     similarly to the dry thermals in a neutral ambient (c.f. \ref{fig:moist_unstrat_V_w_R}). 
     \label{fig:dry_unstable_V_w_R}}
\end{figure}

In figure \ref{fig:dry_unstable_V_w_R}, we compare the location, maximum flow
velocity, radius and volume of dry thermals in unstably and neutrally stratified ambients.
As in moist thermals (section \ref{sec:moist_unstrat}), the volume and
radius (and velocity) of dry thermals in unstably stratified ambients increase with
time faster than dry thermals in a neutraly stratified ambient. Furthermore, as in
moist thermals, the influence of turbulence is larger in accelerating dry thermals 
than in dry thermals in a neutrally stratified ambient. Unlike in
moist thermals, however, the arrowhead shape is not completely destroyed in
turbulent dry thermals here.



\subsection{Dry thermals {in a dry-neutral ambient} at $Re=12600$} \label{sec:high_Re}

\begin{figure}
    \centering
    \includegraphics[width=0.7\columnwidth]{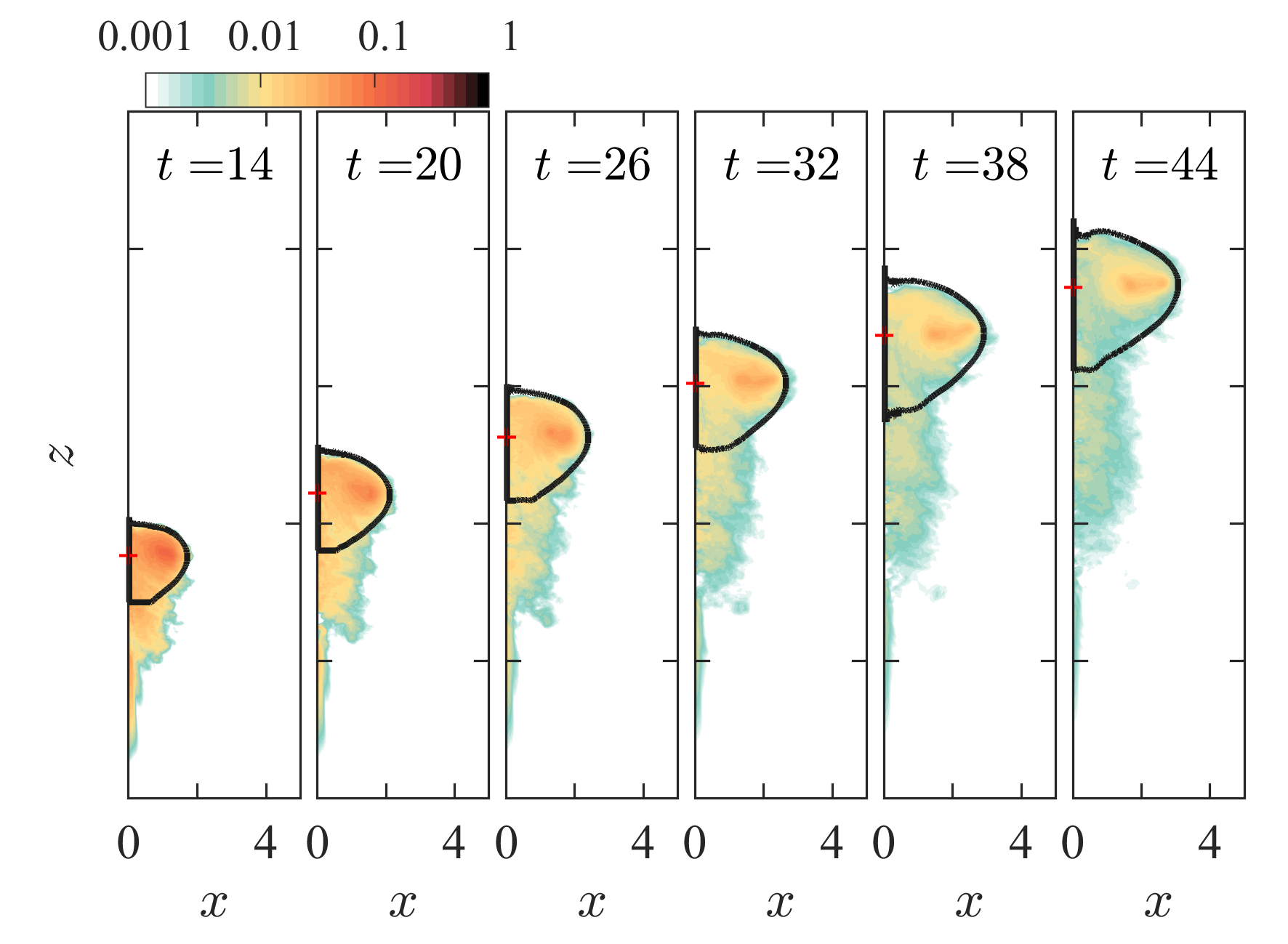}
    \caption{The evolution of a dry turbulent thermal at $Re=12600$. The contour plots shows the aximuthally averaged temperature $\theta$ and as with other cases, the black line represent the thermal boundary. C.f. the laminar and turbulent ($Re=6300$) cases in figure \ref{fig:dry_unstrat}.}
    \label{fig:dry_Re12k}
\end{figure}

Figures \ref{fig:dry_unstrat} show that turbulent dry thermals can evolve such that their vortex cores are not small, with vorticity in their interiors and that they detrain at larger rates. The spread of the vortex cores is greater for an even larger $Re=12600$, as seen from the evolution in figure \ref{fig:dry_Re12k}.
In section \ref{sec:theory}, we hypothesised that the spreading of the vortex core, (or even complete disintegration; see section \ref{sec:moist_unstrat})
may mean that at a sufficiently large Reynolds number, the entrainment could 
change significantly. However, our results at $Re=12600$ are inconclusive. The
mean values of entrainment rate and entrainment efficiency are approximately the 
same as those found for $Re=6300$ 
We note that this is an imperfect test: while finding significant differences would have shown that turbulence plays a role, finding no differences does not rule out a role for turbulence, since the Reynolds number studied here is barely above the mixing transition Reynolds number of $10^4$.\\

\begin{figure}
     \centering
     \begin{subfigure}[b]{0.47\textwidth}
         \centering
        \includegraphics[width=\textwidth]{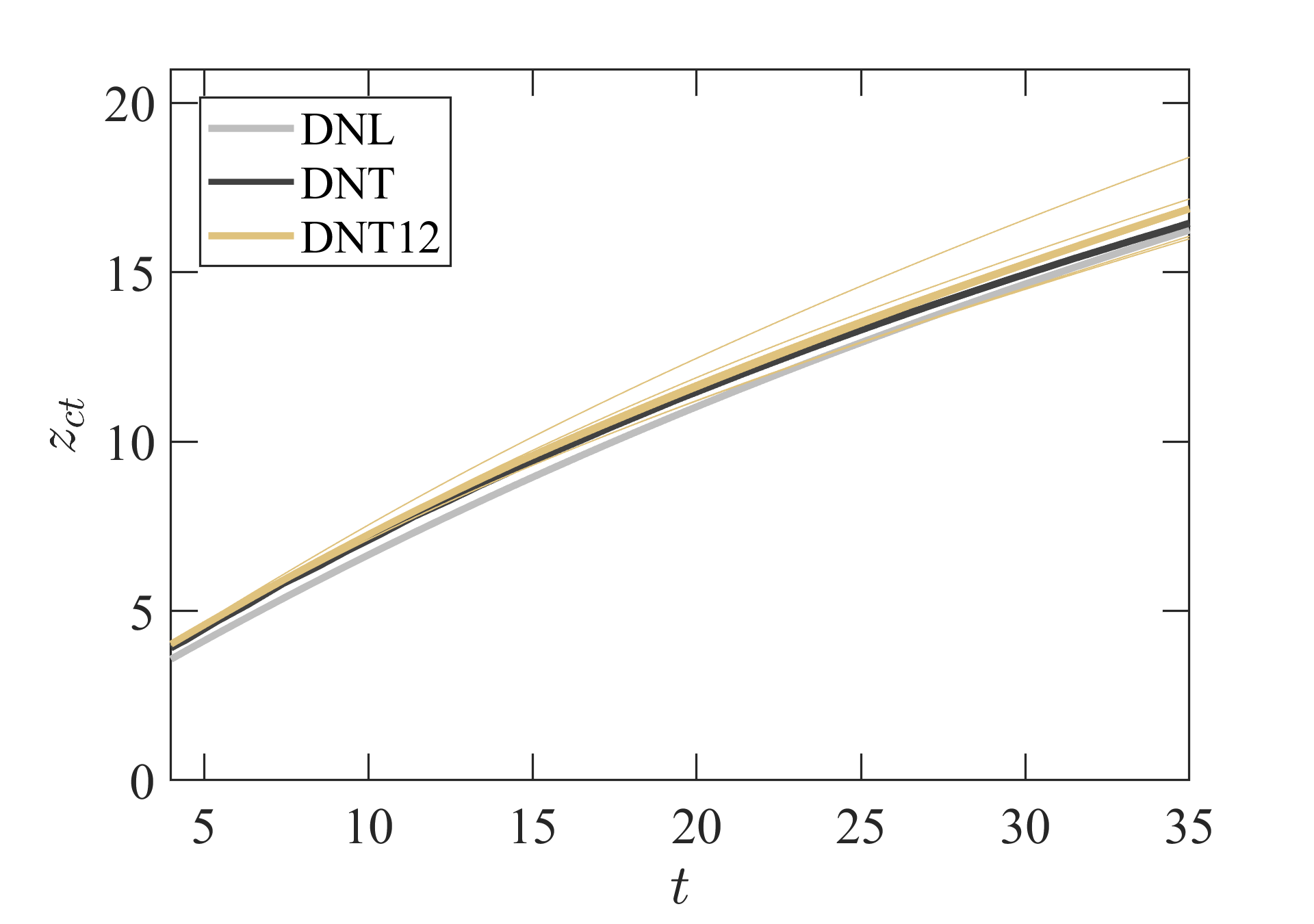}
        \caption{}
     \end{subfigure}
     \hspace{0.1cm}
        \begin{subfigure}[b]{0.47\textwidth}
         \centering
        \includegraphics[width=\textwidth]{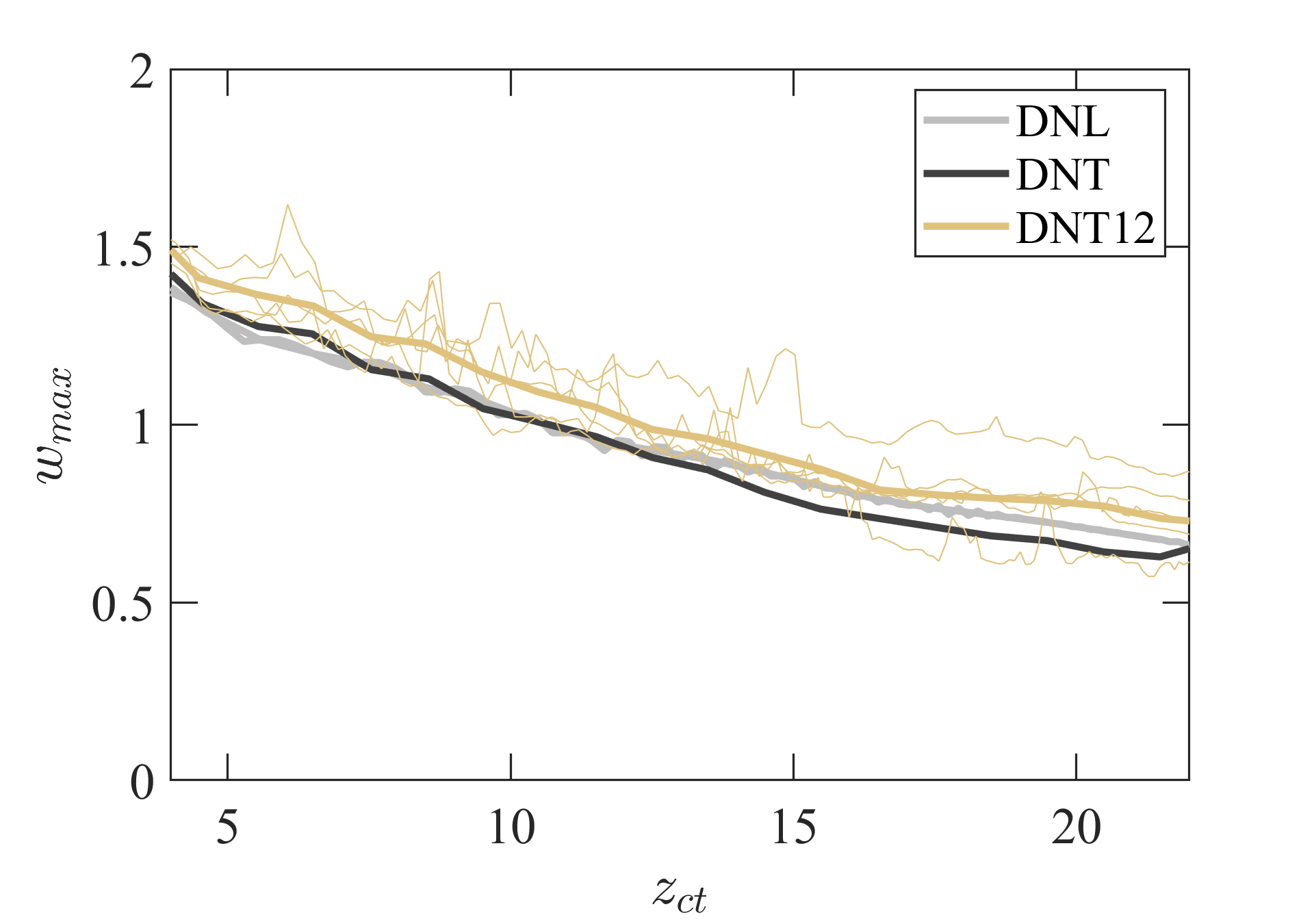}
        \caption{}
     \end{subfigure}
     \hspace{0.1cm}
        \begin{subfigure}[b]{0.47\textwidth}
         \centering
          \includegraphics[width=\textwidth]{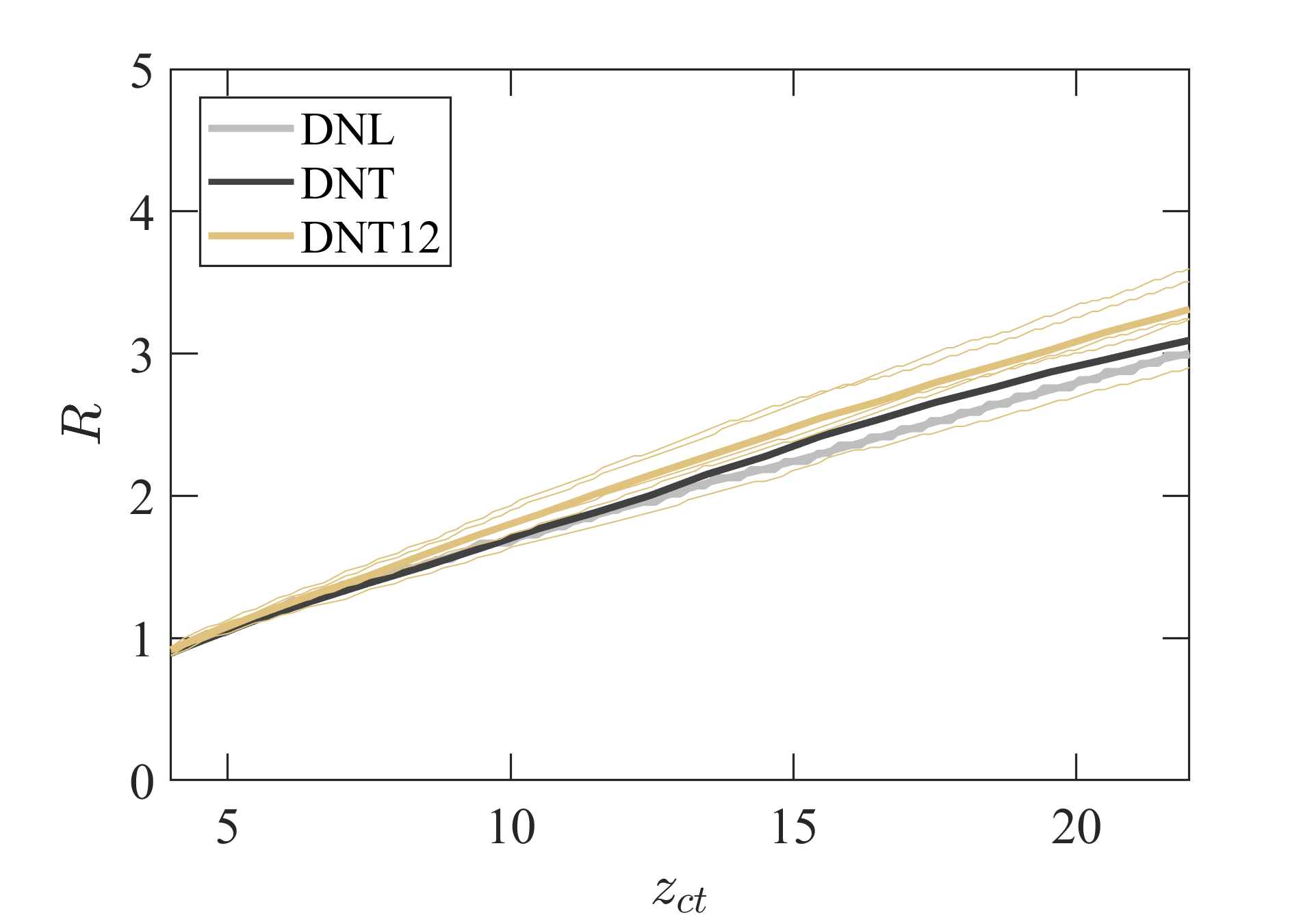}
        \caption{}
     \end{subfigure}
          \hspace{0.1cm}
        \begin{subfigure}[b]{0.47\textwidth}
         \centering
          \includegraphics[width=\textwidth]{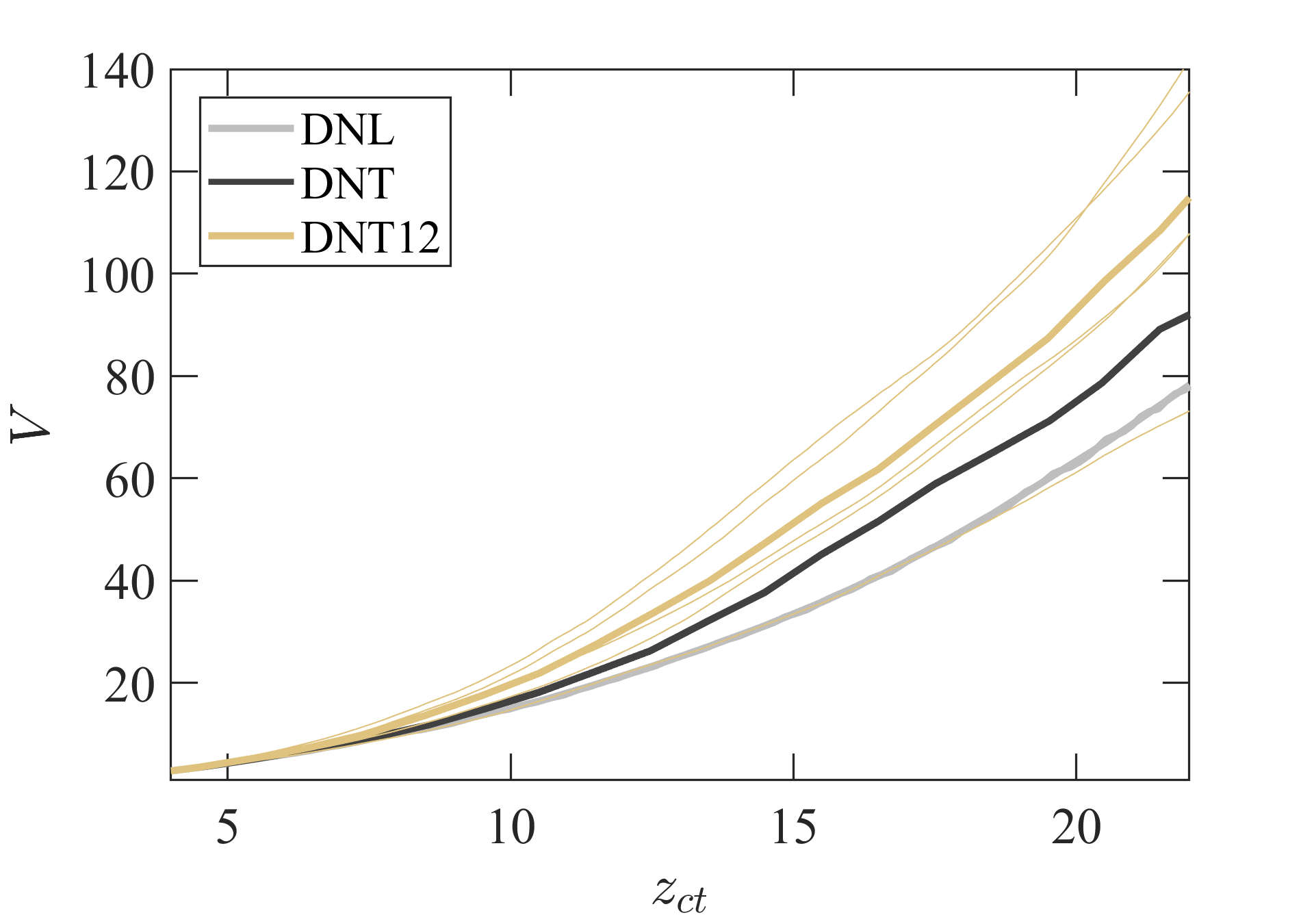}
        \caption{}
     \end{subfigure}
     \caption{ 
     \label{fig:dry_unstrat_V_w_R} The (a) location $z_{ct}$, (b) maximum velocity $w_{max}$,
     (c) radius $R$ and (d) volume $V$ of a laminar dry thermal ($Re=630$) and
     turbulent dry thermals ($Re=6300$ and $Re=12600$) in a neutrally stratified ambient. {In the turbulent case, curves for} individual runs at $Re=6300$ (thin grey lines) {and $Re=12600$} are plotted along
     with the ensemble average.}
   
\end{figure}
The plots of thermal location, maximum velocity, radius and volume in figure
\ref{fig:dry_unstrat_V_w_R} show that thermals at the higher Reynolds numbers rise
at (slightly) smaller velocities, and that the thermal volume and radius increase
faster for larger $Re$. The entrainment rates are therefore larger for larger $Re$.
This is not a qualitative change, however, and simulations at Reynolds number much
greater than $10^4$ are therefore necessary to determine if the effects of 
turbulence are indeed small beyond the mixing transition.


\subsection{Entrainment rate and entrainment efficiency} \label{sec:entrainment}
\begin{figure}
     \centering
  
     \hspace{1cm}
     \begin{subfigure}[b]{0.4\textwidth}
         \centering
         \includegraphics[width=\textwidth]{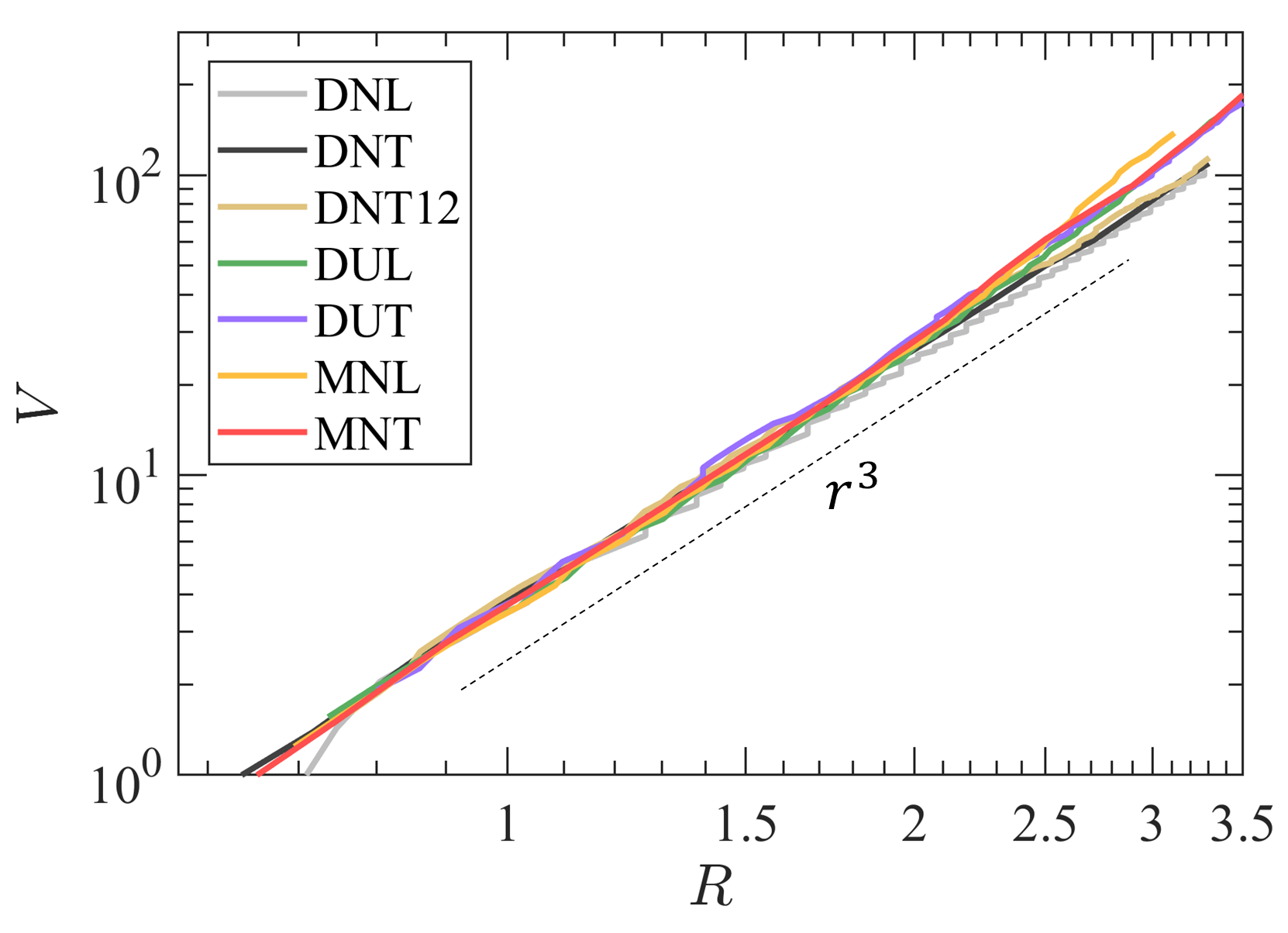}
     \end{subfigure}
     \caption{As thermals rise, the volume scales with the thermal radius as $V \propto R^{3} $, suggesting self-similarity. }
     \label{fig:thermal_volume}
\end{figure}

In sections \ref{sec:dry_unstrat} - \ref{sec:high_Re}, we examined several 
cases of dry and moist thermals, and compared their evolution with the standard
case of dry thermals in a neutrally stratified ambient.
In figure \ref{fig:thermal_volume}, we plot the thermal volume as a function of the thermal radius, showing that $V \sim R^3$ and suggesting that the evolution of the thermals in all these cases is self-similar, {and therefore that analytical or reduced-order models for the evolution of thermals may be possible.}

Figures \ref{fig:moist_unstrat_V_w_R}, \ref{fig:dry_unstable_V_w_R} and
\ref{fig:dry_unstrat_V_w_R} also show that thermals that have an external source of buoyancy--either condensation heating or an unstable ambient stratification--grow fastest in volume, and thus have the largest entrainment rates.
\begin{figure}[H]
     \centering
     \begin{subfigure}[b]{0.4\textwidth}
         \centering
         \includegraphics[width=\textwidth]{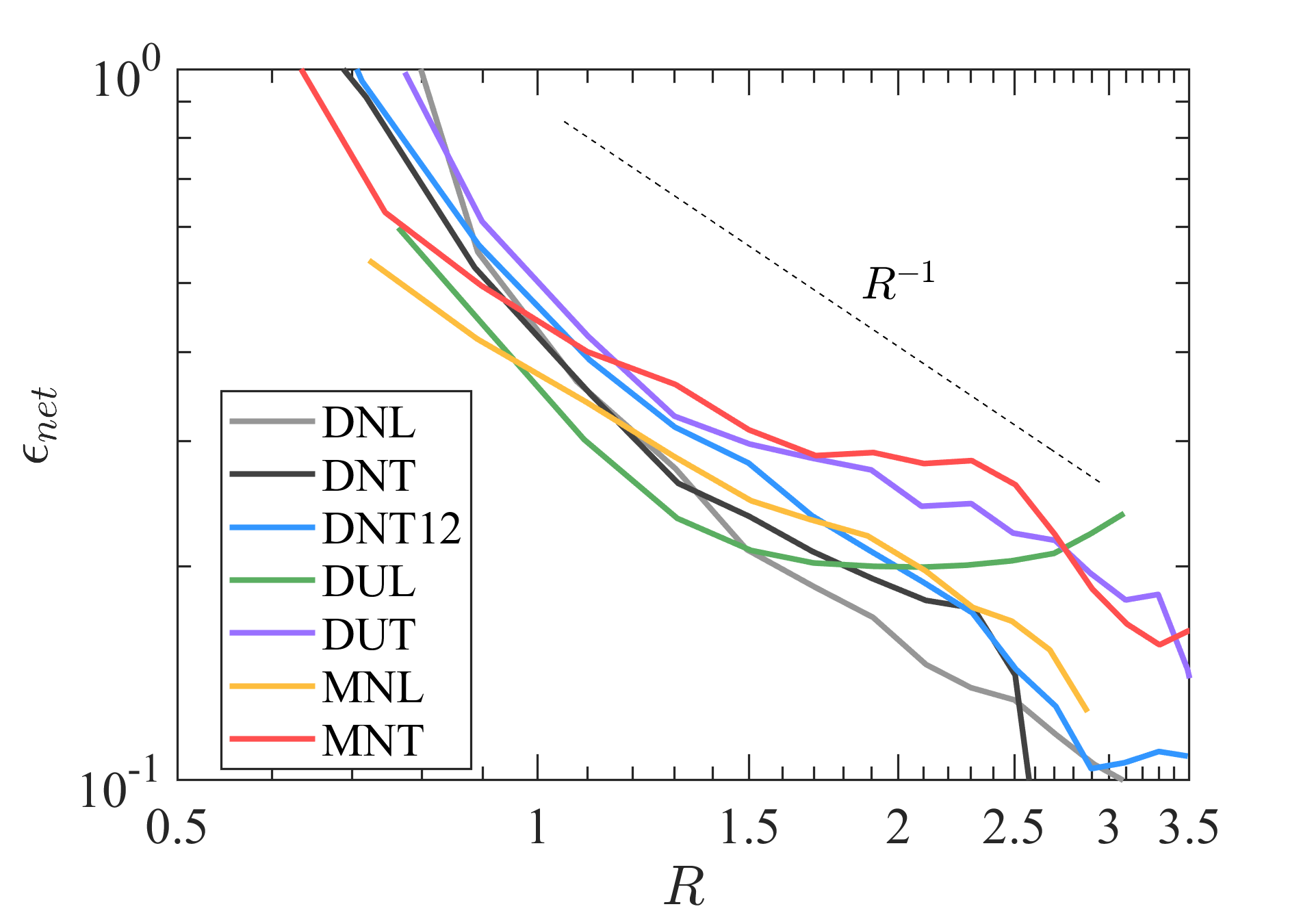}
         \caption{}
     \end{subfigure}
     \hspace{1cm}
     \begin{subfigure}[b]{0.4\textwidth}
         \centering
         \includegraphics[width=\textwidth]{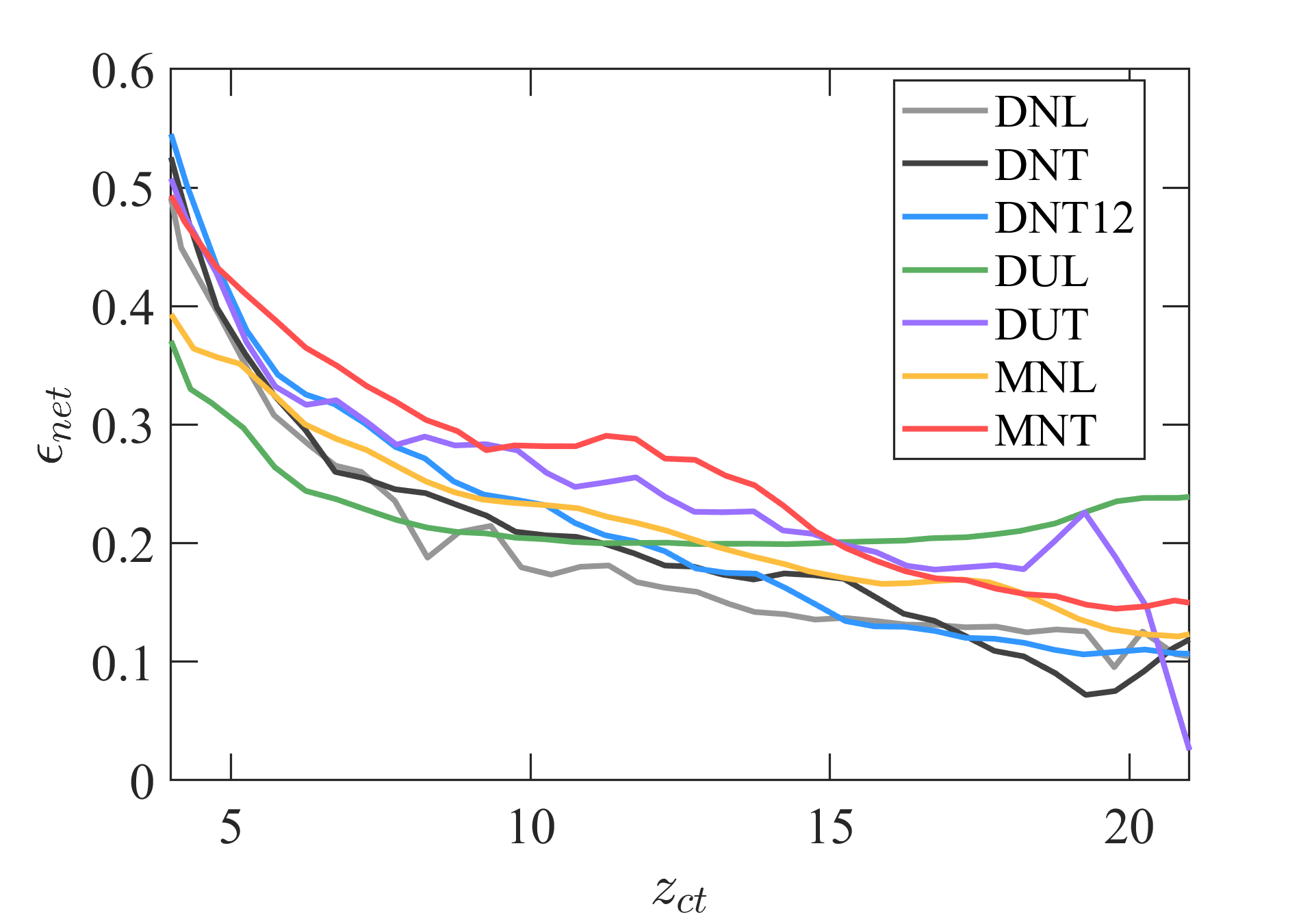}
         \caption{}
     \end{subfigure}
         \hspace{1cm}
     \begin{subfigure}[b]{0.4\textwidth}
         \centering
         \includegraphics[width=\textwidth]{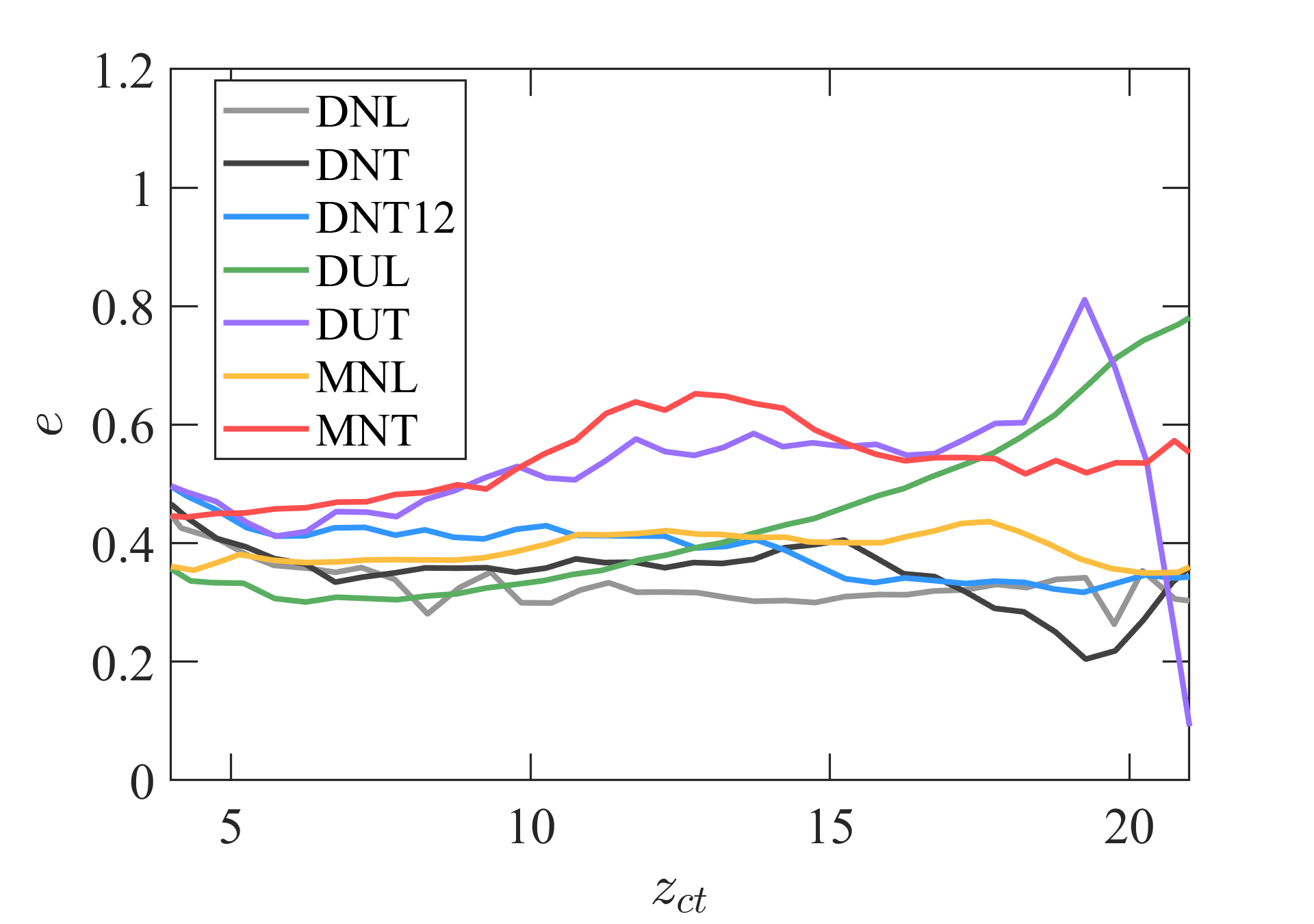}
         \caption{}
     \end{subfigure}
         \hspace{1cm}
     \begin{subfigure}[b]{0.4\textwidth}
        \centering 
        \includegraphics[width=\textwidth]{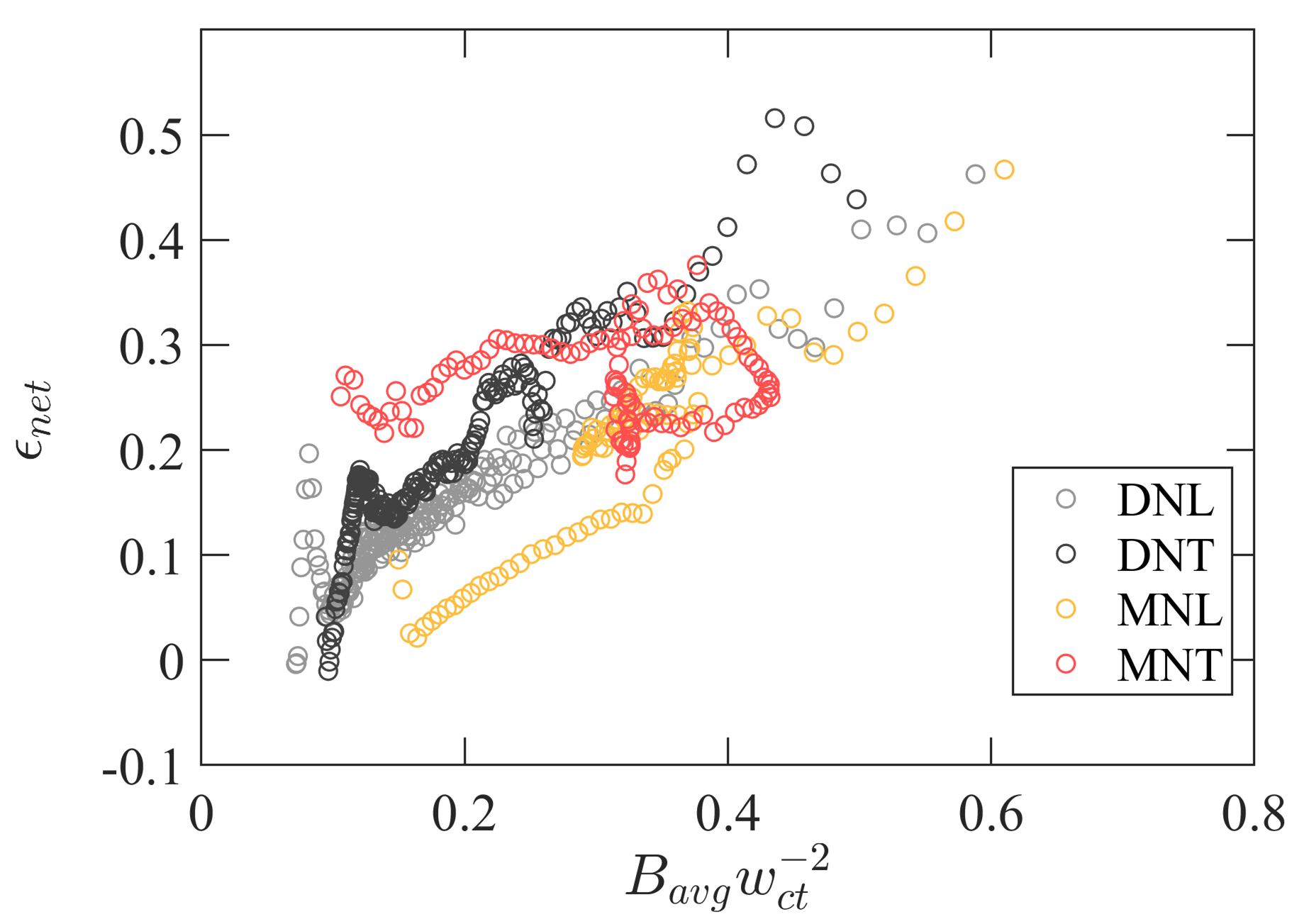}
        \caption{}
     \end{subfigure}
     \caption{(a) The net entrainment rate $\epsilon_{net}$ as a function of the
     thermal radius $R$ for the cases in sections \ref{sec:dry_unstrat} - 
     \ref{sec:high_Re}. The dotted line represents the scaling relation equation 
     \ref{eq:entrainment_rate}. The moist laminar and turbulent cases seem to have
     slightly higher entrainment rates. (b) $\epsilon_{net}$ as a function of the thermal location $z_{ct}$. (c) The entrainment efficiency $e$ 
     (equation \ref{eq:entrainment_efficiency}) plotted as a function of $z_{ct}$.
     The entrainment efficiency in the moist cases is larger than the entrainment efficiency in the corresponding dry cases. 
     {(d) The correlation $\epsilon_{net} \sim B_{avg} w_{ct}^{-2}$ (see text), showing a reasonable collapse. The points are plotted for every 0.3 flow time unit}
     \label{fig:entrainment_combined}}
\end{figure}
{Cumulus cloud entrainment is parameterized in cloud models using
dimensional and scaling relationships, with the entrainment varying as $\epsilon \propto$ $R^{-1}$, $Z_{th}^{-1}$ and $B_{avg} w_{th}^{-2}$  \citep[][]{HernandezDeckers2018}}.
In figures \ref{fig:entrainment_combined}(a,b), we plot the
entrainment rate as a function of the thermal radius $R$ and as a function of the thermal height $z_{ct}$. The curves deviate to different degrees from the $\epsilon_{net} \sim R^{-1}$ for dry thermals. {Figure \ref{fig:entrainment_combined}(a) shows that while $\epsilon_{net}$ $\propto$ $R^{-1}$ holds well for laminar and turbulent dry thermals, the entrainment in MNT and DUT thermals deviates from the $\epsilon_{net} \sim R^{-1}$ relationship especially in the later stages).}

{Both off-source addition of buoyancy to the thermals (either through condensation or through unstable stratification), while the role of turbulence (higher $Re$) in increasing the entrainment rate is marginal.  
This is seen in figure \ref{fig:entrainment_combined}(c), where we plot
the entrainment efficiency in these different cases.}

{Figure \ref{fig:entrainment_combined}(c) also shows that the entrainment efficiency $e$ in the MNL thermal is always marginally higher than the DNL thermal, consistent with the role of buoyancy in inducing entrainment envisaged in LJ19 (see also section \ref{sec:dry_unstrat}). However, this is contrary to \cite{Morrison2021} where  $e$  is lower than the dry thermal at lower levels of thermal assent and thereafter it is similar to dry thermal.}
{The marginal role of turbulence (alone) in entrainment is further corroborated by the observation that the entrainment at a higher (nominal) $Re$ in the DNT12 thermal is only marginally higher than the DNT thermal. However, the entrainment in turbulent thermals with \emph{increasing buoyancy} (MNT or DUT) is significantly higher, suggesting that the interplay of volumetric heating and turbulence is responsible.}
{In figure \ref{fig:entrainment_combined}(d), we plot the entrainment rate $\epsilon$ vs the ratio $B_{avg} / w_{ct}^2$, showing that these are proportional. We note that the  however, we note that the spread is the data is not small. A systematic study of the curves for MNT thermals as the ambient relative humidity is varied will be presented elsewhere.}

\begin{figure}
     \centering
         \centering
         \includegraphics[width=\textwidth]{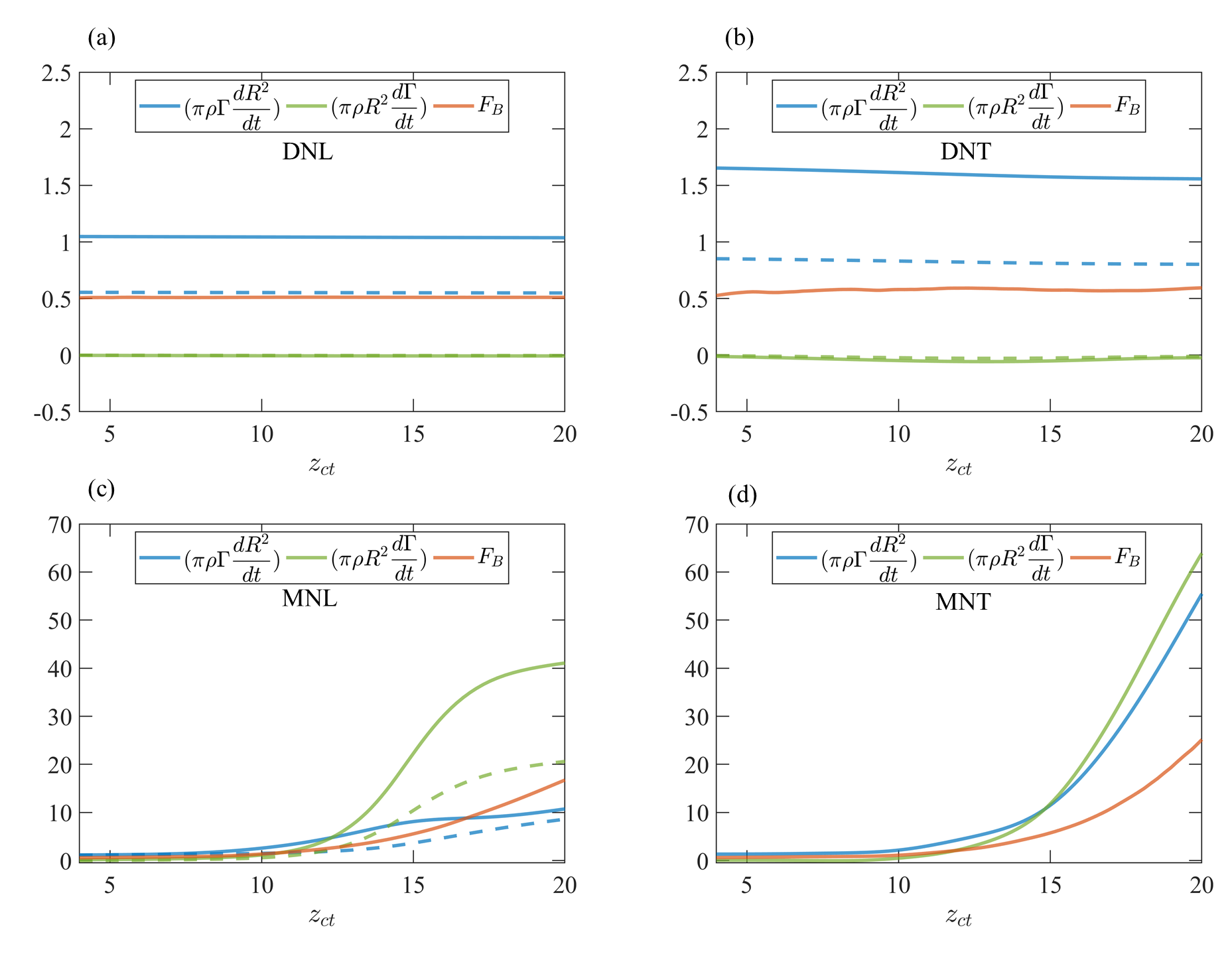}
     \caption{{The terms in the impulse budget, Eq. \ref{eq:dIdt}, as a function of the thermal height $z_{ct}$ for (a) dry laminar (b) dry turbulent, (c) moist laminar and (d) moist turbulent thermals. Terms calculated using the radius of the thermal are plotted with solid lines, while terms calculated using the vortex ring radius (in plots a-c) are plotted with dashed lines. In moist thermals, the buoyancy, the circulation and entrainment terms are all an order of magnitude higher than in dry thermals. Turbulence also leads to a greater entrainment rate in moist thermals. {Note that $\rho=1$ under the Boussinesq approximation.} \label{fig:impulse_budget}}}
\end{figure}

{For a more detailed understanding of the role of condensation heating in entrainment, we plot the entrainment and circulation terms in Eq. \ref{eq:dIdt} in Figures \ref{fig:impulse_budget}. For dry thermals, in agreement with LJ19, the circulation term remains small throughout the  evolution, while the buoyancy term is balanced by the entrainment term. For both laminar and turbulent moist thermals, on the other hand, the circulation term is nonzero. As a result, the individual terms on the left hand side (LHS) of the impulse budget (Eq. \ref{eq:dIdt}) in figure \ref{fig:impulse_budget} are much larger for moist thermals than for dry thermals, and the balance implied by Eq. \ref{eq:dIdt} breaks down for moist thermals.}

\begin{figure}
    \centering
    \includegraphics[width=0.9\columnwidth]{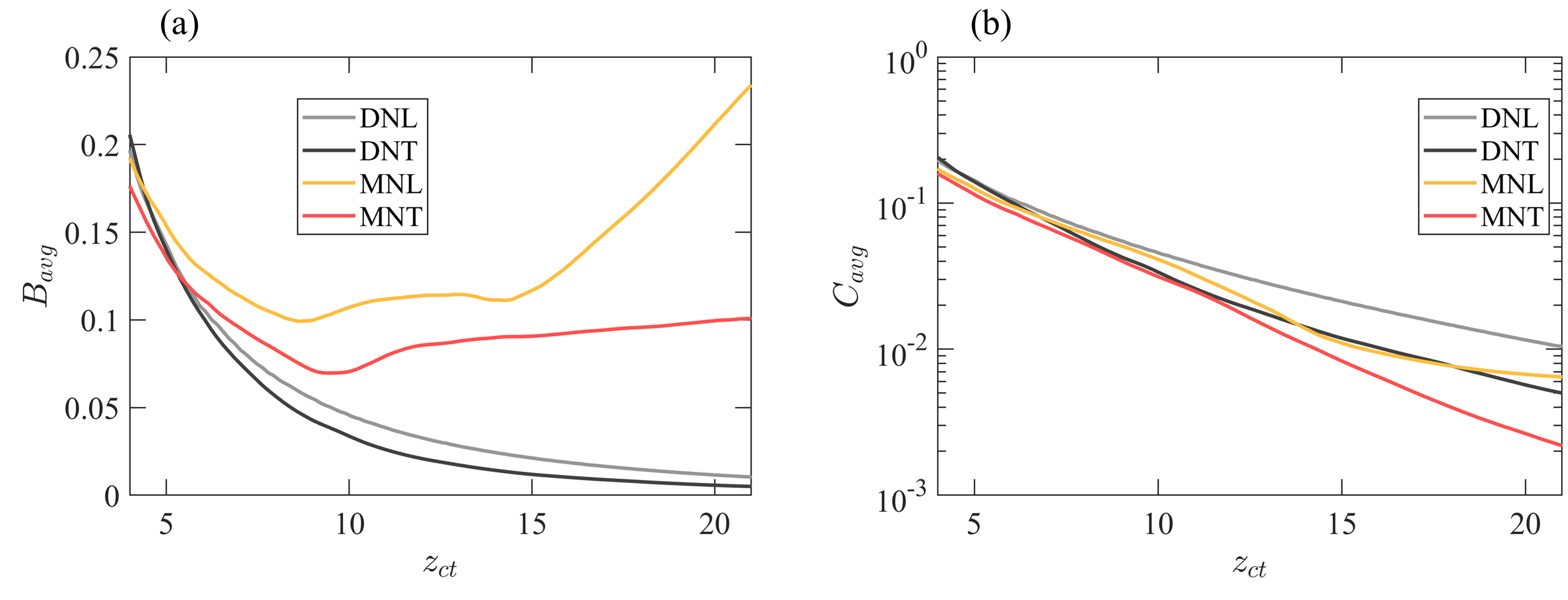}
    \caption{{\label{fig:buoy_avg} (a) The \emph{average} buoyancy $B_{avg}$, and (b) the \emph{average} value of the passive scalar $C$ (see text) in dry and moist thermals as they rise in dry-neutral ambient.}}
\end{figure}

{Furthermore, we see that the buoyancy term $F_B$ is slightly larger for MNT than for MNL thermals. This is due to the fact that while the mean buoyancy $B_{avg}$ in the thermals is lower for MNT thermals, the larger volume leads to a larger value of $F_B = B_{avg} \times V$. This may be seen explicitly from the plots of the average buoyancy $B_{avg}$ in figure \ref{fig:buoy_avg}(a). We note that the mean buoyancy increases for both MNL and MNT thermals due to latent heat release by condensation, but increases much faster for MNL thermals where the entrainment is lower. In dry thermals, where there is no source of buoyancy in the flow, the buoyancy decreases as expected.
Plots of the mean value $C_{avg}$ of a passive scalar $C$, initialised similarly to the temperature $\theta$ and governed by the equation
\begin{equation}
\frac{D C}{Dt}  = \frac{1}{Re} \nabla^2 C, \label{eq:passive_scalar} \\
\end{equation}
which is thus a direct measure of the dilution of the flow, also show that the larger entrainment dilutes the thermal to a greater extent, as expected.}

{From the foregoing arguments and figures \ref{fig:entrainment_combined}--\ref{fig:buoy_avg}, we draw the following conclusions about the effects of condensation heating and turbulence on entrainment: {(a) an increase in buoyancy alone increases the entrainment efficiency $e$ from $0.32$ in DNL thermals to $0.39$ in MNL thermals; (b) an increase in the Reynolds number alone increases $e$ to $0.37$ in DNT thermals and to a similar value in $DNT12$ thermals; (c) the combined effects of heating and turbulence lead to a significant increase in $e$ to a value of $0.54$ in MNT thermals (where the effective Reynolds number is $\approx2\times10^4$.)} This increase in the entrainment is made possible by the breakdown of the vortex rings, and thus equation \ref{eq:dIdt}, for moist thermals.}
{The increased entrainment is consistent with the parametric relationship $\epsilon \sim B_{avg} / w_{th}^2$, as shown in figure \ref{fig:entrainment_combined}(d), because while MNT thermals have smaller average buoyancy $B_{avg}$, they also have smaller velocities $w_{th}$, than MNL thermals.}

{The breakdown of the vortex ring in MNT thermals is accompanied by the generation of intense small scale vorticity, as seen in figures \ref{fig:moist_unstrat_omgt} and \ref{fig:omg_mag_dnt_mnt}(b). The role of this small-scale vorticity, which increases the mixing of the fluid inside the thermal \cite{Bhat1996}, in increasing the entrainment in turbulent moist thermals merits further study.}\\

We also note here that while we see an increase in the entrainment efficiency for moist
thermals over dry thermals, laboratory experiments on dry thermals find a larger
spread of entrainment values ranging from $0.4 - 0.7$ \citep[see e.g.][]{Morton1956, Scorer1957, Johari1992}.

\section{Conclusion} \label{sec:conclusion}
In summary, we have presented results from direct numerical simulations (DNS) 
of laminar and turbulent dry {thermals in neutral and unstable ambients and laminar and turbulent moist thermals in a dry-neutral ambient}. Using a self-consistent and robust definition of the
volume of a thermal, we studied how the properties of thermals--their
velocities, volumes and radii--evolve under these different conditions.
Our results for dry thermals in a neutrally stratified ambient agree with
those of LJ19, and we showed that turbulence, at least for {$Re\lesssim10^4$}, only
causes a small increase in the entrainment rate.\\

We found that thermals whose buoyancy increases with time, either because of
condensation  heating, or because of unstable ambient stratification, entrain
at \emph{higher} rates than dry thermals in a neutral ambient, in contrast with the 
findings of \cite{Morrison2021}, and also in contrast with some earlier numerical
and  experimental studies that found a decrease in entrainment upon heat addition {\cite{Bhat1996,Basu1999}}.
We showed that the influence of turbulence is in fact greater in these accelerating
thermals, and that the change in entrainment due to the combined action of buoyancy
and turbulence is greater than the change in entrainment due to either buoyancy or turbulence individually. 
{We argued that the intense small scale vorticity 
generated in moist thermals and the resulting scalar mixing that occurs
in the shear layers of thermals may play a role in this increase in the
entrainment rate. Therefore, simulations of moist thermals at Reynolds numbers significantly higher than the mixing transition Reynolds number of $Re=10^4$ \cite{DIMOTAKIS2000} may be essential to understand the role of turbulence in cumulus entrainment.}

Our results suggest several avenues for future research. The role that baroclinic torques and vorticity play in increasing the entrainment rate in moist thermals relative to dry thermals are the subject of an ongoing study. {Results from such studies may be useful in devising parameterisations for the entrainment that include terms for the interaction between turbulence and volumetric heating, building on previous parameterisations of the influences of shear and baroclinic effects \citep{kaye2008turbulent}.}
Furthermore, while we find that both condensation heating and unstable stratification lead to increased entrainment, the differences between these cases  (e.g. in the generation of small scale vorticity) may be studied to understand what role evaporative cooling at the edges of the thermal plays in entrainment. Such studies would build on previous studies that have addressed the role of shear turbulence in scalar mixing {\cite{Baker1984,heus2008}}. \\

Condensation heating, as we have seen, increases buoyancy while stable
stratification would decrease buoyancy. The combined effects of condensation 
heating and stable stratification, therefore, would depend on the relative 
magnitudes of these effects. This has bearing on the transition from shallow 
to deep convection \citep[e.g.][]{wu2009}.\\

Cumulus clouds are known to more closely resemble a series of thermals than
steady plumes. Fully resolved studies of how ensembles of thermals interact, 
therefore, would help better understand how clouds behave in the Earth's
atmosphere. Studies along these lines using LES have recently
\citep[][]{Morrison2020,Peters2020} been conducted. Such  studies could pave the way towards more reliable weather and climate simulations.\\
\\

\section*{Appendix A: Calculation of Thermal Volume} \label{sec:AppendixA}
As noted in section \ref{sec:tracking}, the volume of the thermal has to be 
consistently defined before the entrainment rate can be calculated. Following
\cite{Romps2015a,Lecoanet2019,Morrison2021}, we use the volume bounded by the
dividing streamline $\psi=0$ in a frame of reference moving with the thermal.
This therefore requires the velocity of the thermal to first be calculated.
The following are the steps involved.

(i) First, we azimuthally average the instantaneous axial and radial velocity and temperature to obtain $\bar{w}(r)$, $\bar{u}_r(r)$, and $\bar{\theta}(r)$ respectively.\\
(ii) These azimuthal averages are used to find the location of the centroid 
$z_{ct}$ of the thermal
\begin{equation}
    z_{ct}=\frac{\sum_{{\bar{w}^*}>0.6}{\bar {w}^* z}}{\sum_{{\bar{w}^*}>0.6} \bar{w}^*}, 
    \label{eq:zct}
\end{equation}
where $\bar{w}^* = \bar{w} / w_{max}$ with $w_{max}$ the maximum flow velocity in 
the domain. \\
(iii) The velocity of the thermal is computed as 
\[
w_{th} =\frac{dz_{ct}}{dt},
\] 
where $z_{ct}$ is computed at intervals of $\Delta t = 0.4$. As noted in 
\cite{Glezer1990,Lecoanet2019}, the volume of the thermal is sensitive to the 
$w_{th}$. We therefore perform a third order Savitsa-Golay curve fit to the $z_{ct}$
date before computing $w_{th}$\\
(iv) The streamfunction $\psi$ is obtained using the flow velocity in the frame of reference of the thermal, $(\bar{u}_r, \bar{w} - w_{th})$, 
\begin{equation}
    \partial_r\psi\ =\ 2\pi r(w-w_{th})\ \textrm{ and } \ \partial_z\psi\ =\ 2\pi r(u_r)
\end{equation}
and the dividing streamline $\psi =0$. In some cases, the method identifies the more
than one closed loop, and these have to be manually removed.\\
(v) The thermal volume is then given by 
\[
V = \int \pi r^2(z) dz
\]
where the integral is calculated using composite Simpsons rule.\\

\begin{figure}
     \centering
     \begin{subfigure}[b]{0.32\textwidth}
         \centering
         \includegraphics[width=\textwidth]{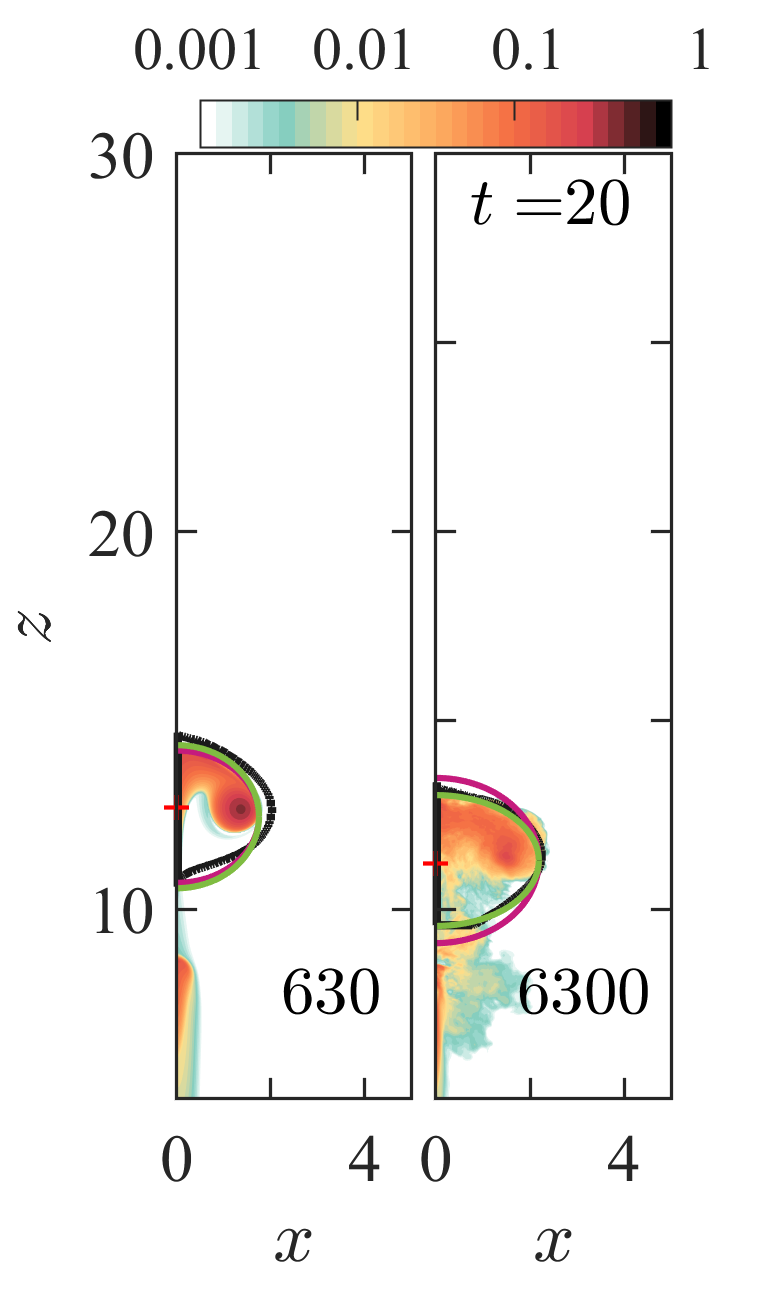}
     \end{subfigure}
     \hspace{0.1cm}
        \begin{subfigure}[b]{0.32\textwidth}
         \centering
         \includegraphics[width=\textwidth]{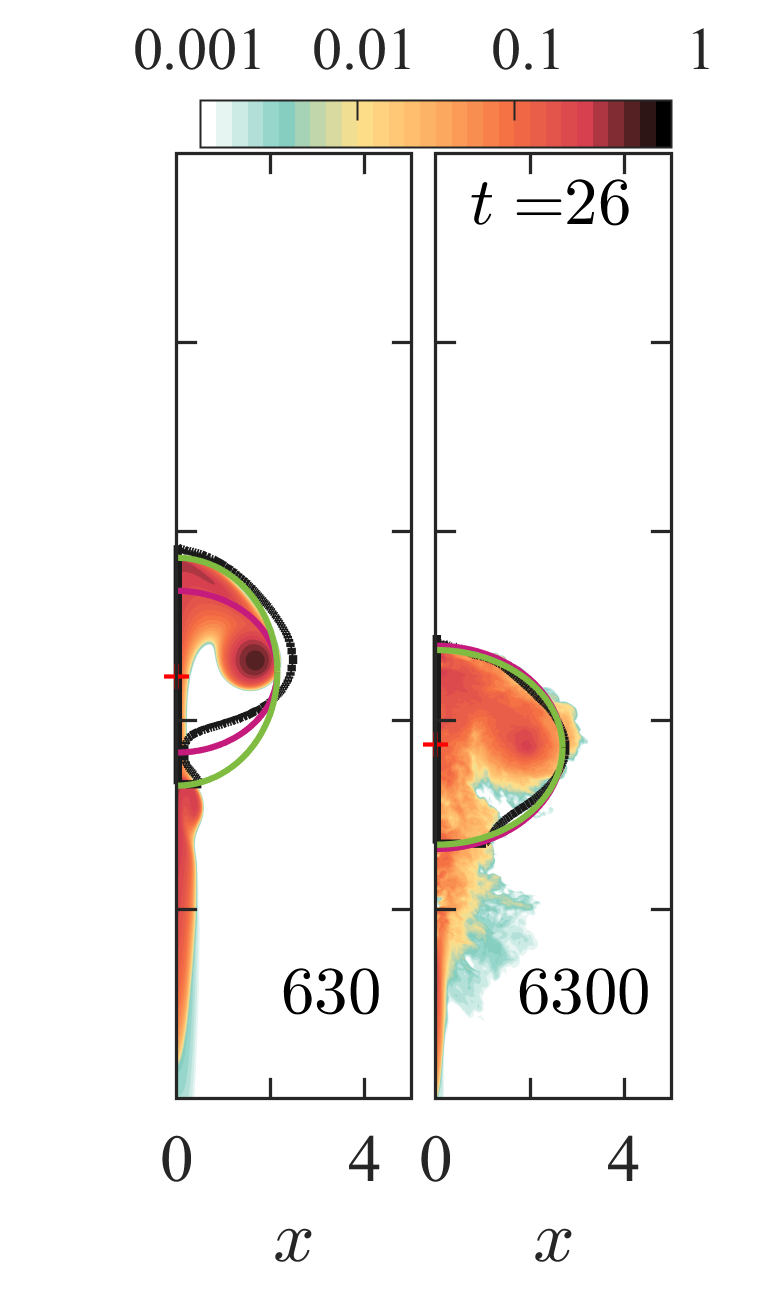}
     \end{subfigure}
    \hspace{0.1cm}
     \begin{subfigure}[b]{0.32\textwidth}
         \centering
         \includegraphics[width=\textwidth]{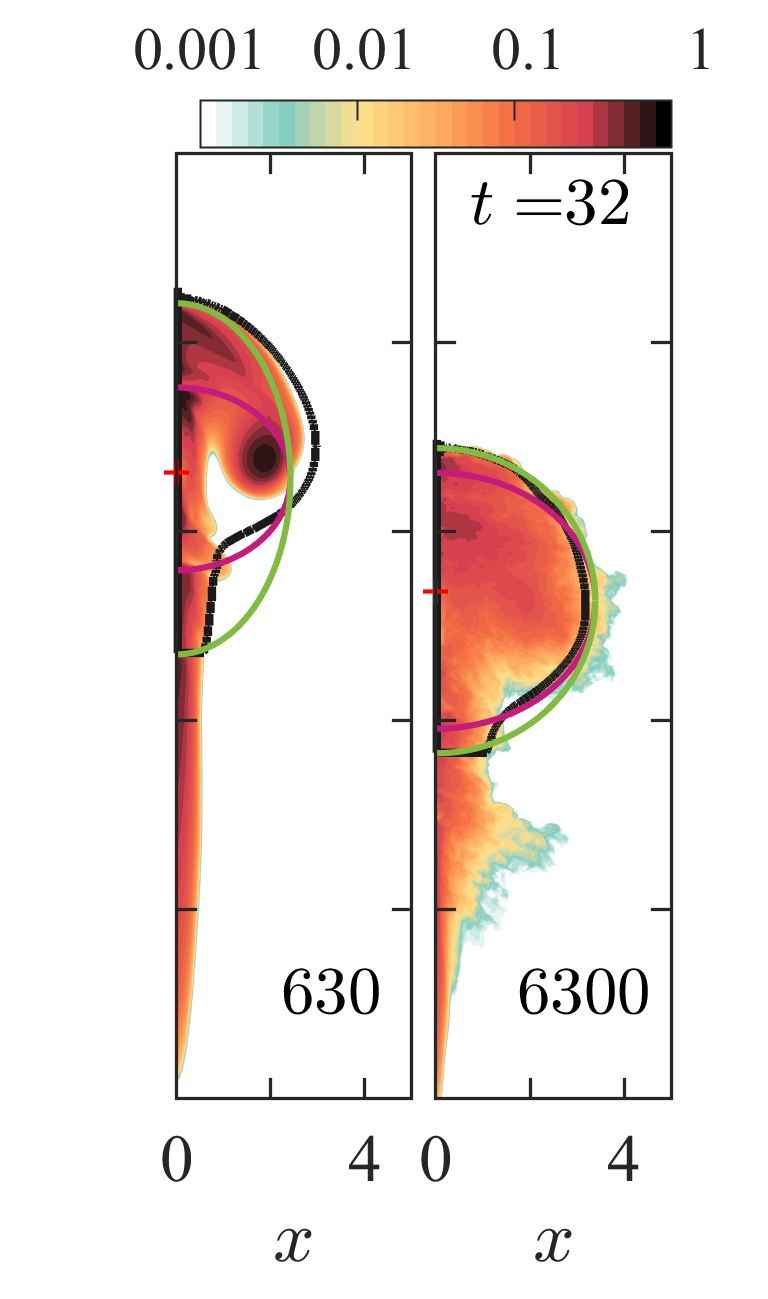}
     \end{subfigure}
     \caption{Identification of coherent thermals using the steps from Appendix A.
     The parameters are the same as in figure \ref{fig:moist_unstrat_u}.
     The dividing streamline ($\psi=0$, black curve) delineates the volume of the thermal from the ambient. The dividing streamline is superimposed on filled contours of the azimuthally averaged temperature, showing that the using the streamfunction allows for nearly autonomous boundary detection, with minimal manual intervention. The pink curve shows a spherical volume drawn with the thermal centroid as its centre and radius $R$ (defined in section \ref{sec:tracking}). The green curve is a spheroidal volume with the distance from thermal top to the centroid as its major axis and $R$ as its minor axis. These three methods, variously used in the literature \citep[e.g.][]{HernandezDeckers2016,HernandezDeckers2018,Lecoanet2019,Morrison2021} predict similar thermal volumes for the turbulent case while the differences are noticeable in the laminar case.}
     \label{fig:psi_moist_unstrat}
\end{figure}

This method is compared with two simpler methods that only involve locating
the centroid of the thermal and the top of the thermal in figure \ref{fig:psi_moist_unstrat}.

\section*{Appendix B: The role of ambient vapour in moist thermals}
Consider a saturated parcel of air of volume $V_0 = 1$ at an altitude $z_0=0$ with $\theta=0$ such that the buoyancy is $B_0 = 0$ rising adiabatically--i.e. without entrainment--to an altitude $z_2$. The conservation of energy requires
\be
\theta + L_2 r_s(z=0) = L_2 = \theta(z=z_2) + L_2 r_s(z=z_2),
\ee
giving the adiabatic temperature $\theta_{ad} = L_2 (1 - r_s(z=z_2)) \geq 0$, and the resulting buoyancy, 
\be
B_{ad} = (L_2 - r^0) \left( 1 - r_s(z_2) \right) + r^0 \chi (1-s_\infty) r_s(z_2)  \geq 0
\ee

Consider now a case where the parcel entrains an equal mass (or, in the Boussinesq approximation, an equal volume) of ambient air from an altitude $z_1$ before rising to $z=z_2$. The conservation of energy again gives
\be
\theta^* + L_2 r_s(z_2) = L_2 + s_\infty L_2 r_s(z_1),
\ee
giving 
\be
\theta^* = L_2 \left( \frac{1 + s_\infty r_s(z_1)}{2} - r_s(z_2) \right),
\ee
and
\be
B^* = (L_2 - r^0) \left[ \frac{1+s_\infty r_s (z_1)}{2} - r_s(z_2) \right] + r^0 \chi \left[(1-s_\infty) r_s(z_2) \right]  \geq 0
\ee
The latter expressions show that for sufficiently large $s_\infty$
\begin{enumerate}
    \item $\theta^* \geq 0$, and thus $B^* \geq 0$; and
    \item $B^* \leq B_{ad}$.
\end{enumerate}

\section*{Acknowledgements}
SR is supported under the Swedish Research Council grant no. 638-2013-9243. The Nordic Insitution for Theoretical Physics (Nordita) is partially supported by Nordforsk. The support and the resources provided by PARAM Yukti Facility under the National Supercomputing Mission, Government of India at the Jawaharlal Nehru Centre for Advanced Scientific Research (JNCASR) Bangalore are gratefully acknowledged. We are grateful to Dr. Akhilesh Prabhu for discussions and help with the code, and to Prof. Rama Govindarajan for comments on the manuscript. \tb{We are grateful to the late Prof. Roddam Narasimha, for the original idea for this work.}

\bibliographystyle{unsrt}

\end{document}